\documentclass[10pt, letterpaper]{article}
\usepackage[top = 0.85in, left = 2.25in, footskip = 0.75in]{geometry} 
\usepackage[shortlabels]{enumitem} 
\usepackage{amsmath, amssymb}
\usepackage{mathtools} 
\usepackage{changepage}

\usepackage{textcomp, marvosym}

\usepackage{cite}

\usepackage{nameref, hyperref}

\usepackage[right]{lineno}

\usepackage[nopatch = eqnum]{microtype}
\DisableLigatures[f]{encoding = *, family = *}

\usepackage[table]{xcolor}

\usepackage{array}

\newcolumntype{+}{!{\vrule width 2pt}}

\newlength\savedwidth

\usepackage{ragged2e} 
\justifying 
\usepackage[aboveskip = 2pt, labelfont = bf, labelsep = period, singlelinecheck = off]{caption} 

\makeatletter
\renewcommand{\@biblabel}[1]{\quad#1.}
\makeatother

\usepackage{lastpage, fancyhdr, graphicx}
\usepackage{epstopdf}
\fancyheadoffset[L]{2.25in}
\fancyfootoffset[L]{1.75in} 
\begin{document}
\vspace*{0.125in}


\begin{flushleft}

{\Large
    \textbf\newline{AI-powered simulation-based inference of a genuinely spatial-stochastic model of early mouse embryogenesis}
}
\newline
\\
Michael A. Ramirez-Sierra\textsuperscript{1,2},
Thomas R. Sokolowski\textsuperscript{1*}
\\
\bigskip
\textbf{1} Frankfurt Institute for Advanced Studies (FIAS), Ruth-Moufang-Straße 1, 60438 Frankfurt am Main, Germany
\\
\textbf{2} Goethe-Universität Frankfurt am Main, Faculty of Computer Science and Mathematics, Robert-Mayer-Straße 10, 60054 Frankfurt am Main, Germany
\\
\bigskip

* ramirez-sierra@fias.uni-frankfurt.de

\end{flushleft}



\section*{Abstract}

Understanding how multicellular organisms reliably orchestrate cell-fate decisions is a central challenge in developmental biology. This is particularly intriguing in early mammalian development, where early cell-lineage differentiation arises from processes that initially appear cell-autonomous but later materialize reliably at the tissue level. In this study, we develop a multi-scale, spatial-stochastic simulator of mouse embryogenesis, focusing on inner-cell mass (ICM) differentiation in the blastocyst stage. Our model features biophysically realistic regulatory interactions and accounts for the innate stochasticity of the biological processes driving cell-fate decisions at the cellular scale. We advance event-driven simulation techniques to incorporate relevant tissue-scale phenomena and integrate them with Simulation-Based Inference (SBI), building on a recent AI-based parameter learning method: the Sequential Neural Posterior Estimation (SNPE) algorithm. Using this framework, we carry out a large-scale Bayesian inferential analysis and determine parameter sets that reproduce the experimentally observed system behavior. We elucidate how autocrine and paracrine feedbacks via the signaling protein FGF4 orchestrate the inherently stochastic expression of fate-specifying genes at the cellular level into reproducible ICM patterning at the tissue scale. This mechanism is remarkably independent of the system size. FGF4 not only ensures correct cell lineage ratios in the ICM, but also enhances its resilience to perturbations. Intriguingly, we find that high variability in intracellular initial conditions does not compromise, but rather can enhance the accuracy and precision of tissue-level dynamics. Our work provides a genuinely spatial-stochastic description of the biochemical processes driving ICM differentiation and the necessary conditions under which it can proceed robustly.


\section*{Author summary}

Our study presents a spatial-stochastic model for the gene regulatory network (GRN) and the signaling pathway governing cell-fate differentiation during early mouse embryogenesis, specifically at the blastocyst stage. Departing from biophysics-based models of gene regulation, we perform stochastic simulations of the biochemical processes driving early mouse embryogenesis both at the cell and tissue level. Combining these simulations with state-of-the-art AI-aided inference techniques, we successfully parameterize our model, replicating key experimental observations and providing mechanistic insights into the biochemical interactions giving rise to them. Thanks to the stochastic nature of our approach, we quantify the high robustness of ICM specification to various kinds of noise. Altogether, we provide a deeper understanding of the intricate mechanisms driving early cell-fate decisions in mouse embryogenesis, highlighting the synergy of local cellular and broader tissue-scale interactions that shape development.


\section*{Introduction}

The process of maintaining cellular plasticity while concurrently directing functional differentiation of cells is a cornerstone paradigm in early biological development. A fundamental question connected to this is how a complex multicellular organism can robustly emerge from a single cell, despite the inherent stochasticity (or noise) in the biochemical processes driving cell specification.

In mammalian development, the cell signaling and fate specification dynamics during the preimplantation stage of mouse embryogenesis serve as key catalysts for understanding this core paradigm. These processes have been extensively investigated through both experimental and theoretical approaches \cite{cang_multiscale_2021, krupinski_simulating_2011, rossant_blastocyst_2009, li_maternal_2010, tosenberger_multiscale_2017, tosenberger_computational_2019, habibi_transcriptional_2017, arias_molecular_2013, herberg_computational_2015, miyamoto_pluripotency_2015, nissen_four_2017, bessonnard_icm_2017, stanoev_robustness_2021, robert_initial_2022, mathew_mouse_2019, liebisch_cell_2020, zhu_synthetic_2022}, and rely on dynamic cross-interactions between gene expression, molecular signaling at the tissue level, and mechanical cues essential for tissue remodeling \cite{zhu_principles_2020, plusa_common_2020, zernicka-goetz_cleavage_2005, saiz_coordination_2020, ryan_lumen_2019, saiz_growth-factor-mediated_2020, simon_making_2018, allegre_nanog_2022, meilhac_active_2009, chen_tracing_2018, schultz_oocyte--embryo_2018, niwayama_tug--war_2019, fischer_transition_2020, chan_integration_2020, yanagida_cell_2022}.

The preimplantation stage in mouse embryos involves two pivotal events transforming the zygote into a blastocyst composed of three distinct cell types \cite{rossant_blastocyst_2009, bessonnard_gata6_2014, allegre_nanog_2022, simon_making_2018, schrode_regulation_2015}. Initially, cell segregation results in the formation of the trophectoderm (TE), which contributes to the embryonic part of the placenta, and the inner cell mass (ICM), an uncommitted bipotent progenitor tissue. Subsequently, a second cell-fate decision within the ICM leads to the differentiation of epiblast (EPI) and primitive endoderm (PRE) tissues. The EPI, marked primarily by NANOG expression, is pluripotent and gives rise to all embryo-proper tissues. In contrast, the PRE, marked primarily by GATA6 expression, contributes to the formation of extraembryonic supporting tissues. Notably, EPI cells secrete FGF4, a signaling ligand that promotes PRE cell specification. The FGF/ERK signaling pathway therefore plays a crucial role in regulating the balance between EPI and PRE cells. Interestingly, FGF4 acts within an external feedback loop that ultimately inhibits its own production, downregulating it in PRE cells \cite{schroter_fgfmapk_2015, bessonnard_icm_2017, saiz_growth-factor-mediated_2020}. The mechanisms underlying the precise and timely distribution of FGF4 across the ICM, critical for appropriate cell lineage specification, remain elusive \cite{raina_cell-cell_2021, raina_intermittent_2022}.

For successful embryogenesis, the TE, EPI, and PRE lineages in the blastocyst must be segregated in specific proportions and within a narrow developmental time window \cite{plachta_oct4_2011, bessonnard_icm_2017, allegre_nanog_2022}. This specification occurs over approximately 48 hours between embryonic day (E) 2.5 and E4.5 \cite{saiz_growth-factor-mediated_2020, bessonnard_icm_2017}. By E3.0, when blastocyst formation begins, ICM cells co-express NANOG and GATA6. In the further course, EPI and PRE cells emerge asynchronously from ICM cells in a spatially stochastic manner, influenced by the short-range FGF4 signal and regulated by auto- and paracrine feedback loops \cite{krawczyk_paracrine_2022, allegre_nanog_2022, raina_cell-cell_2021}. A mutually exclusive gene expression profile emerges, with EPI cells exhibiting high NANOG and low GATA6 levels, and vice versa in PRE cells.

This process is driven by a gene-regulatory network characterized by self-activation and mutual repression of NANOG and GATA6, following a common motif for generating multistable expression states \cite{burda_motifs_2011, sokolowski_mutual_2012, sokolowski_deriving_2023, majka_stability_2023, majka_stable_2023}, and is refined by the external FGF4 feedback. The spatio-temporal segregation of EPI and PRE lineages is subsequently coordinated through a mechanical cell-sorting process \cite{yanagida_cell_2022}. The final spatially ordered pattern thus is an emergent property of the differentiating tissue, in stark contrast to systems which employ morphogen gradients for high-precision system-wide coordination of developmental trajectories \cite{gregor_stability_2007, bollenbach_precision_2008, little_formation_2011, richards_spatiotemporal_2015, smith_role_2016, ellison_cellcell_2016, zagorski_decoding_2017, verd_dynamic_2017}, although layers of interacting genes downstream of the gradients can acquire self-organizing and scaling capabilities after their position-dependent activation \cite{vakulenko_size_2009, kicheva_coordination_2014, raspopovic_digit_2014, almuedo-castillo_scale-invariant_2018, verd_damped_2018, morales_embryos_2021, majka_stable_2023, nikolic_scale_2023}.

Experimental studies have sought to characterize the dynamics and biophysical parameters of ICM specification, focusing on internal and external developmental cues, molecular mechanisms underpinning gene transcription and mRNA translation, and key signaling pathway components \cite{ochiai_stochastic_2014, ochiai_genome-wide_2020, habibi_transcriptional_2017, thompson_extensive_2022, krawczyk_paracrine_2022}. Nevertheless, this remains challenging, as the involved biochemical species are present at low mRNA and protein abundance, leading to significant biological noise. Here biological noise is categorized as either intrinsic (related to the discrete and stochastic nature of biochemical reactions and molecular transport within cells) or extrinsic (referring to external fluctuations affecting all cells). This noise strongly impacts the reliability of developmental dynamics, particularly during early development \cite{gonze_modeling-based_2018, tkacik_many_2021, lin_stochastic_2018, lin_central_2016, vandevenne_rna_2019, pantazis_transcription_2012, bezeljak_stochastic_2020}, and hampers experimental measurements at the same time.

\newpage 

For these reasons, mathematical and computational models have established themselves as potent tools capable of providing mechanistic insights into biochemical noise control strategies. However, existing models of EPI-PRE specification primarily employ deterministic methods, treating noise as a secondary feature \cite{robert_initial_2022, bessonnard_gata6_2014, tosenberger_multiscale_2017, stanoev_robustness_2021, cang_multiscale_2021}. These models do not fully capture the stochastic nature of transcription, translation, biochemical signaling, among other processes, and therefore may not accurately reflect the impact of noise on early embryo development.

More recently, deep-learning-based approaches have been successfully used for both pattern recognition and reconstruction of ICM organoid data, employing experimental and synthetic datasets for training \cite{dirk_recognition_2023}. Although these approaches have good predictive power for both cell-fate spatial composition and determination, they do not reveal full mechanistic insights into the gene regulatory interactions that orchestrate ICM patterning.

In this study, we investigate cell specification during blastocyst formation in the early mouse embryo under truly stochastic conditions. In order to elucidate mechanisms that enable robust and replicable cell-type proportioning in the ICM, we have constructed a spatial model that allows for exact simulation of the stochastic dynamics of key biochemical species in the developing ICM. This model incorporates three distinct diffusive-signaling modes between neighboring cells: autocrine, paracrine, and intermembrane ligand-exchange (akin to juxtacrine) signaling via FGF4.

Departing from traditional event-driven algorithms for simulating the Reaction-Diffusion Master Equation (RDME) in a spatial context, we devised a scheme for simulating the stochastic evolution of core species governing early ICM development at the cellular and tissue scales. This approach leverages the Simulation-Based Inference (SBI) framework, combining our spatial-stochastic simulator with an advanced AI-based inference technique: the Sequential Neural Posterior Estimation (SNPE) algorithm \cite{greenberg_automatic_2019, deistler_truncated_2022}. This allowed us to perform millions of individual stochastic simulations of our spatial system and to use these data to infer parameter sets that align with desired system behavior.

By integrating biochemically realistic spatial simulations with SBI, we learned parameters that accurately replicate key experimental observations of mouse ICM cell-lineage differentiation, including its temporal dynamics based on reported lifetimes of the involved biochemical species. Successful parametrization of our explicitly stochastic model enabled us to explore the effects of various sources and degrees of system perturbations, providing insights into the potential role of noise in the functional development of mouse blastocyst cell populations.

Our findings suggest that the ICM system, naturally biased towards a default ``raw'' state, requires tissue-level coordination of fate differentiation. Specifically, we demonstrate that: (1) early emergence of the EPI lineage is essential for stimulating the PRE fate; (2) intercellular communication enhances the robustness of ICM fate differentiation compared to a purely cell-autonomous process; (3) the target cell-lineage proportions arise independently of the number of cells in the system; (4) moderate variability in cellular initial conditions can enhance the accuracy of establishing the proper cell fate ratio; and (5) ICM plasticity and its sensitivity to exogenous FGF4 are coordinated in time.

Our study not only highlights the early mouse embryo's resilience to biological noise, but also underscores the existence of critical windows in developmental timing that dictate cell plasticity and responsiveness to neighboring signaling molecules. It exemplifies that AI-driven simulation-based inference can be instrumental in uncovering mechanistic details of system-wide coordination of noise control in highly stochastic developmental systems, applicable beyond the specific embryonic system studied here.

\newpage 


\section*{Results}

\graphicspath{{./Manuscript_Graphics/}} 


\subsection*{Multiscale spatial-stochastic model of mouse blastocyst development: from ICM progeny to EPI and PRE lineages}

We constructed a biophysics-based model of mouse embryo blastocyst formation that features a stochastic-mechanistic description of its intracellular gene-regulatory network and its intercellular diffusion-based signaling processes. It focuses on the differentiation of epiblast (EPI) and primitive endoderm (PRE) lineages from the inner cell-mass (ICM) progenitor population. This differentiation is driven by mutual regulatory interactions between the NANOG and GATA6 genes which act as primary markers of the EPI and PRE fates, respectively. It also accounts for their interactions with FGF4 via FGF receptors and the ERK signaling cascade, which implements a sensing mechanism for external FGF4; the FGF4 proteins are secreted by cells that commit to the EPI fate. We describe our model and simulation methods in detail in section \nameref{subsection:computational_model} of the Methods part.

For computational feasibility, our model represents the developing tissue as a static two-dimensional (2D) lattice. This structural representation preserves the essential characteristic of the mouse blastocyst as a densely packed assembly of pluripotent cells that can spatially communicate among each other. Similar geometric assumptions have been successfully utilized in previous ICM models \cite{tosenberger_multiscale_2017, stanoev_robustness_2021, robert_initial_2022}. Each voxel of the lattice symbolizes an individual embryonic cell, providing an accessible foundation to analyze cell-cell communication modes (for more details, see \nameref{subsubsection:tissue_scale_model}).

Derived from established Reaction-Diffusion Master Equation (RDME) simulation schemes, our approach computes realistic molecular count time series using accurate lifetimes for essential biochemical species (detailed in \nameref{subsubsection:cell_scale_model}). This results in an authentic reproduction of the temporal dynamics of ICM fate specification, as seen in Fig~\ref{fig1}. We opted for an event-driven scheme in order to maximize computational efficiency of our simulations, which is a necessary precondition for applying the Simulation-Based Inference (SBI) framework to it, as described next.

\subsection*{Exploring model parameter space via SBI}

In computational modeling of complex biological systems, inferring parameter sets that reproduce experimental observations is a key challenge. This usually requires specific domain knowledge, as the selection of an appropriate parameter search method is chiefly influenced by the particular properties of the problem at hand \cite{baker_mechanistic_2018}. One significant hurdle in developmental biology is the lack of a versatile, general-purpose inference method suitable for high-dimensional stochastic models, which typically require large and comprehensive datasets with fine-grained resolution. This expands both in terms of the features measured (e.g., gene expression levels at the single-cell scale rather than bulk measurements) and the temporal resolution (e.g., time-series data capturing dynamics at all relevant scales). However, most experimental studies are unable to simultaneously measure all system variables required for such in-depth mechanistic representation. To bridge this gap, Biologically-Informed Neural Networks (BINNs) and Simulation-Based Inference (SBI) frameworks have emerged as powerful strategies addressing several of these modeling and inferential challenges \cite{cranmer_frontier_2020, lagergren_biologically-informed_2020, seyboldt_latent_2022, perez_efficient_2022}. Particularly advantageously, SBI allows to learn multidimensional parameter sets that comply with a prescribed behavior formulated in terms of lower-dimensional utility functions, thus mitigating the limitations posed by the scarcity of detailed quantitative data.

\begin{adjustwidth}{-1.75in}{0in} 
\includegraphics[width = 7.5in, height = 8.75in]{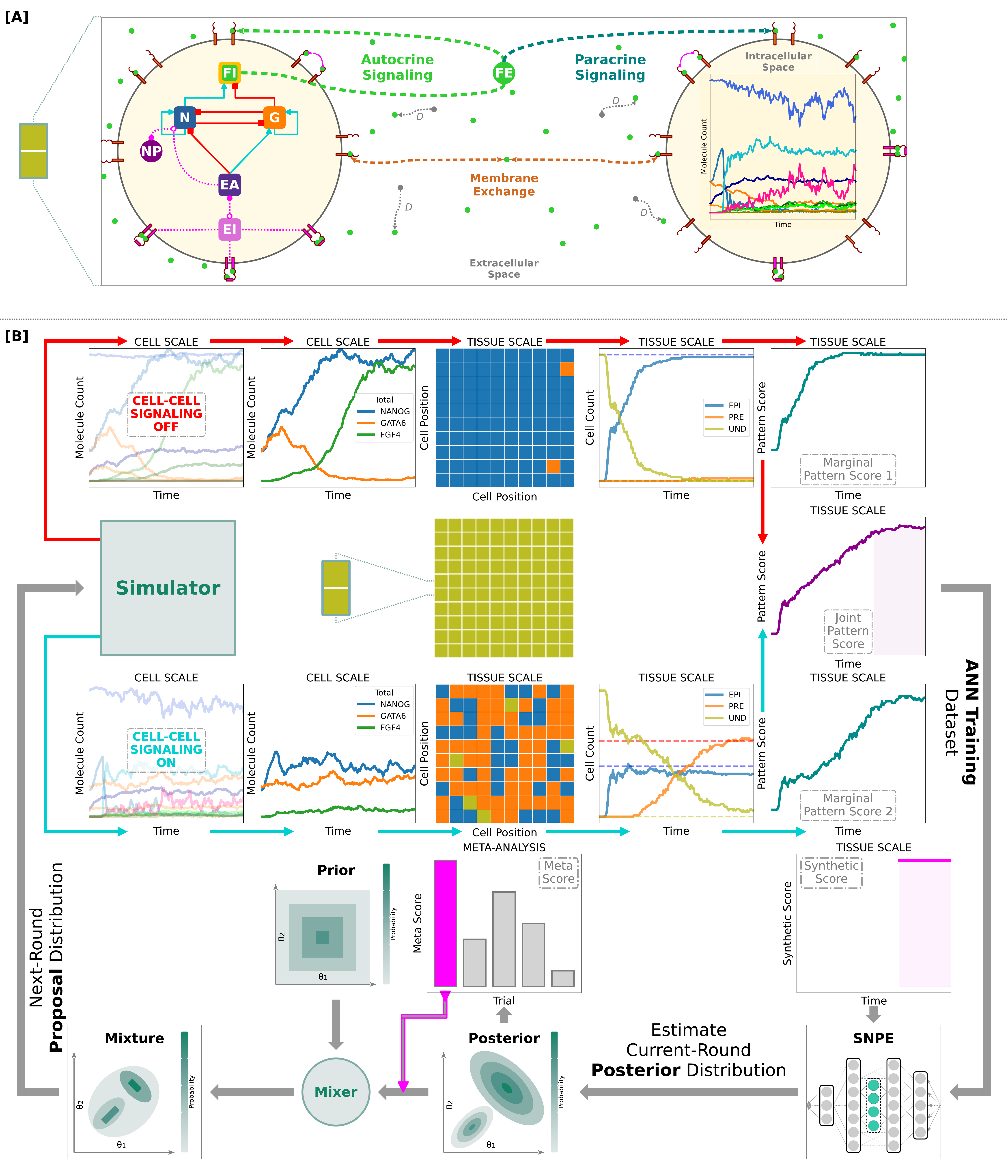} \centering 
\end{adjustwidth} 
\begin{figure}[hpt!]
\begin{adjustwidth}{-1.75in}{0in} 
\caption{{\bf Workflow summary: GRN motif, cell signaling model, and inference framework.} {\bf [A]} The developing ICM is represented by a static spatial lattice of biochemical reaction volumes (cells) coupled via FGF4. Each cell contains a core gene-regulatory network (GRN) featuring mutual repression between the genes \textit{Nanog} (N) and \textit{Gata6} (G), and their self-activation. Both N and G regulate the expression of internal FGF4 (FI). External FGF4 (FE) can either diffuse to a neighboring cell (paracrine signaling), bind to the membrane of the origin cell (autocrine signaling), or be exchanged between neighboring cell membranes. FGF receptors transmit the sensed FGF4 signal back to the core GRN by activating ERK (EI~$\leftrightarrow$~EA). {\bf [B]} Pipeline of data generation and analysis. Key stages of parameter inference (columns 1 through 5). Initially, all cells display the undifferentiated (UND) fate (row 2 column 3). Rows 1 and 3 show simulations without and with FGF4 signaling activated, respectively. All generated stochastic trajectories are processed by the same steps: (I) resampling data onto a regular time grid and calculating relevant system observables at cell scale (total NANOG, GATA6, and FGF4 levels); (II) determining the lineage for each cell at every time point (EPI, PRE, or UND); (III) summing up the corresponding total cell count for each fate at tissue scale; (IV) constructing the (joint) pattern score time series. The map between simulation parameters and resultant score time series is used for training a deep neural density estimator via the sequential neural posterior estimation (SNPE) algorithm (row 5 column 5), which directly estimates the parameter posterior distribution (row 5 column 3) conditioned on a target observation (row 4 column 5). Multiple posterior estimates are produced with the same training set, selecting the best learned distribution conditional on the target observation by analyzing a ``meta score'' distribution (row 4 col 3). This summary/meta score is calculated per posterior, and it relies on the maximum a posteriori (MAP) estimate of all the model parameters. In general, the next-round prior does not need to be the current-round posterior: it is plausible to obtain a well-informed next-round mixture distribution (row 5 col 1). Several iterations of the workflow are performed until the meta score surpasses an arbitrarily prescribed level. For additional information about this data-processing pipeline, please see \nameref{subsection:inference_framework}.}
\label{fig1}
\end{adjustwidth} 
\end{figure}

Here we use the SBI framework \cite{tolley_methods_2023, hashemi_amortized_2023, stillman_generative_2023} to establish a comprehensive parameter exploration workflow for our spatial-stochastic model of mouse ICM fate decisions. Notably, our approach does not depend directly on fitting quantitative experimental data. Instead, it primarily uses qualitative observations to reconstruct system behavior and infer parameter distributions complying with it in a holistic manner. At heart, our approach fuses a novel AI-based technique, specifically the Sequential Neural Posterior Estimation (SNPE) algorithm \cite{greenberg_automatic_2019, deistler_truncated_2022}, with concepts from classical inference and optimization strategies \cite{schnoerr_approximation_2017, franzin_landscape-based_2023}. The key steps of our approach are summarized by the following workflow:
\begin{enumerate}[(1)]
    \item Construct an objective function that quantitatively represents the system dynamics as a time-varying ``score'', indicating the progression towards a desired or ideal state. For the ICM model studied here, this function traces the deviation from the desired cell-fate ratio (see \nameref{subsubsection:score} for details).
    \item Simulate the model many times, each with a different parameter value vector sampled from a suitable prior distribution.
    \item Evaluate the objective function using the data obtained from these simulations.
    \item Train an artificial neural network (ANN) to learn the posterior distribution of the model parameters, effectively using the ANN as a surrogate for the simulator to approximate the mapping between parameter values and objective function scores.
    \item Define a target score time series that represents the ideal system behavior, using it as a synthetic observation.
    \item Estimate the maximum-a-posteriori (MAP) parameter value that aligns with the desired system behavior represented by the target score time series, using the posterior distribution predicted by the ANN conditioned on the given target score.
    \item Conduct multiple iterations starting from step (2) until the resulting score time series aligns with the target score time series.
\end{enumerate}
For a more detailed description of these steps, see Fig~\ref{fig1} and \nameref{subsection:inference_framework}. In a companion study \cite{ramirez-sierra_comparing_2024} (in preparation), we compare this SNPE-guided methodology against a strategy inspired by a classical optimization algorithm, specifically simulated annealing; we find that the AI-based approach surpasses its classical counterpart in performance, assuming equal allocation of computational resources. This superiority can be traced back to the innate flexibility of the ANN and the absence of a native parameter interpolation procedure in simulated annealing.

\subsubsection*{Reproduction of tissue-level features defines a score function}

We explored ICM differentiation guided by three pivotal characteristics which informed and constrained our parameter inference workflow.

The first key feature is the high reproducibility of the EPI and PRE cell-fate proportions in the blastocyst \cite{bessonnard_gata6_2014, simon_making_2018, saiz_coordination_2020}. Despite varying reports on its precise ratio, here we adopted a typically reported value of $2:3$ \cite{saiz_asynchronous_2016}. Note that minor deviations from this ratio are unlikely to significantly change the inferred posterior distributions or other findings of our model.

Secondly, blastocyst formation occurs within approximately 1.5-2 days of development, starting around E2.75 \cite{plusa_common_2020, saiz_coordination_2020, allegre_nanog_2022}. The EPI and PRE populations must reach and sustain the required fate proportions 8 to 12 hours before the end of the preimplantation period (around E4.75). This requirement serves as another optimization criterion for our model, regardless of the actual underlying cell-differentiation mechanism.

The third key feature is the role of spatial coupling via FGF4 in fate determination. In its absence, almost the entire ICM population assumes the EPI fate, which impedes the exit from naive cellular pluripotency \cite{bessonnard_icm_2017, thompson_extensive_2022}. Correspondingly, only few cells adopt the PRE fate, a phenomenon attributable to intrinsic stochasticity at the gene expression level. This highlights the critical importance of local cell-cell signaling for correct pattern formation, and the inherent bias of the ICM population towards a naive pluripotent state.

All these key features together were incorporated into the design of the ``score function'' used in our SBI approach, in order to infer parameter distributions complying with them. The mathematical definition of the score function is described in Methods section \nameref{subsubsection:score}.

\subsubsection*{Comparative analysis of parameter sets for complementary wild-type and mutant models via SBI}

Exploiting our SBI workflow, we successfully identified two distinct sets of model parameters that shed light on crucial aspects of the underlying biological problem. These two parameter sets correspond to two complementary models: one representing a wild-type system with functional cell-cell communication via FGF4/ERK, referred to as ``Inferred-Theoretical Wild-Type'' or ITWT, and another mimicking a mutant system without FGF/ERK signaling (more precisely, without cell-cell FGF4 signaling), referred to as ``Reinferred-Theoretical Mutant'' or RTM. The RTM is a direct adaptation from the ITWT, but with reinferred core GRN interaction parameters.

Fig~\ref{fig2} summarizes the key differences between the ITWT and RTM systems, contrasting the respective inferred parameter values, i.e. the maximum-a-posteriori-probability (MAP) estimates, for both the central intracellular GRN components (top row) and the extracellular signaling topology (bottom row). The most notable difference between the ITWT and RTM models is the reversal of the relationship between the self-activation thresholds of \textit{Nanog} and \textit{Gata6}. For the ITWT, the half-saturation threshold of \textit{Nanog} self-activation (\textit{Nanog}\_NANOG) is lower/stronger than the half-saturation threshold of \textit{Gata6} self-activation (\textit{Gata6}\_GATA6). However, both models exhibit a well-balanced mutual repression between \textit{Nanog} and \textit{Gata6}, albeit with varying interaction strengths. We present a complete overview of the full model parameter distributions inferred in this study, and a first approximation of the model parameter sensitivities, in the \nameref{section:supporting_information}, Figs \ref{fig101} and \ref{fig102}.

\begin{adjustwidth}{-1.75in}{0in} 
\includegraphics[width = 7in, height = 7in]{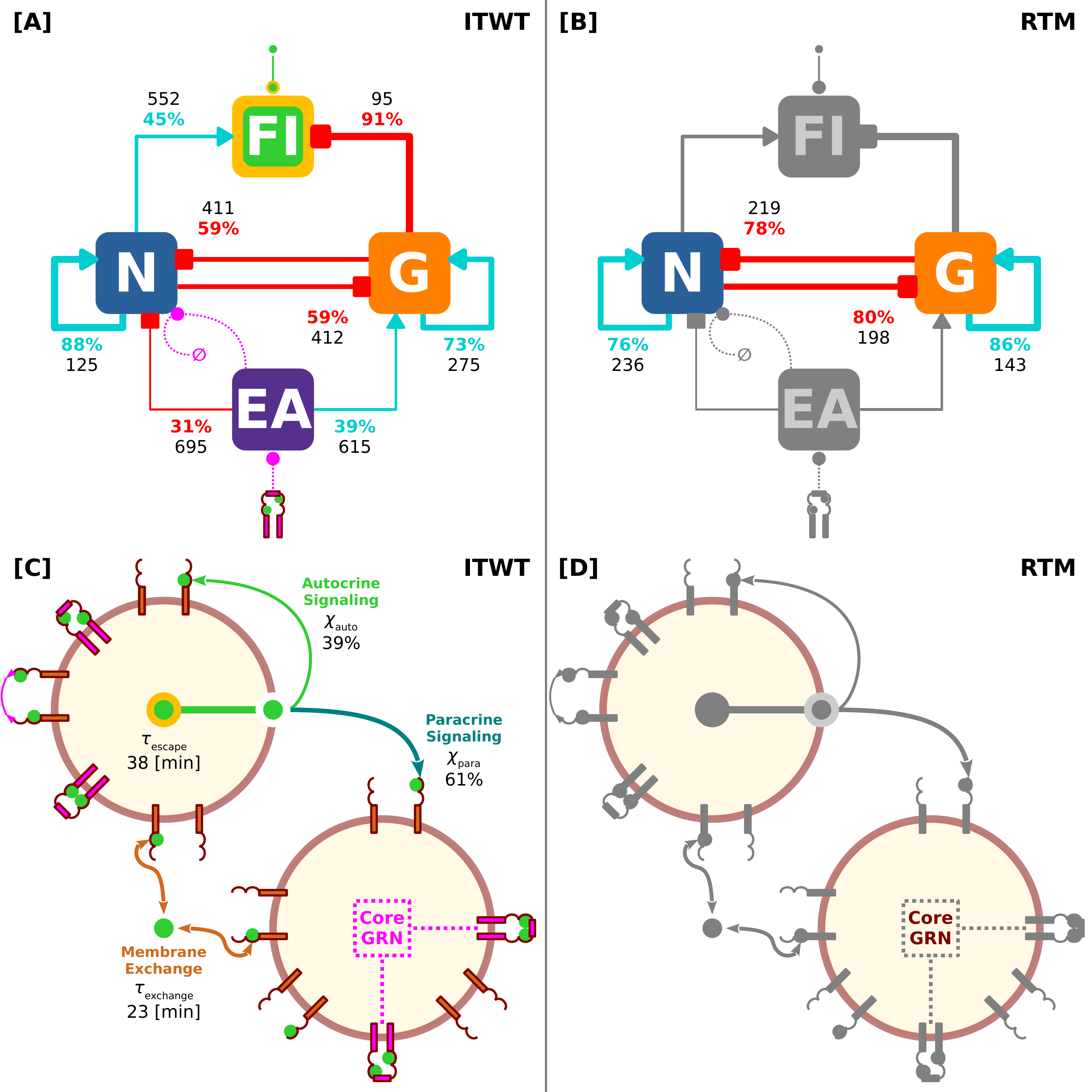} \centering 
\end{adjustwidth} 
\begin{figure}[hpt!]
\begin{adjustwidth}{-1.75in}{0in} 
\caption{{\bf Summary of inferred central GRN and signaling model parameters (ITWT versus RTM).} Two parameter sets were learned for distinct systems capable of recapitulating the correct final ratio between the emerging ICM lineages (EPI and PRE). The wild-type-like system ITWT (left column) with functional cell-cell signaling via FGF4 (green color), and the mutant-like system RTM (right column) for which FGF4 signaling was inhibited. Black-colored numbers show the inferred values (MAP estimates) of the relevant parameters defining the central intracellular GRN (top row) and the extracellular signaling topology (bottom row) components. Red- and cyan-colored percentage values indicate relative interaction strengths calculated via the formula $\nu = 100(\omega-\theta)/\omega$. Here $\omega$ is the length of the respective parameter range used for the inference scheme, and $\theta$ is the MAP estimate of the particular parameter. Cyan and red colors refer to gene activation and repression, respectively.}
\label{fig2}
\end{adjustwidth} 
\end{figure}

Both models correctly recapitulate the final ratio (of cell counts) between the two emerging ICM lineages (EPI and PRE) of the fully-formed mouse blastocyst. However, the ITWT relies on a cell non-autonomous mechanism in which spatial coupling via FGF4 is essential for accurate and precise lineage specification. In contrast, the RTM relies on a cell autonomous mechanism, which is purely probabilistic. We explore this mechanistic difference and its consequences in more detail below (sections \nameref{subsection:intercellular_communication} and \nameref{subsection:robust_cell_fate_proportions}). In the following, however, we focus on the ITWT, as it incorporates the necessary FGF4 signaling confirmed by experiments.

\subsection*{EPI and PRE fates emerge robustly at tissue scale in spite of high single-cell variability}

Our simulations were initiated with all cells in an undifferentiated (UND) state with a balanced distribution of cellular resources. At cell scale, the first part of a typical stochastic trajectory reveals the dynamic interplay between NANOG and GATA6 proteins, with FGF4 protein expression being adjusted in response to the levels of these two pivotal regulators. After about 12 hours, on average, the initial symmetry between NANOG and GATA6 is broken, as their proteins embark on divergent expression paths; this is exemplified for the case of high GATA6 and low NANOG final expression in Fig~\ref{fig3}[A, B]. As time progresses, the individual cells predominantly commit to a PRE or an EPI fate, accompanied by a decrease or increase in FGF4 expression, respectively. This differentiation is clearly depicted at the tissue scale, with cells categorically aligning into one of three fates: UND, EPI, or PRE (Fig~\ref{fig3}[B]).

The variability in the expression profiles of NANOG, GATA6, and FGF4 at the cell scale is significant, as shown by the large standard deviation in molecular counts across different cells and simulations (Fig~\ref{fig3}[C]). However, this cell-scale variability does not translate into high variability of the cell fate ratio at the tissue scale, which instead exhibits remarkable robustness, with a significantly lower standard deviation (Fig~\ref{fig3}[D]). This observation alone underscores the system's ability to integrate and manage cellular variability, ensuring consistent and reliable outcomes in the differentiation process across the tissue.

\subsection*{EPI cells precede and are necessary for specification of PRE cells}

Transitioning from this foundation of robust differentiation, the model predicts an early and critical onset of the EPI lineage, occurring around 2 hours into the simulated trajectories, with the PRE lineage emerging only about 12 hours later. The emergence of the first EPI cells is tightly clustered within a narrow time frame between 2 and 3 hours of development (gray region in Fig~\ref{fig4}[A]); this temporal behavior is consistent across a wide range of initial conditions (detailed in \nameref{subsection:computational_experiments}). Following this timely commitment, the EPI population expands rapidly, attaining, on average, 75\% of its target proportion within the initial 4 hours (see Fig~\ref{fig3}[D]). This leads to elevated FGF4 levels among the newly specified EPI cells, enabling the distribution of FGF4 across the ICM. Only then PRE cells begin to appear in significant numbers within a broader time window, ranging from 8 to 17 hours (refer to the gray region of Fig~\ref{fig4}[B]), implying an approximate 7-hour delay between the emergence of the first EPI and PRE cells, in line with recent experiments \cite{allegre_nanog_2022}. Subsequently, the PRE cell population gradually increases, reaching its target proportion at around the 40-hour mark (Fig~\ref{fig3}[D]).

Nascent PRE cells are coordinated by EPI cells via differential expression of \textit{Fgf4}, and the expression profile of \textit{Gata6} is a principal indicator of the onset of PRE cells. This coordination is controlled by nuances in fate-specific FGF4 distributions. Such nuances not only control the emergence of PRE lineage, but they are also key for EPI- and PRE-fate maintenance \cite{bessonnard_icm_2017, raina_cell-cell_2021}. Tight control of \textit{Nanog} expression in EPI cells is a requirement for escaping naive pluripotency during the implantation stage \cite{abranches_stochastic_2014, xenopoulos_heterogeneities_2015, kale_nanog-perk_2022}.

\begin{adjustwidth}{-1.75in}{0in} 
\includegraphics[width = 6.25in, height = 6.25in]{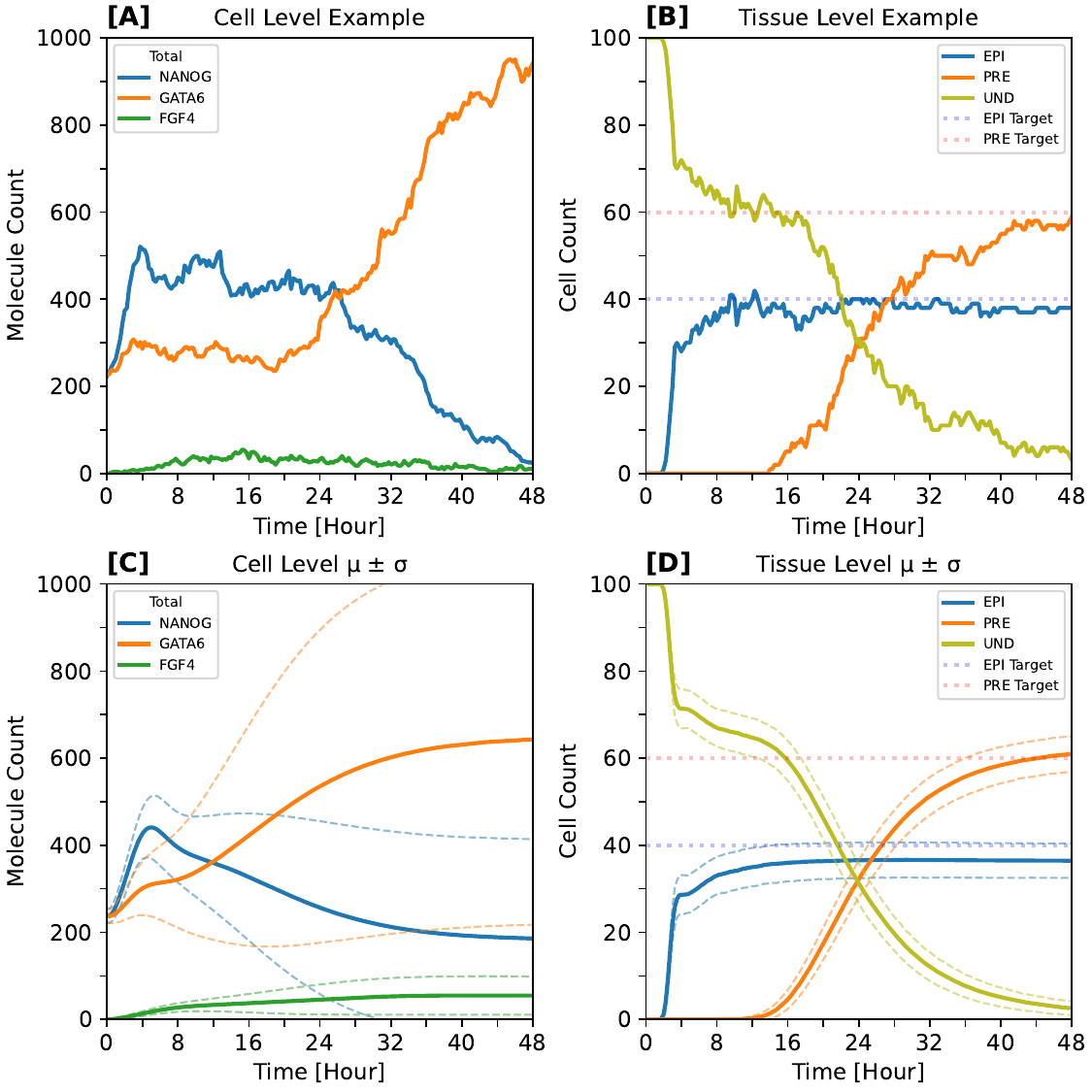} \centering 
\end{adjustwidth} 
\begin{figure}[hpt!]
\begin{adjustwidth}{-1.75in}{0in} 
\caption{{\bf Stochastic trajectories at cell and tissue scale.} The ITWT system shows excellent agreement with various experimentally observed characteristics of ICM cell specification. All simulations start with all cells in the undifferentiated (UND) state, having well-balanced cellular resources, before committing exclusively to EPI or PRE fates. {\bf [A]} Example cell-level stochastic trajectory, randomly taken from a 100-cell (10{\texttimes}10 grid) simulation. Within the first 24~h, the cell remains in the UND fate while the FGF4 protein level increases. At around 24~h, a symmetry-breaking event occurs, as the NANOG and GATA6 protein levels take divergent expression paths. Within the last 24~h, the cell acquires the PRE fate while FGF4 levels decrease again. {\bf [B]} Example tissue-level stochastic trajectory of cell fate counts in a 10{\texttimes}10 cell-grid system. Each cell is categorized into one of three possible fates: UND, EPI, or PRE. With progressing time, UND cells reduce in number while EPI and PRE cell counts settle close to prescribed constant target levels (dotted lines); see Methods section \nameref{subsection:computational_experiments} for details of cell-fate classification. {\bf [C]} Typical cell-level behavior of NANOG (blue), GATA6 (orange), and FGF4 (green) total protein levels. {\bf [D]} Typical tissue-level behavior of EPI (blue), PRE (orange), and UND (olive) cell-fate counts. Solid lines represent means, dashed lines represent standard deviations around means. Statistics are computed from a batch of 1000 simulations of a 100-cell tissue system.}
\label{fig3}
\end{adjustwidth} 
\end{figure}

Note that in our model the observed precedence of EPI emergence over PRE cells arises naturally as a predictive result, without any explicit incorporation into the modeling or inference procedures. The mechanism driving the delayed emergence of the opposing cell fates can be summarized as follows: first, stochastic self-activation of \textit{Nanog} triggers the EPI fate specification program in a subset of the ICM cells. This, in turn, promotes the progressive differentiation of other unspecified cells into the PRE fate when they sense FGF4, which is only released by cells where \textit{Nanog} reached substantial expression levels. A crucial precondition of this mechanism is the lower self-activation threshold of \textit{Nanog} compared to \textit{Gata6}. Despite the inherent stochasticity of the differentiation process at the single-cell level, coordination via FGF4 makes it appear deterministic at the tissue level.

\subsection*{Expression profiles reveal strong linear correlations among key regulatory proteins}

\begin{figure}[ht!]
\begin{adjustwidth}{-1.75in}{0in} 
\includegraphics[width = 7.5in, height = 5in]{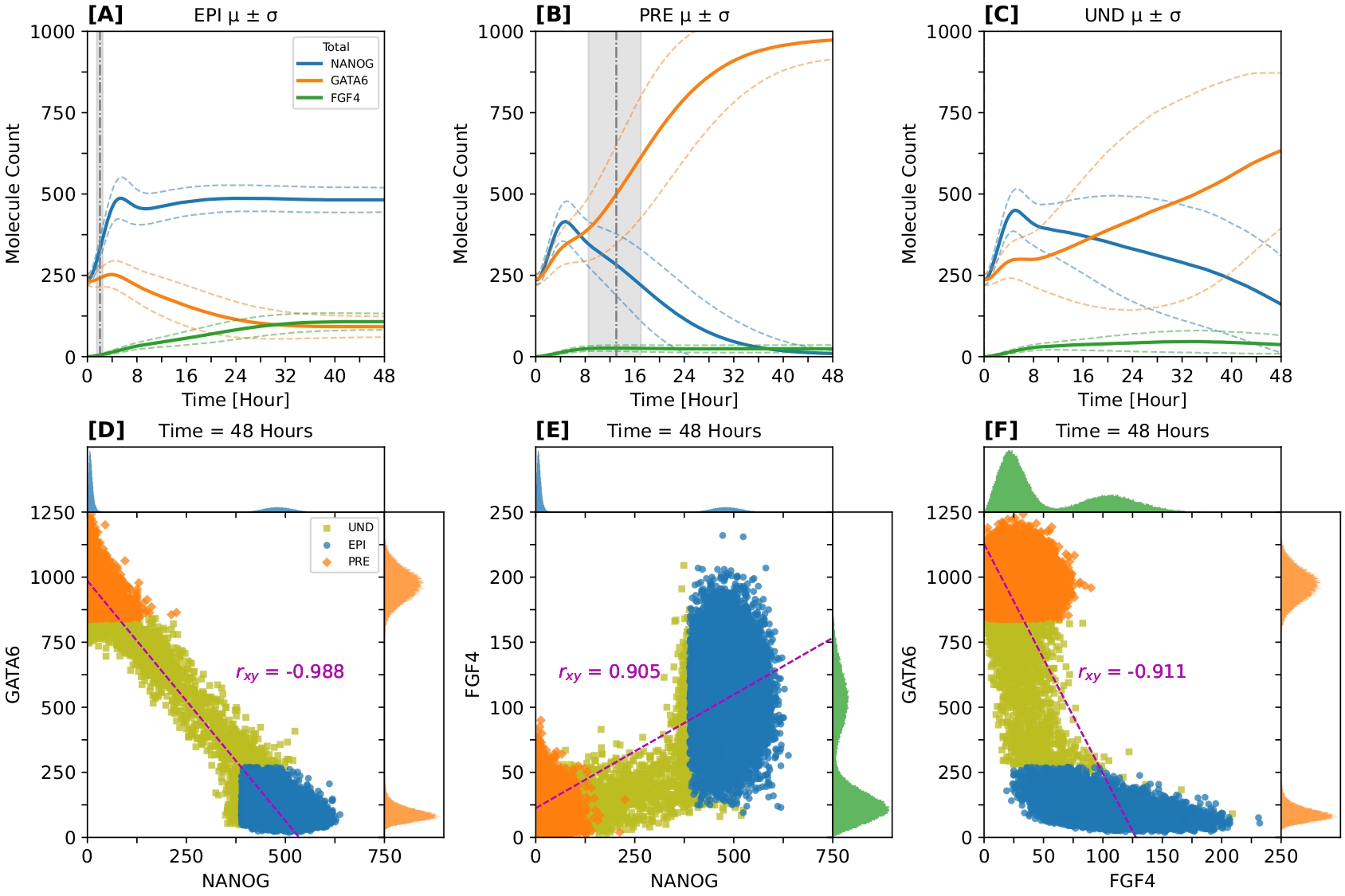} \centering 
\end{adjustwidth} 
\begin{adjustwidth}{-1.75in}{0in} 
\caption{{\bf EPI cells precede and are necessary for specification of PRE cells. Expression profiles reveal strong linear correlations among key proteins.} Upper row shows typical cell-level protein count time evolution for each cell-fate category. Cell fates were identified at 48~h. For each fate, cell-level dynamics were traced from last to initial simulation data point. {\bf [A-C]} Solid lines represent mean behavior, dashed lines represent standard deviations around means. Gray regions accentuate time intervals of first differentiated cell appearance for the given fate (vertical dash-dotted line = mean time). Statistics are computed from a batch of 1000 simulations. Only the most relevant proteins are shown: FGF4 (green), NANOG (blue), GATA6 (orange). {\bf [D-F]} Pairwise relationships among the most important proteins at 48~h. Inner panels show different colors representing the protein-pair relationship for the respective cell fate: UND (olive), EPI (blue), PRE (orange). Outer panels show protein count histograms. Data points come from a batch of 1000 simulations. \emph{r\textsubscript{xy}} = Pearson's product-moment correlation coefficient. For details of cell-lineage classification, see Methods section \nameref{subsection:computational_experiments}.}
\label{fig4}
\end{adjustwidth} 
\end{figure}

Our simulations show that the copy number distributions of the three key proteins NANOG, GATA6, and FGF4 are clearly bimodal by the 48-hour mark, as depicted in Fig~\ref{fig4}[D-F]. At the beginning of every simulation, we use a well-defined initial condition distribution (ICD) which restricts the protein and mRNA expression levels to a region where all the cells start with the undifferentiated (UND) fate. This guarantees symmetric splitting of initial resources on average and prevents any systematic fate bias at the simulation start. Despite the innate randomness of the ICDs employed for simulating, early variability does not have adverse effects in the final lineage proportions. Instead, contrasting expression profiles slowly emerge and are clearly visible by the last simulation time point (48 hours).

A notable observation from our simulation data are strong linear correlations among these key regulatory proteins, which emerge in spite of the nonlinear regulatory interactions between them. Specifically, we identify a pronounced negative linear correlation between NANOG and GATA6, as well as between GATA6 and FGF4 (refer to Fig~\ref{fig4}[D, E]). Conversely, a strong positive linear correlation is observed between NANOG and FGF4 (see Fig~\ref{fig4}[F]). These robust linear relationships mirror findings reported in experimental studies, affirming the validity of our simulation approach \cite{saiz_asynchronous_2016, demot_cell_2016, fischer_transition_2020, plusa_common_2020}.

\subsection*{Intercellular communication via FGF4 functionally improves ICM differentiation robustness by 10-20\% compared to a purely-binomial baseline scenario} \label{subsection:intercellular_communication}

In order to investigate the robustness to noise in ICM specification, we assessed whether and to which extent tissue-level coupling via FGF4 is capable of reducing variability in the acquired cell fates. To this end we compared the variability observed in our Inferred-Theoretical Wild-Type (ITWT) model simulations to an entirely cell-autonomous fate decision-making scenario. In such hypothetical ``Purely Binomial'' (PB) scenario, the cell lineage distribution is supposed to follow a binomial pattern, as each cell's fate, either EPI or PRE, is determined independently of others. The comparison was carried out by analyzing the coefficient of variation (CV) of 48-hour fate proportions across 13 different system sizes ($\eta = [5, 10, 15, 25, 35, 50, 65, 75, 85, 100, 150, 225, 400]$ cells) for the two cases.

For the first case, we calculated the coefficient-of-variation (CV\textsubscript{1}) as the ratio of sample standard deviation to sample mean. For the second case, the hypothetical PB scenario, the corresponding measure (CV\textsubscript{0}) is simply the coefficient of variation of the binomial distribution, with standard deviation being a function of mean fate numbers and total cell count. Interestingly, for the EPI and PRE fates (excluding the UND category) CV\textsubscript{1} is consistently lower than CV\textsubscript{0}, indicating of noise reduction due to FGF4 signaling (see Fig~\ref{fig5}[A]). The ratio $CV_{0}/CV_{1}$ suggests that the ITWT model outperforms the PB model by 10-20\%, implying fewer incorrectly specified cells (inset of Fig~\ref{fig5}[A]).

To further validate the role of cell-cell communication in enhancing patterning robustness, we also compared the ITWT model to the Reinferred Theoretical Mutant (RTM). The RTM lacks cell-cell signaling (see Fig~\ref{fig2}, Fig~\ref{fig101}, and Methods section), reproducing the prescribed cell-fate ratio (on average) with a purely cell-autonomous patterning mechanism. We asked whether binomial noise emerges naturally in this system. Indeed, we found that the CV of the RTM model is comparable to that of a PB model (Fig~\ref{fig5}[C]). Both systems adhere to the same power law, with negligible differences across system sizes (inset of Fig~\ref{fig5}[C]).

\clearpage

In conclusion, our findings demonstrate that cell-to-cell communication via FGF4 diffusion, encoding local environmental variations, enhances ICM fate differentiation robustness by approximately 10-20\% compared to a purely cell-autonomous scenario. This finding corroborates the notion that ICM differentiation is a tissue-level process, where tissue-scale signaling feedback via FGF4 plays a functional role in mitigating cell-fate decision noise. At the same time, it is in line with previous studies highlighting the benefit of spatial coupling for noise reduction in developing tissues \cite{erdmann_role_2009, sokolowski_mutual_2012, sokolowski_optimizing_2015, ellison_cellcell_2016, smith_role_2016, fancher_fundamental_2017, sokolowski_deriving_2023}.

\subsection*{Robust cell-fate proportions are independent of cell-grid size} \label{subsection:robust_cell_fate_proportions}

Recent experiments suggest that robust control in the EPI to PRE lineage ratio does not depend on the absolute size of these populations \cite{saiz_asynchronous_2016, simon_making_2018}. Resilience of the mouse embryo to variations in ICM size, as reflected by alterations in total cell number, was found both in vivo and in silico \cite{nissen_four_2017, saiz_growth-factor-mediated_2020}. Nonetheless, there remains a debate on whether a critical embryo size is essential for proper blastocyst lineage segregation \cite{rossant_blastocyst_2009, yeh_capturing_2021, stanoev_robustness_2021, stanoev_robust_2022}. To assess the impact of absolute tissue size (cell number) on ICM specification, we analyzed cell-fate proportions and associated noise levels across various tissue sizes, hypothesizing that smaller cell numbers might correlate with increased noise in fate decisions.

We conducted 1000 simulations for each of 13 distinct cell grid sizes, ranging from 5 to 400 cells in total, for both the Inferred-Theoretical Wild-Type (ITWT) and the Reinferred Theoretical Mutant (RTM) models.

We find that noise intensity scales with system size with a power law in both models, as illustrated in Fig~\ref{fig5}[A, C]. This indicates that noise diminishes predictably as system size increases.

Moreover, our simulations reveal a universal mean value ($\mu$) for cell-fate proportions, consistent across all different cell grid sizes for both the ITWT and RTM. Despite this, the two models show, respectively, unique characteristics in commitment times, standard-deviation magnitudes, and independence of fate choice among cells (see Fig~\ref{fig5}[B, D]).

The commitment time discrepancies between the ITWT and RTM models can be attributed to their distinct mechanisms. The ITWT augments probabilistic differentiation with tight control via FGF4 signaling, which makes it more resilient against perturbations. This mechanism requires initial random emergence of a portion of the EPI population, which subsequently coordinates other undifferentiated cells towards specific fates based on local neighborhood information, thus globally regulating EPI-PRE proportions. Early commitment to the EPI fate and subsequent emission of FGF4 is crucial to this process. In contrast, the RTM relies purely on stochastic differentiation, and is more sensitive to perturbations. This system lacks regulation beyond the inherent genetic program at the cellular level and does not integrate tissue-neighborhood information, with fate commitment timing primarily dictated by target protein levels for EPI and PRE markers.

In sum, our findings suggest no critical cell number for accurate ICM fate specification; however, its precision increases with the (square root of the) cell number, while spatial coupling via FGF4 can reduce the noise magnitude by $\sim{20}\%$ compared to a purely cell-autonomous mechanism.

\begin{adjustwidth}{-1.75in}{0in} 
\includegraphics[width = 6.75in, height = 6.75in]{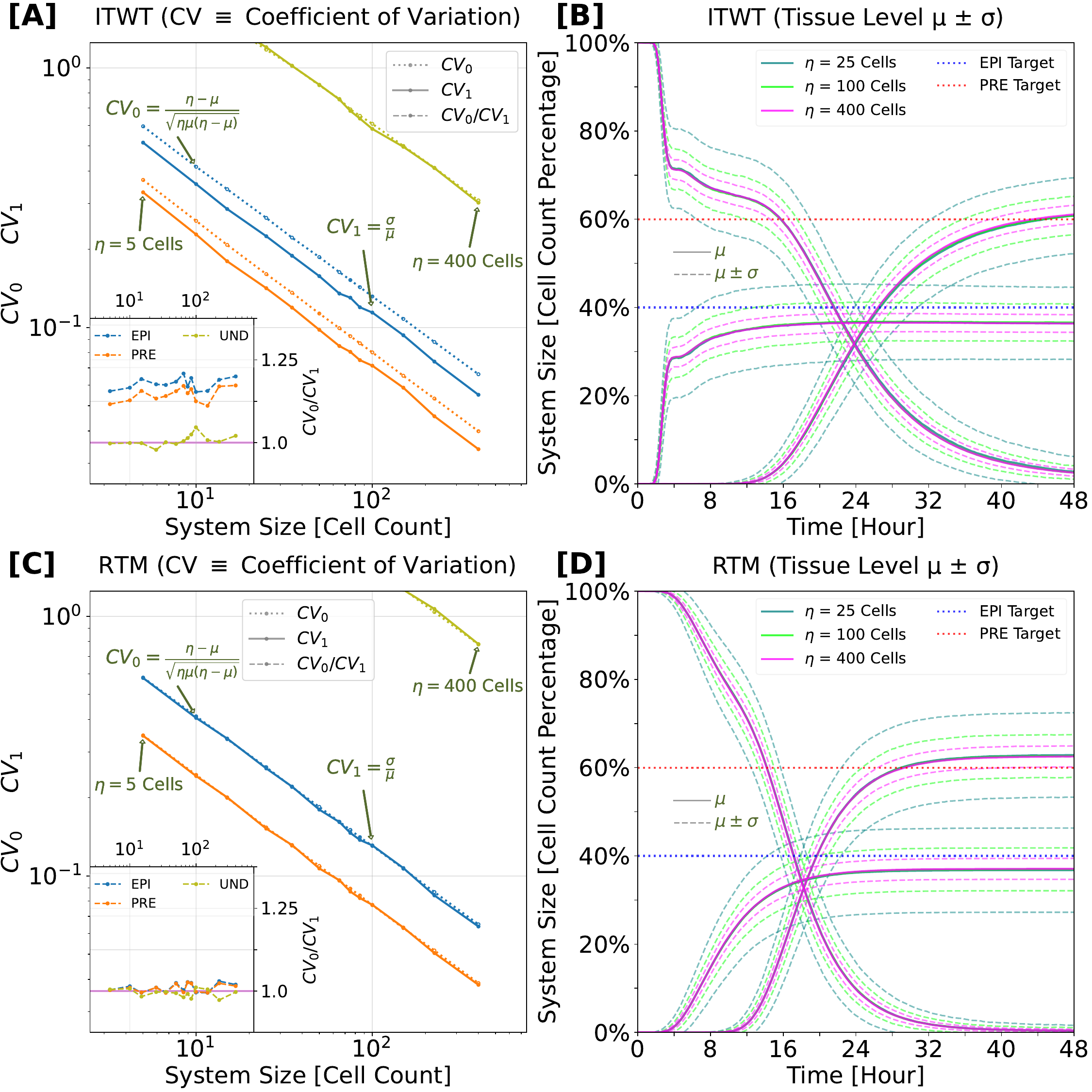} \centering 
\end{adjustwidth} 
\begin{figure}[hpt!]
\begin{adjustwidth}{-1.75in}{0in} 
\caption{{\bf Noise at tissue level: intercellular communication improves ICM differentiation robustness by 10-20\% compared to purely-binomial baseline scenario.} {\bf [A]} Comparison between inferred-theoretical wild-type (ITWT $\sim$ CV\textsubscript{1}) and purely-binomial (PB $\sim$ CV\textsubscript{0}) systems. The main plot shows the coefficient of variation (CV) as a function of the system size for each cell fate (colors as in Fig~\ref{fig4}). The inset shows the ratio between CV\textsubscript{0} and CV\textsubscript{1}, highlighting systematically lower fate specification error in the ITWT compared to the PB system. {\bf [C]} Comparison between reinferred-theoretical mutant (RTM $\sim$ CV\textsubscript{1}) and purely-binomial (PB $\sim$ CV\textsubscript{0}) systems. Colors and symbols as in [A]. The inset plot shows that fate specification errors in the spatially uncoupled RTM system are of the same magnitude as in the PB scenario. Axes use logarithmic (base 10) scale. For each system size, the data point is computed from a batch of 1000 simulations. {\bf [B, D]} Typical time evolution of normalized cell fate counts for three example system sizes ($\eta \in [25, 100, 400]$ cells). Solid lines represent means, dashed lines represent standard deviations around means.}
\label{fig5}
\end{adjustwidth} 
\end{figure}
\clearpage

\subsection*{Autocrine- and paracrine-signaling modes play reciprocal roles in robust cell-cell communication}

One key characteristic of the mouse blastocyst is the overwhelming dominance of EPI cell fates when FGF4 production is inhibited or related loss-of-function mutations are applied. In such cases, almost all cells commit to the EPI fate by the time of implantation, as they are unable to exit naive pluripotency due to the absence of mechanisms controlling precise \textit{Nanog} expression, leading to adverse developmental outcomes \cite{bessonnard_icm_2017, thompson_extensive_2022, kim_erk1_2014, deathridge_live_2019, kale_nanog-perk_2022}.

In agreement with this (and as demanded by the imposed score function), when FGF4 signaling is impeded in our simulations, cells initially co-express EPI and PRE fate-specific markers, but eventually only a small subset adopts the PRE fate \cite{bessonnard_icm_2017, raina_cell-cell_2021}. This leads to NANOG upregulation and commitment to the EPI fate in the majority of ICM cells \cite{saiz_growth-factor-mediated_2020, allegre_nanog_2022}.

\begin{figure}[hpt!]
\begin{adjustwidth}{-1.75in}{0in} 
\includegraphics[width = 7.5in, height = 1.875in]{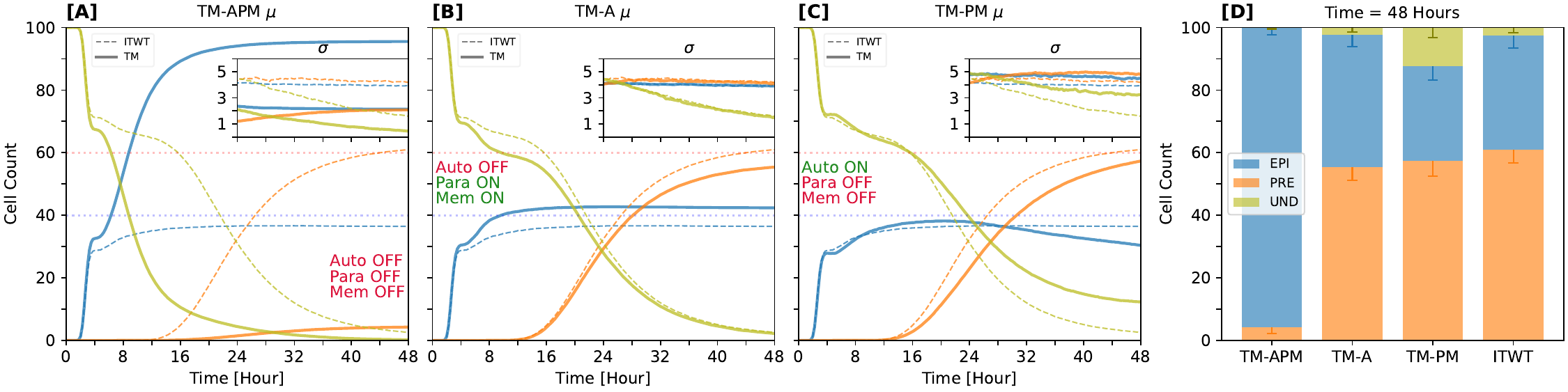} \centering 
\end{adjustwidth} 
\begin{adjustwidth}{-1.75in}{0in} 
\caption{{\bf Autocrine- and paracrine-signaling modes play reciprocal roles in robust cell-cell communication.} The figure shows the effects of perturbing autocrine- and paracrine-signaling modes in theoretical mutant (TM) systems. {\bf [A]} The TM-APM (inhibition of autocrine, paracrine, and membrane-exchange signaling) represents the loss-of-function phenotype of experimental FGF/ERK pathway mutants. {\bf [B, C]} The complementary TM-A and TM-PM portray the importance of individual feedback modes (paracrine plus membrane exchange and autocrine signaling only, respectively; see also inset text). Solid lines represent mean and standard deviation values ($\mu$ and $\sigma$) for a given TM; dashed lines represent $\mu$ and $\sigma$ for the ITWT. {\bf [D]} Cell-lineage allocation at the 48~h time point. All TMs are compared against the inferred-theoretical wild type (ITWT). For all compared systems, statistics were calculated from 1000 independent simulation runs.}
\label{fig6}
\end{adjustwidth} 
\end{figure}

To dissect the roles of different communication modes in ICM specification, we modified the Inferred-Theoretical Wild-Type (ITWT) model to interrupt specific components of the signaling pathway. This way we created three ``theoretical mutants'' implementing the following signaling scenarios: complete absence of FGF4 (TM-APM), lack of autocrine signaling (TM-A), and absence of paracrine signaling and membrane-to-membrane exchange (TM-PM). Each modification affects cell fate determination differently, as shown in Fig~\ref{fig6}[A-C].

As expected, the TM-APM model exhibits a strong bias towards the EPI fate (Fig~\ref{fig6}[A, D]). Initially, its dynamics parallel those of the ITWT system, but as the simulation progresses, the EPI fate predominates, with only a minor fraction of cells adopting the PRE fate.

The TM-A model displays a decreased accuracy of the ICM specification mechanism due to the absence of self-regulation, though the overall precision of the system remains unaffected (Fig~\ref{fig6}[B, D]). Adjustments in other signaling components could potentially correct for this, but these also would likely reduce the system's ability to buffer against dynamic signaling perturbations.

In the TM-PM model, the critical role of paracrine communication and, to a lesser extent, FGF4 membrane exchange becomes evident (Fig~\ref{fig6}[C, D]). Eliminating these communication modes disrupts the maintenance of target cell-lineage ratios, even when autocrine signaling is preserved. During initial simulation phases, differentiation seems normal, but over time, a significant fraction of cells remains undifferentiated, and the EPI lineage fails to sustain its population. Exchanging FGF4 with neighboring cells therefore is crucial for correct cell-fate specification, once again underpinning the importance of its tissue-scale coordination.

In summary, both autocrine and paracrine signaling are integral to ICM differentiation and maintenance. Autocrine signaling ensures the accuracy of fate specification, while paracrine signaling, along with membrane exchange, maintains lineage proportions, enhances precision, and promotes cellular homeostasis.

\subsection*{ICM cells produce similar local and global neighborhood features: lineage ratios are preserved at both scales} \label{subsection:neighborhood}

We next asked whether the tissue-level spatial coupling via FGF4 leads to specific signatures in the emerging spatial distribution of cell fates.

A central challenge in developmental biology is the precise characterization of spatial patterns, such as the ``salt-and-pepper'' arrangement reported for the ICM, which has often been described informally in the literature \cite{bessonnard_gata6_2014, demot_cell_2016, tosenberger_computational_2019, cang_multiscale_2021, schardt_adjusting_2023}. The term typically implies a random distribution, yet randomness in a mathematical context can take various forms. Recent studies have endeavored to rigorously define this pattern using experimental and theoretical approaches \cite{liebisch_cell_2020, fischer_transition_2020, forsyth_iven_2021, dirk_recognition_2023, fischer_salt-and-pepper_2023}. Here we understand the ``salt-and-pepper'' pattern as an archetype in which each individual cell-fate decision is independent of the cell fates of its neighbors, which implies a multinomial distribution of cell fates in every tissue neighborhood.

\clearpage

The dynamically growing ICM is also shaped by cellular division and intercellular forces, which can lead to local fate clustering and compositional variability, as reported in prior studies \cite{fischer_transition_2020, cang_multiscale_2021, yanagida_cell_2022}. In such scenario, a multinomial or ``salt-and-pepper'' distribution is not expected in the first place. However, here our static cell arrangement isolates the problem from these factors, and allows for an analysis that focuses solely on the influence of cell-cell communication on the spatial distribution of cell fates. We therefore asked to which extent the spatial cell distribution in our simulated ICM system agrees with or deviates from a multinomial baseline.

\begin{figure}[hpt!]
\begin{adjustwidth}{-1.75in}{0in} 
\includegraphics[width = 7.5in, height = 3.75in]{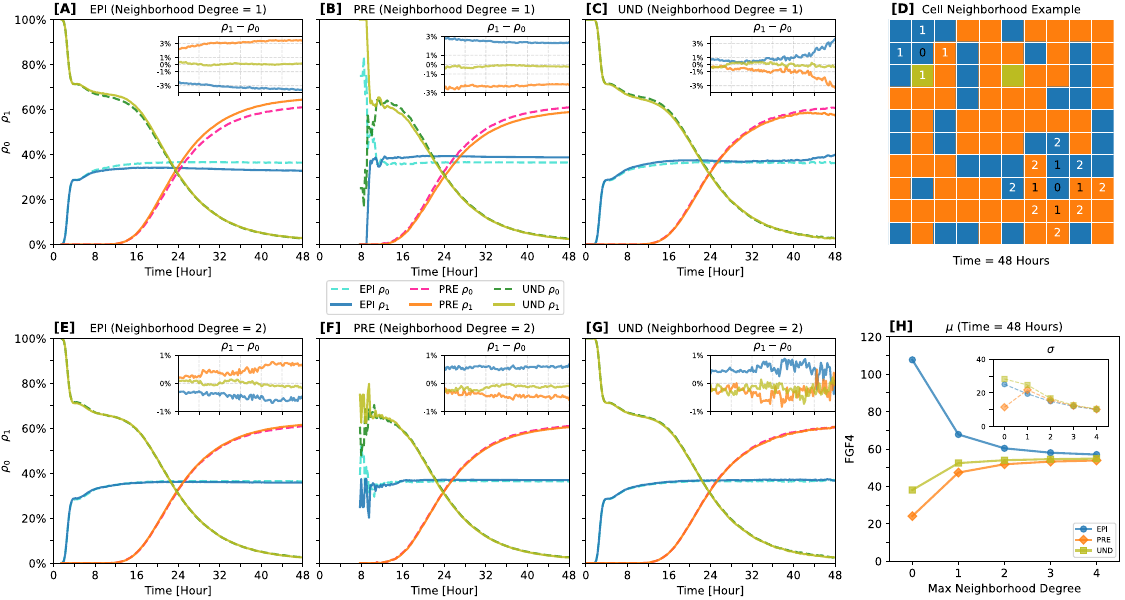} \centering 
\end{adjustwidth} 
\begin{adjustwidth}{-1.75in}{0in} 
\caption{{\bf Exploring self-similarity and spatial dynamics of simulated ICM neighborhoods.} {\bf [A-C, E-G]} Temporal evolution of spatial correlations among acquired cell fates. Panels [A-C] (upper row) and [E-G] (lower row) show cell-neighborhood degrees 1 and 2, respectively. Here $\rho_{1}$ and $\rho_{0}$ represent mean cell-neighborhood composition values split by fate: $\rho_{1}$ is directly computed from the spatio-temporal distribution of cell fates in ITWT simulations; in contrast, $\rho_{0}$ represents a multinomial baseline model assuming that cell fates emerge in a spatially uncoupled fashion from independent trials leading to acquisition of one of three cell fates with fixed occurrence probability. The latter is estimated by averaging cell fate proportions at every time point across simulations, discarding any spatial information. {\bf [D]} Example of cell neighborhood at 48-h time point. {\bf [H]} Mean and standard deviation of FGF4 protein count as functions of maximum cell-neighborhood degree. FGF4 levels quickly become uncorrelated beyond nearest-neighbor cell distance.}
\label{fig7}
\end{adjustwidth} 
\end{figure}

To this end, we first determined the neighborhood composition in our Inferred-Theoretical Wild-Type (ITWT) model, focusing on the three cell fate categories: EPI, PRE, and UND (Fig~\ref{fig7}[A-C, E-G]). For each cell within a specific category, we included neighbors up to a predetermined degree (first or second-degree neighbors, as seen in Fig~\ref{fig7}[A-C, E-G] and Fig~\ref{fig7}[D]; see\nameref{subsubsection:tissue_scale_model} for the details of neighborhood stratification). We then compared the resulting cell-fate arrangements to the distributions resulting from artificially generated systems, in which the cell fates were sampled from a multinomial distribution with the same proportions as extracted from the simulated data. This comparison was carried out for all simulated times.

A subtle discrepancy, especially at the first-degree neighborhood level, emerged between the non-cell-autonomous system model ($\rho_{1}$) and the cell-autonomous, multinomial model ($\rho_{0}$) after 24 hours of simulation (inset plots of Fig~\ref{fig7}[A-C]). This difference, about \textpm 3\% in average neighborhood composition, is significant when considering their sampling distributions (standard errors are around \textpm 0.5\%). However, when comparing to the variation between simulations (with a \textpm 10\% standard deviation), this discrepancy becomes less significant.

Furthermore, we analyzed the FGF4 distribution across five independent neighborhood degrees ([0, 1, 2, 3, 4]; Fig~\ref{fig7}[H]). After 48 hours, FGF4 predominantly remained concentrated near its source (EPI cells) but also spread to more distant neighboring cells at a notably lower molecular count. This finding aligns with experiments reporting that, in artificial systems, FGF4 signaling proteins stabilize around their source cells at a single cell-distance length scale \cite{raina_cell-cell_2021}. It suggests that diffusive coupling balances the overall FGF4 level across the system, essentially acting as a quorum signal that reflects the proportion of FGF4-producing cells at the tissue scale.

\clearpage

Our findings indicate that both at the local and global scales, the spatial distribution of cell fates is similar, irrespective of whether cell differentiation is coordinated at the tissue level, as in our ITWT system, or purely cell-autonomous. This observation seems counter-intuitive, given the necessity of cell-cell signaling for proper lineage establishment. However, this may be part of a strategy to withstand strong perturbations (like drastic changes in cell population) for which cell fate proportions must be maintained locally but coordinated globally, such that neighborhood characteristics are preserved at both scales. This results in a seemingly irregular pattern that nevertheless preserves cell-fate ratios in a spatially homogeneous fashion.

\subsection*{Increased variability in initial conditions enhances developmental accuracy while sustaining its precision}

We next assessed how variability in initial condition distributions (ICDs) affects the accuracy and precision of cell-fate specification in our ITWT model; here, ``accuracy'' is defined as the closeness of the average simulated EPI-PRE proportions to the target ratios, and ``precision'' refers to the variability of these proportions among simulation ensembles.

\clearpage
\begin{adjustwidth}{-1.75in}{0in} 
\includegraphics[width = 5.25in, height = 5.25in]{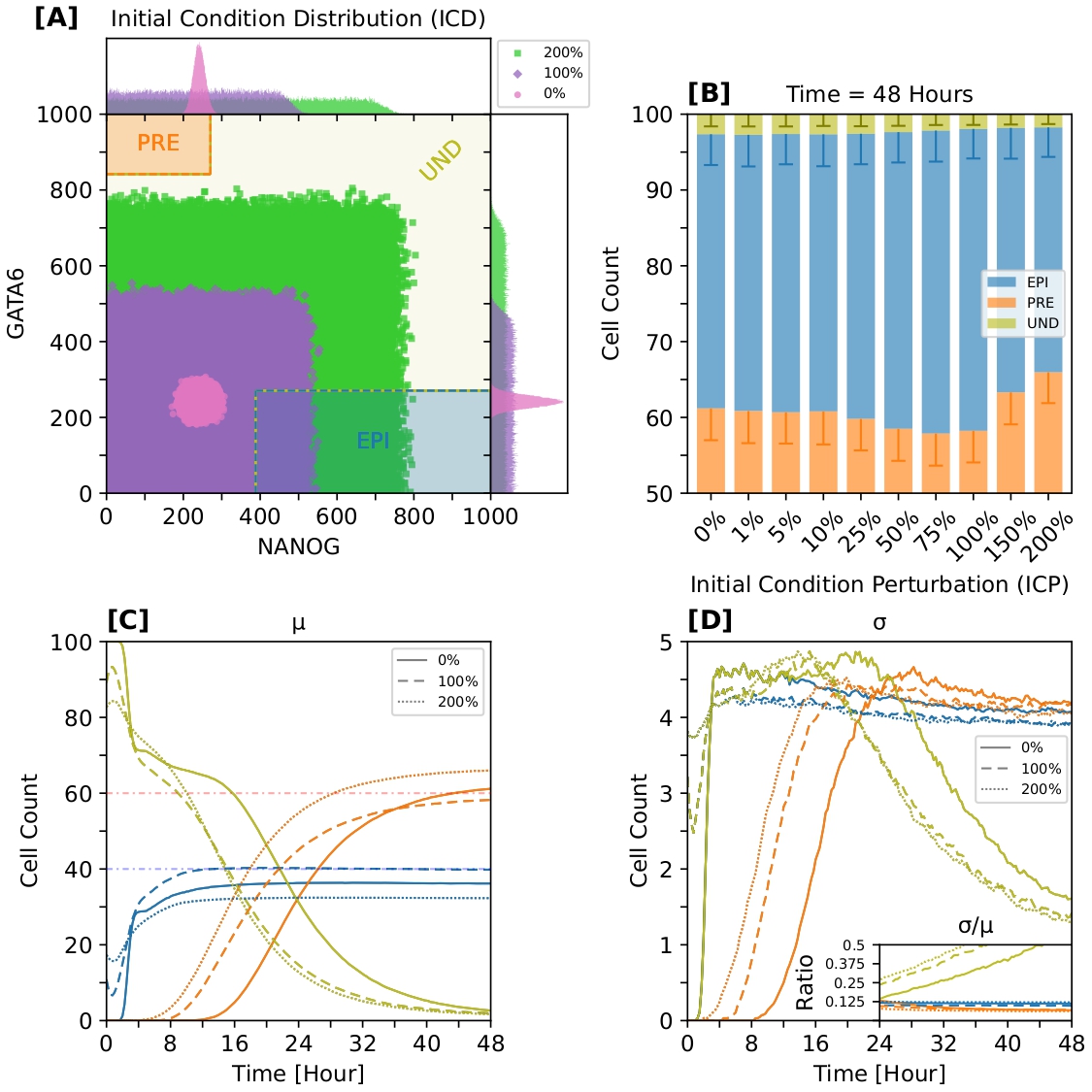} \centering 
\end{adjustwidth} 
\begin{figure}[hpt!]
\begin{adjustwidth}{-1.75in}{0in} 
\caption{{\bf Initial-condition variations enhance patterning accuracy and simultaneously sustain its precision.} Robustness to initial-condition perturbations (ICPs): uniform perturbation of protein (NANOG, GATA6) and mRNA (\textit{Nanog}, \textit{Gata6}) initial resources. {\bf [A]} Example initial-condition distributions (ICDs) with increasing variability (see legend) for NANOG and GATA6 proteins (analogous mRNA data not shown for simplicity). Interior scatter plots show sampling intervals; exterior plots show sampling distributions. The 0\% distribution defines the unperturbed baseline case. For each ICD, data points come from 1000 independent simulations. {\bf [B]} Mean cell-fate composition (bars) with standard deviation (error bars) for all tested variability intensities. Bar height ranges from 50 to 100 cells. {\bf [C]} Time trajectories of mean cell-fate counts for selected ICD variability intensities. Dotted lines specify target proportions for EPI and PRE fates, respectively. {\bf [D]} Time trajectories of cell-fate count standard deviations for selected perturbation intensities. Inset plot displays temporal coefficient of variation ($CV = \sigma{\slash}\mu$) for the last 24~h. Colors represent different cell fates. Statistics were calculated from batches of 1000 independent simulations per ICD.}
\label{fig8}
\end{adjustwidth} 
\end{figure}

Our approach was guided by two aspects: firstly, the broad inferred posterior distributions of the parameters governing the ICDs, and their moderate sensitivity to value changes (Fig~\ref{fig102}[D, H]); secondly, similar assessments that were carried out in previous models of mouse-blastocyst, where mainly the ICD variance was modulated \cite{tosenberger_multiscale_2017, stanoev_robustness_2021}. Taking this into account, we generated 1000 stochastic trajectories for 10 distinct ICDs, modifying their variance from 0\% to 200\% compared with the baseline value (see Fig~\ref{fig8}[A] for examples); the details of ICD modulation are described in \nameref{subsection:computational_experiments}.

We analyzed both the mean cell count (accuracy) and the corresponding standard deviation (precision) for each cell fate (Fig~\ref{fig8}[C, D]).

Remarkably, increasing the variance of the ICD can, in some cases, positively influence cell-fate specification in the ITWT system. For example, in the 100\% initial condition perturbation (ICP) scenario, the accuracy of EPI/PRE specification improves notably compared to the baseline (contrast dashed to solid lines in Fig~\ref{fig8}[C]), while its precision remains unaffected (Fig~\ref{fig8}[D] and inset).

Between the 25\% and 75\% ICPs we observe a systematic reduction of the PRE populations in favor of the EPI populations (Fig~\ref{fig8}[B]). This can be attributed to the fact that with increasing ICP strength a larger subset of cells is initially biased towards the EPI fate. This trend is inverted as we proceed to stronger ICPs (150\% and 200\%), since now the initially available \textit{Nanog} abundance quickly induces FGF4 production, which promotes the PRE fate. Notably, while different ICPs thus alter the cell-specification accuracy, the corresponding precision remains similar for all levels of ICP strength (error bars in Fig~\ref{fig8}[B] and inset plot of Fig~\ref{fig8}[D]).

Our findings indicate that when the ICD is perturbed within normal ranges expected from full protein induction, cell-specification accuracy can be improved without compromising its precision. However, if the ICD is perturbed beyond this, the excess initial resources negatively affect the accuracy, while precision remains unchanged. These observations align with various previous studies which highlighted that stochasticity can play a constructive role in biological systems \cite{paulsson_stochastic_2000, swain_intrinsic_2002, eldar_functional_2010, tsimring_noise_2014, lin_central_2016, vandevenne_rna_2019, tkacik_many_2021, cuesta_bernoulli_2022, benzinger_synthetic_2022, frei_genetic_2022, briat_noise_2023}.

\subsection*{Cell-fate assignment remains robust when less than 25\% of cells start with perturbed initial conditions} \label{subsection:initial_condition_noise}

Having established that moderate increases in the variability of initial condition distributions (ICDs) can be beneficial for robust cell-fate specification, we now turn our attention to understanding the limits of this robustness by examining the system's response to different formats of ICD perturbations. In these tests, while maintaining a constant tissue size of 100 cells, a varying number of cells (ranging from 1 to 100) are randomly selected for ICD modifications.

\begin{figure}[ht!]
\begin{adjustwidth}{-1.75in}{0in} 
\includegraphics[width = 7.5in, height = 3.75in]{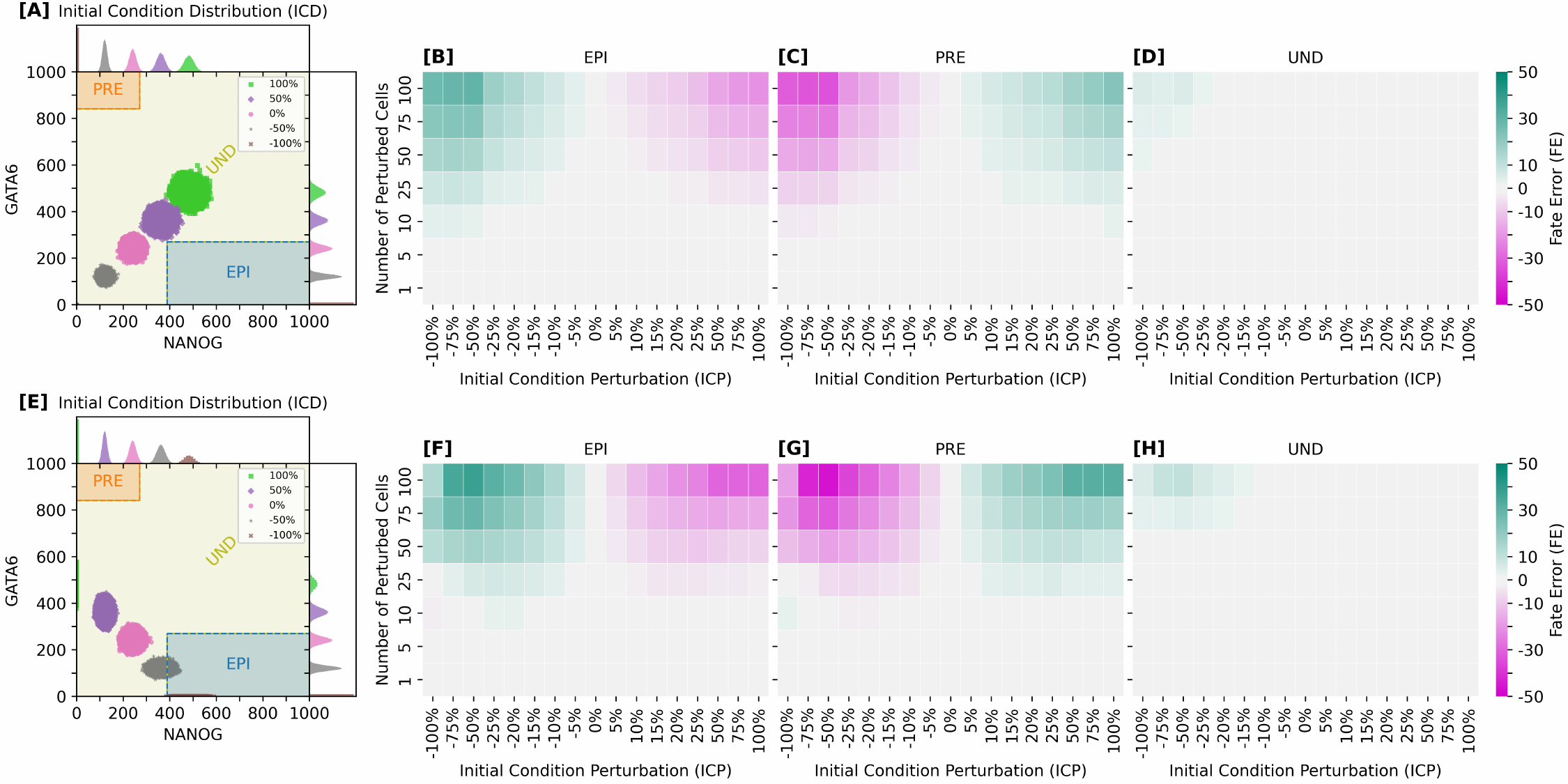} \centering 
\end{adjustwidth} 
\begin{adjustwidth}{-1.75in}{0in} 
\caption{{\bf Resilience to perturbations is mainly determined by number of affected cells: patterning remains robust when less than 25\% of cell population is perturbed.} Robustness to linear initial condition perturbations (ICPs) of protein and mRNA initial resources (NANOG-GATA6 and \textit{Nanog}-\textit{Gata6}). Upper row [A-D] shows the effect of equally perturbing resources of both genes. Lower row [E-H] shows adversarial perturbation of the resources, i.e. linearly progressing perturbation magnitude from 200\% NANOG and 0\% GATA6 (\textminus{100}\% ICP scenario) to 0\% NANOG and 200\% GATA6 (100\% ICP scenario). {\bf [A, E]} Initial condition distributions (ICDs) for NANOG and GATA6 proteins (analogous mRNA distributions not shown for simplicity). Interior (scatter) plots show sampling intervals. Exterior plots show sampling distributions. For each ICD, data points come from 1000 independent simulations. {\bf [B-D, F-H]} Quantifying the effect of equal [B-D] and adversarial [F-H] ICPs for different intensities (molecular count percentage with respect to baseline scenario: ICP = 0\%), and varying number of perturbed cells for all different cell-fate categories (EPI, PRE, UND). For each fate category, color saturation indicates the fate error, i.e. the mean difference of cell-fate counts between the perturbed and unperturbed baseline systems. Each matrix entry (perturbation pair) is calculated from 1000 independent simulations.}
\label{fig9}
\end{adjustwidth} 
\end{figure}

The first perturbation scheme linearly modifies both NANOG- and GATA6-related resources in tandem, with the new mean ICD values ranging from 0\% to 200\% of the typical initial resources (Fig~\ref{fig9}[A-D]). For clarity, we have labeled these scenarios based on their deviation from the standard ICD, such as \textminus{100}\%, 0\% (reference), and 100\%.

The second perturbation scheme involves a negative linear correlation between NANOG- and GATA6-related resources, creating scenarios where one of these resources is initially dominant (Fig~\ref{fig9}[E-H]). The range of adjustments spans from 200\% NANOG and 0\% GATA6 to 0\% NANOG and 200\% GATA6, again compared to the unperturbed initial conditions.

We quantified the deviations from the typical ICD behavior using the relative Fate Error (FE), which measures the discrepancy in ICM specification accuracy for each cell fate (UND/EPI/PRE) between the perturbed scenarios and the reference (0\% ICP) scenario at 48 hours; see also caption of Fig~\ref{fig9} for details about FE.

Our findings from both schemes indicate that when more than 25\% of the cell population is affected by the ICD disturbances, the system experiences significant deviations in lineage distributions, regardless of perturbation strength. This suggests the existence of a critical threshold ratio of perturbed cells, beyond which the system's resilience is notably compromised. When this threshold is exceeded, profound gene expression imbalances emerge across the ICM population.

Both types of perturbation, whether implying scarcity or over-abundance of initial resources, result in similar FE values suggesting a potential correction mechanism that can overcome highly irregular cellular initial conditions. Notably, no significant deviation is observed when less than 25\% of the cell population undergoes perturbation. Here the system demonstrates remarkable robustness, underscoring the tissue-level coordination inherent in the ICM specification process.

\subsection*{Cell plasticity and FGF4-sensitivity are time-window dependent}

Temporal modulation of FGF4 concentration and the corresponding shift in cell plasticity are key aspects of ICM cell differentiation, with numerous studies documenting their influence \cite{bessonnard_icm_2017, raina_fgf4_2021, raina_cell-cell_2021, yeh_capturing_2021, allegre_nanog_2022}. To examine whether our model can replicate the experimentally observed time-dependent responsiveness to FGF4 level changes, we introduced controlled perturbations of FGF4 in our simulations. This involved adding extra FGF4 molecules to each simulated cell at specific time points, mimicking the effect of exogenous FGF4.

We assessed the response to FGF4 perturbations in two models (ITWT and TM-FGF4) and at two distinct final simulation times, 48 and 96 hours. The second time point, while biologically irrelevant, was introduced to test whether manipulation of FGF4 levels can alter the typical time scale on which cell-fate commitment converges. Exogenous FGF4 addition was carried out at predetermined time points: $\textrm{simulation time} = [0, 4, 8, 12, 16, 24, 32, 40]$ hours, with the amount of added FGF4 molecules determined by a mix of Poissonian and binomial distributions (details in \nameref{subsection:computational_experiments}).

\begin{figure}[ht!]
\begin{adjustwidth}{-1.75in}{0in} 
\includegraphics[width = 7.5in, height = 3.75in]{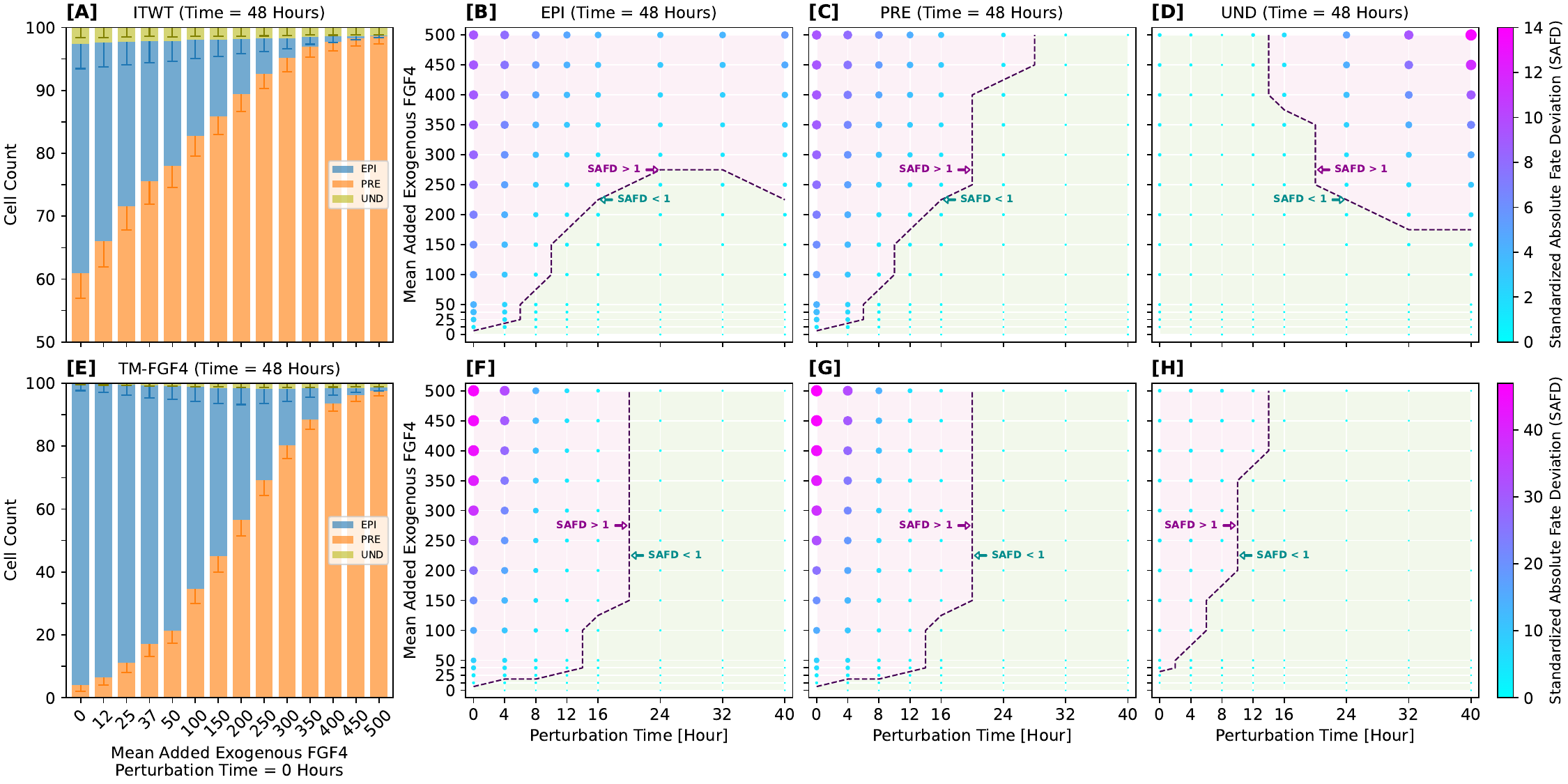} \centering 
\end{adjustwidth} 
\begin{adjustwidth}{-1.75in}{0in} 
\caption{{\bf Robustness to varying amounts of exogenous FGF4 added at distinct time points quantified by SAFD at 48~h: cell plasticity and FGF4 sensitivity are time-window-dependent.} {\bf [A]} Final cell-fate composition (at 48~h) of the inferred-theoretical wild-type (ITWT) system, for all tested FGF4 protein perturbation magnitudes and FGF4 addition at simulation start (0~h). Bar height ranges from 50 to 100 cells. {\bf [B-D]} Effect of different FGF4 perturbation magnitudes and FGF4 addition times on cell-fate composition at 48~h, quantified by Standardized Absolute Fate Deviation (SAFD) for all three fate categories (EPI, PRE, UND) in the ITWT system. SAFD (color represents amplitude) refers to the distance from the reference (or null) configuration (no exogenous FGF4 at 0~h) to every alternative perturbation magnitude and addition time. SAFD is defined as the absolute difference between the mean of reference and perturbed proportions, rescaling this absolute difference by the standard deviation of the reference proportion. The separating hyperplane indicates the boundary between the regions with distance $<= 1$ and distance $> 1$. {\bf [E]} Final cell-fate proportions (at 48~h) of the theoretical mutant system lacking FGF4 production (referred to as TM-FGF4), for all tested FGF4 protein perturbation magnitudes and FGF4 addition at simulation start (0~h). Bar height ranges from 0 to 100 cells. {\bf [F-H]} Effect of different FGF4 perturbation magnitudes and FGF4 addition times on cell-fate composition at 48~h (quantified by SAFD) in the TM-FGF4 system. All bars and data points are calculated from separate batches of 1000 independent simulations for each perturbation.}
\label{fig10}
\end{adjustwidth} 
\end{figure}

Note that the TM-FGF4 system represents a theoretical mutant with blocked FGF4 production, mimicking a full loss-of-function phenotype for the \textit{Fgf4} gene; this means that any FGF4 originates in these systems from the external addition.

We quantified the ICM patterning robustness to varying amounts of exogenous FGF4 using the Standardized Absolute Fate Deviation (SAFD), which measures the distance, in terms of mean cell-fate proportions, from the null configuration (no exogenous FGF4 at 0 hours) to every alternative perturbation-magnitude-and-addition-time configuration; see also caption of Fig~\ref{fig10} for details about SAFD.

Our analysis at the 48-hour mark shows that the TM-FGF4 model is highly responsive to exogenous FGF4 added at the start of the simulation (addition time = 0 hours). Depending on the FGF4 count, it is possible to manipulate the lineage ratios and rescue the PRE fate (Fig~\ref{fig10}[E]), which is in line with recent in vitro experiments \cite{raina_cell-cell_2021}.

\clearpage
\begin{figure}[ht!]
\begin{adjustwidth}{-1.75in}{0in} 
\includegraphics[width = 7.5in, height = 3.75in]{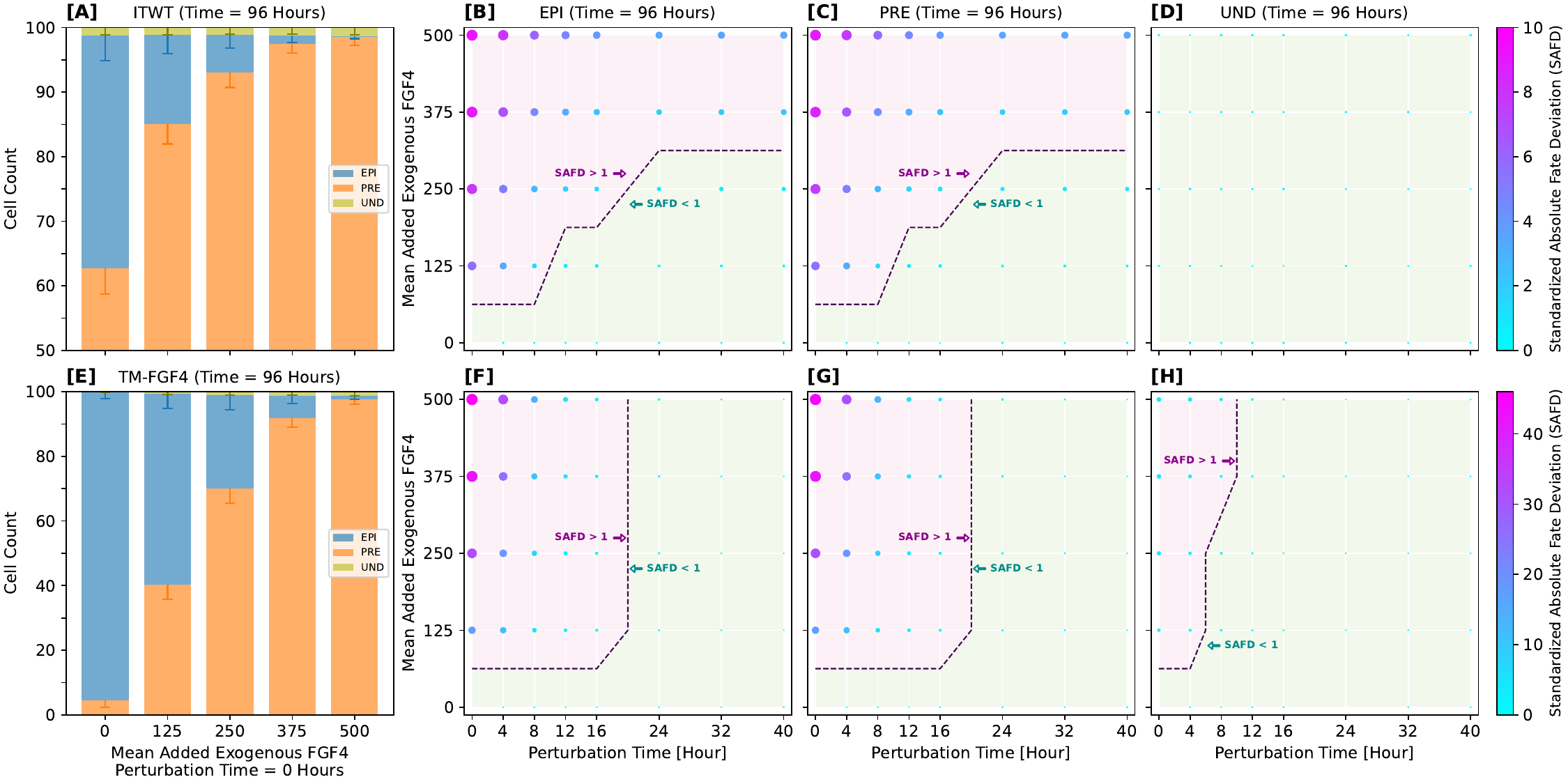} \centering 
\end{adjustwidth} 
\begin{adjustwidth}{-1.75in}{0in} 
\caption{{\bf Robustness to varying amounts of exogenous FGF4 added at distinct time points quantified by SAFD at 96~h: cell-fate composition perturbed by non-endogenous FGF4 addition remains largely unchanged beyond 48~h.} {\bf [A]} Final cell-fate composition (at 96~h) of the Inferred-Theoretical Wild-Type (ITWT) system, for all tested FGF4 protein perturbation magnitudes and FGF4 addition at simulation start (0~h). Bar height ranges from 50 to 100 cells. {\bf [B-D]} Effect of different FGF4 perturbation magnitudes and FGF4 addition times on cell-fate composition at 96~h (quantified by SAFD) in the ITWT system. {\bf [E]} Final cell-fate proportions (at 96~h) of the theoretical mutant system lacking FGF4 production (referred to as TM-FGF4), for all tested FGF4 protein perturbation magnitudes and FGF4 addition at simulation start (0~h). Bar height ranges from 0 to 100 cells. {\bf [F-H]} Effect of different FGF4 perturbation magnitudes and FGF4 addition times on cell-fate composition at 96~h (quantified by SAFD) in the TM-FGF4 system. See caption of Fig~\ref{fig10} for definition of SAFD.}
\label{fig11}
\end{adjustwidth} 
\end{figure}

However, we identify an end-of-plasticity time point around 24 hours, after which the system becomes insensitive to additional FGF4 and locks into its pre-existing cell lineage proportions (Fig~\ref{fig10}[F-H]). In the ITWT system, cell fate ratios can also be manipulated by varying the FGF4 amount, although no distinct end-of-plasticity time point is observed in this model (Fig~\ref{fig10}[A-D]).

Extending the simulation to 96 hours, the TM-FGF4 exhibits similar behavior as in the 48-hour case (Fig~\ref{fig11}[E-H]). However, the ITWT system now displays a gradual loss of cell plasticity; beyond the 32-hour mark, additional FGF4 does not significantly alter final fate ratios, provided the average amount of added FGF4 remains below 250 molecules per cell (Fig~\ref{fig11}[A-D]).

In summary, our simulations with exogenously administered FGF4 underscore that ICM cell plasticity is confined to a specific time window. The ICM population's transient sensitivity to external FGF4 allows for the maintenance of EPI and PRE lineage proportions under normal conditions. Nevertheless, the balance between these two fates can be influenced by the timing and concentration of exogenous FGF4, showcasing the nuanced interplay between external factors and intrinsic developmental processes.



\section*{Discussion}


The specification of the inner cell mass (ICM) lineages is a pivotal process in mouse blastocyst formation and an important paradigm in tissue development. During this process, two distinct cell lines, the epiblast (EPI) and the primitive endoderm (PRE), differentiate in a reliable manner without any dependency on maternal inputs, unfolding from the zygote in a completely self-organizing fashion. To this end, the embryo orchestrates multiple subprocesses at two ancillary spatio-temporal scales: at the single-cell level, complex regulatory interactions concertedly calibrate genetic programs, partly responding to membrane-receptor mediated feedback loops that can couple them to neighboring ICM cells; at the tissue level, globally conserved features materialize driven by biochemical signaling throughout the system.


The stochastic character of gene expression dynamics, together with the relatively small number of cells forming the early mouse embryo, make ICM differentiation an inherently noisy process. Therefore, correct progression of ICM specification not only depends on a stochastic surge of the EPI and PRE fates, but also leans on reliable cellular maturation of these arising lineages. Maintaining a well-balanced ratio between EPI and PRE populations is particularly important in this context, as breaking this balance can have significant physiological implications for the early mouse embryo \cite{saiz_asynchronous_2016, saiz_coordination_2020}. A successful conclusion of these processes requires mechanisms that make the developing tissue robust against intrinsic noise and extrinsic perturbations. Recent studies have shown that an FGF-mediated cell-cell communication mechanism constitutes a necessary precondition for the robust emergence of the two distinct ICM lineages, seemingly adding a deterministic dimension to this process \cite{saiz_growth-factor-mediated_2020, raina_cell-cell_2021}.

The inherent presence of randomness in ICM differentiation is evidenced by the significant cell-cell heterogeneity observed in experiments \cite{simon_making_2018, ochiai_genome-wide_2020, robert_initial_2022, allegre_nanog_2022}, and reproducing these characteristics sets a benchmark for any faithful model of this system. Therefore, a genuinely stochastic modeling approach that realistically incorporates the noisy dynamics of gene regulatory networks (GRNs) and cellular signaling pathways is necessary for understanding cell fate specification during early mouse development, as well as for quantifying its robustness.

Several phenomenological models have been proposed for blastocyst formation in the mouse embryo \cite{bessonnard_gata6_2014, tosenberger_multiscale_2017, saiz_growth-factor-mediated_2020, stanoev_robustness_2021}, but they are primarily deterministic in nature, and as such do not allow for a rigorous investigation of the implications of noise emerging from the basic processes driving this developmental process, neither for quantification of its robustness. To correctly capture the inherent randomness in ICM differentiation, we developed a biophysics-rooted spatial-stochastic model simulated via the Reaction-Diffusion Master Equation (RDME) formalism, and embedded it into a Simulation-Based Inference (SBI) framework building on recent advancements in Machine Learning (ML). Our multi-cellular model mechanistically describes the biochemical ICM patterning dynamics and its accuracy using biologically realistic lifetimes for the involved molecular species, and provides a biophysically realistic implementation of the mesoscopic processes generating noise at the cell level. Using this combined framework, we inferred multiple parameter distributions that inform our model both in wild-type-like and several mutant-like conditions.


Our inferred theoretical wild-type (ITWT) model recovers key experimental findings, specifically: high reproducibility of the EPI and PRE lineage proportions, the observed timescale of blastocyst formation, amounting to 1.5-2 days of embryonic development, and the indispensability of FGF4-mediated signaling for appropriate ICM patterning \cite{bessonnard_gata6_2014, saiz_asynchronous_2016, allegre_nanog_2022}. In the absence of FGF4 coupling, the simulated blastocyst fails to establish correct cell fate proportions, displaying a strong bias towards the EPI fate.

We thus argue that, given a default naive pluripotent state in ICM-like systems, successful cell-fate specification necessitates a tissue-level coordinating mechanism. Previous experimental and simulation results also underscore the importance of cell-cell signaling in maintaining reproducible lineage proportions globally while facilitating correct pattern formation locally \cite{fischer_transition_2020, saiz_growth-factor-mediated_2020, raina_cell-cell_2021}. Interestingly, our analysis of cell neighborhood composition reveals that the communication range of FGF4, though essentially limited to nearest neighbors, suffices for ensuring effective signaling. This observation aligns with recent findings from both in vitro and in silico studies \cite{raina_cell-cell_2021, dirk_recognition_2023}, highlighting the nuanced roles of local signaling dynamics in complex tissue patterning.


We find that system size (number of cells) does not significantly influence the accuracy of attaining the correct lineage proportions. This is in line with previous studies reporting that the mouse blastocyst exhibits resilience to ICM size variations, maintaining consistent patterning irrespective of cell number \cite{nissen_four_2017, saiz_growth-factor-mediated_2020}. Although, conversely, we find the precision of cell specification to be system-size dependent, our results suggest that cell-fate misspecification can be reduced down to $\sim{10}\%$ when the system size surpasses $\sim{50}$ cells, which is comparable to the typical number of cells in the ICM around E3.5. Moreover, our simulations show that increased variability in initial conditions at the cellular level does not necessarily constitute a detriment for the tissue-level dynamics. Instead, increased variability in initial molecular resources, if not excessive, can enhance the accuracy while sustaining the precision of cell-fate specification.

Our ITWT model also successfully recapitulates the temporal sensitivity of the ICM to exogenous FGF4. We observe a specific time window during which the ICM can respond to external FGF4, with the ability to adjust lineage proportions depending on the timing and dosage of the addition. This finding aligns with recent experimental observations \cite{raina_cell-cell_2021} and highlights the importance of timing in developmental processes.


A paramount challenge in biophysical mechanistic modeling is the estimation of parameter values allowing the constructed model to faithfully recapitulate the characteristics of the considered biological system. This task becomes particularly complex for spatial-stochastic and mechanistic models due to the need for analyzing behavior across numerous independent simulation samples, significantly increasing the computational demands for navigating their vast parameter spaces \cite{wang_massive_2019, perez_efficient_2022}.

In response to this challenge, our approach integrates an AI-powered Simulation-Based Inference (SBI) method with traditional machine learning techniques, specifically employing the Sequential Neural Posterior Estimation (SNPE) algorithm \cite{greenberg_automatic_2019, cranmer_frontier_2020, deistler_truncated_2022}. This strategy leverages simulation data to efficiently traverse parameter space, incorporating both direct analysis of simulation outcomes and qualitative data to identify parameter distributions that align with the expected behaviors of the system, encoded in high-level, low-dimensional utility functions (``target scores'').

We utilized a state-of-the-art SBI toolbox \cite{tejero-cantero_sbi_2020}, which facilitated the integration of the SNPE algorithm into our workflow. This allowed us to train artificial neural networks (ANNs) for predicting model parameter sets capable of reproducing the targeted ICM patterning behavior of several model variants, corresponding to both wild-type and mutant systems.

Our methodology imposes minimal constraints on the inference problem by leveraging only essential experimental observations. This strategy prevents model overfitting, allowing for the extrapolation of system behaviours spanning multiple spatial and temporal scales not directly observed in experimental data.

With this approach, we underscore the distinctions between machine learning models, which prioritize universal prediction at the expense of modeling interpretability, and mechanistic models, which focus on exploratory hypotheses to uncover causal relationships at the expense of modeling fidelity \cite{baker_mechanistic_2018, torregrosa_mechanistic_2021}. By merging these paradigms, we demonstrate that despite the scarcity of detailed quantitative experimental measurements, the flexibility and predictive capabilities of ANNs can aid generating full-featured quantitative predictions by imposing key empirical qualitative observations to mechanistic biophysical models. To our knowledge, our work constitutes the first application of an AI-powered SBI framework to spatial-stochastic modeling in developmental biology.

While here we focused on a minimal spatial geometry for targeted assessment of the interplay between biochemical stochasticity and spatial coupling, ICM development is influenced by important additional factors, such as cell divisions and mechanical interaction among cells. Future elaborations of our framework will incorporate suitable tissue-scale dynamics, which will integrate the stochastic dynamics of single-cell gene expression and inter-cellular signaling with the constant remodeling of the tissue geometry.

This will enable the study of how cell neighborhoods varying both in time and space influence ICM lineage differentiation, while exploiting recently recorded tissue structural data \cite{fischer_transition_2020, forsyth_iven_2021, yanagida_cell_2022, fischer_salt-and-pepper_2023}.

The extended framework will also feature the other two important constitutive elements of the developing mouse blastocyst, namely the blastocoel (blastocyst cavity) and the trophectoderm (TE), exposing an interesting research direction, as recent experimental evidence strongly suggests that the expansion of the mouse blastocyst lumen could play a critical role in stimulating ICM fate differentiation. This is thought to occur through an interplay of mechanical clues and position-specific induction of gene-expression, possibly mediated by FGF4 molecules deposited in the blastocoel \cite{ryan_lumen_2019, shahbazi_mechanisms_2020}.

In conclusion, our AI-parameterized model underscores the complexity and robustness of the EPI-PRE lineage specification, generating unique insights into the interplay of stochasticity, biochemical feedbacks, tissue-level signaling mechanisms, and system size (number of cells) in ICM development. These findings not only deepen our understanding of the developing early mouse embryo but also provide a comprehensive framework for exploring similar controlled stochastic processes in related biological systems.


\section*{Materials and methods} \label{section:materials_methods}

\subsection*{Computational model of mouse blastocyst (ICM cell differentiation)} \label{subsection:computational_model}

The model comprises two fundamental building blocks. The first submodel (cell level) consists of the GRN (NANOG-GATA6-FGF4) coordinating the ICM cell specification process. The second submodel (tissue level) describes the cell-cell signaling dynamics. Unlike other existing models \cite{cang_multiscale_2021, stanoev_robustness_2021, tosenberger_multiscale_2017, robert_initial_2022}, our modeling approach does not integrate the notion of noise as a purely extrinsic component. Instead of an arbitrary noise source, we employ a mesoscale description which incorporates noise as an intrinsic component. Thus, noise plays an essential role for faithfully simulating the temporal evolution of our biological system model.

Indeed, the presence of noise in biophysical models is deemed central for discerning the main features of gene regulatory processes \cite{kar_exploring_2009, ng_stochastic_2018, vandevenne_rna_2019}. Conventionally, noise is separated into intrinsic and extrinsic categories \cite{swain_intrinsic_2002, eldar_functional_2010}. While it is problematic to give a clear delimitation of these two categories, here we provide a general interpretation of their scope within the context of our study.

Intrinsic noise arises from the nature of biochemical reaction and diffusion events; i.e., discrete molecules randomly diffuse and randomly react when a collision occurs between each other. As such, intrinsic noise commonly refers to local fluctuations within basic gene regulation mechanisms; e.g., transcription and translation. Extrinsic noise originates from cellular environment variations or changes. Hence, extrinsic noise typically alludes to global factors systematically affecting all cells but irregularly propagating across cellular mechanisms; for example, cell cycle timing and cellular resource partitioning. Nevertheless, recent experimental and theoretical works argue for treating both noise categories as inseparable entities \cite{sherman_cell--cell_2015, justman_explicit_2015, de_jong_gene_2019, bartz_progress_2023}.

When there is a large number of molecules at play, a biochemical dynamics model typically follows a deterministic formulation: reaction rates are represented by constant functions, species amounts are represented by concentrations (continuous-time functions), and it primarily follows an ordinary differential equation (ODE) scheme. By contrast, when there is a small number of molecules at play, a stochastic formalism takes precedence.

Generally, stochastic biochemical dynamics models are formulated as continuous-time Markov chains (CTMCs); i.e., continuous-time discrete state-space Markov processes. Numerous mathematical and computational methods have been developed for analysis and simulation of such stochastic formulations \cite{gillespie_exact_1977, elf_fast_2003, munsky_multiple_2007, anderson_modified_2007, gillespie_perspective_2013, simoni_stochastic_2019, erban_stochastic_2020, gupta_deepcme_2021}. These techniques methodically incorporate stochasticity, which is relevant for understanding the effects of noise on cell-cell variability.

\clearpage

A biochemical reaction network involves multiple reactions (edges) and species (vertices or nodes); a CTMC is the most common model of such a network. Particularly, biophysical systems can be abstracted using the Chemical Master Equation (CME) formalism; Eq~(\ref{eq:cme}) \cite{coulier_multiscale_2021, gupta_deepcme_2021}.
\begin{equation} 
\frac{\partial{P}}{\partial{t}}(\boldsymbol{x},t{\mid}\boldsymbol{x}_{0},t_{0}) = \sum_{j=1}^{M}a_{j}(\boldsymbol{x}-\boldsymbol{v}_{j},t)P(\boldsymbol{x}-\boldsymbol{v}_{j},t{\mid}\boldsymbol{x}_{0},t_{0})-a_{j}(\boldsymbol{x},t)P(\boldsymbol{x},t{\mid}\boldsymbol{x}_{0},t_{0})
\label{eq:cme}
\end{equation}
Where: $\boldsymbol{x}$ is the state vector of the system $\boldsymbol{X}$ (CTMC); $\boldsymbol{x} = \boldsymbol{X}(t) = [X_{1}(t),\ldots,X_{N}(t)]$; there are $N$ biochemical species ($i \in \{1,\ldots,N\}$); each entry of $\boldsymbol{x}$ represents the copy number of a given biochemical species $S_{i}$; $P(\boldsymbol{x},t{\mid}\boldsymbol{x}_{0},t_{0})$ is the time-dependent probability density function of $\boldsymbol{x}$; $\boldsymbol{x}_{0}$ is the initial state vector; $t_{0}$ is the initial time; there are $M$ reactions ($j \in \{1, \ldots, M\}$); $a_{j}$ is the nominal rate of reaction $R_{j}$; $v_{j}$ is the state-change vector (set of stoichiometric coefficients) of reaction $R_{j}$; $a_{j}(\boldsymbol{x},t)$ is the propensity function (effective rate) of reaction $R_{j}$ when the system $\boldsymbol{X}$ is in state $\boldsymbol{x}$ at time $t$.

The full GRN implemented by our simulator includes all the molecular species relevant for the developmental system dynamics, together with several auxiliary (computational) species; this procedure facilitates the inclusion of all important molecular relationships and the tracking of crucial model variables. An extensive list of species, relations, and nomenclature guidelines is available from the corresponding simulator scripts; see [SUPPORTING INFORMATION]. For ease of exposition, Table~\ref{table1} presents only the actual biochemical species considered for our model.

\begin{table}[ht!]
\caption{{\bf Summary of biochemical species.}}
\centering
\begin{tabular}{ | l | r | }
\hline
\rowcolor{lightgray}
Name & Description \\
\hline
\textit{Nanog} & \{Gene, Promoter\} \\
\textit{Gata6} & \{Gene, Promoter\} \\
\textit{Fgf4} & \{Gene, Promoter\} \\
\textit{Nanog} mRNA & Messenger RNA \\
\textit{Gata6} mRNA & Messenger RNA \\
\textit{Fgf4} mRNA & Messenger RNA \\
NANOG & Protein \\
P-NANOG & \{Protein, Phosphorylated NANOG\} \\
GATA6 & Protein \\
FGF4 & Protein \\
FGFR & \{Protein, Fibroblast Growth Factor Receptor (1/2)\} \\
M-FGFR-FGF4 & \{Protein, FGFR-FGF4 Monomer Complex\} \\
D-FGFR-FGF4 & \{Protein, FGFR-FGF4 Dimer Complex\} \\
I-ERK & \{Protein, Inactive ERK (1/2)\} \\
A-ERK & \{Protein, Active ERK (1/2)\} \\
\hline
\end{tabular}
\begin{flushleft}
Note. Our computational model has more than 50 species, but we only introduce in this table the species directly related to the biological problem. The other purely-computational species are necessary for correctly analyzing and tracing the complex GRN simulated dynamics.
\end{flushleft}
\label{table1}
\end{table}

Within the CME framework, a well-mixed or reaction-limited system is the main assumption; i.e., molecular diffusion is relatively fast compared to the speed of any biochemical reaction. The most popular method to simulate models following the CME formulation is the Stochastic Simulation Algorithm (SSA), a scheme introduced and rigorously proven to be physically relevant by the late Daniel T. Gillespie \cite{gillespie_exact_1977, gillespie_rigorous_1992}.

Correspondingly, molecular diffusion speed can guide the choice of a biochemical dynamics representation. Fast diffusion is synonym with spatially-uniform distribution of resources; a well-mixed or homogeneous environment. Slow diffusion is synonym with spatial correlation and other spatial factors, which creates a heterogeneous environment.

While there exist multiple techniques tackling different spatial and temporal scales \cite{andrews_detailed_2010, gillespie_perspective_2013, gupta_spatial_2018, engblom_stochastic_2019, sokolowski_egfrd_2019}, we aimed for balance between computational efficiency and biophysical realism. Consequently, we have followed the formalism of the Reaction-Diffusion Master Equation (RDME); Eq~(\ref{eq:rdme}) \cite{hellander_reaction-diffusion_2012, barrows_parameter_2023}.
\begin{equation} 
\begin{gathered}
\begin{aligned}
\frac{\partial{P}}{\partial{t}}(\boldsymbol{x},t{\mid}\boldsymbol{x}_{0},t_{0}) & = \mathcal{R}P(\boldsymbol{x},t{\mid}\boldsymbol{x}_{0},t_{0}) + \mathcal{D}P(\boldsymbol{x},t{\mid}\boldsymbol{x}_{0},t_{0})
\\
\mathcal{R}P(\boldsymbol{x},t{\mid}\boldsymbol{x}_{0},t_{0}) & = \sum_{j=1}^{M}\sum_{k=1}^{L}a_{j}(\boldsymbol{x}_{k}-\boldsymbol{v}_{j},t)P(\boldsymbol{x}_{1},\ldots,\boldsymbol{x}_{k}-\boldsymbol{v}_{j},\ldots,\boldsymbol{x}_{N},t{\mid}\boldsymbol{x}_{0},t_{0}) \\
& \quad -a_{j}(\boldsymbol{x}_{k},t)P(\boldsymbol{x},t{\mid}\boldsymbol{x}_{0},t_{0})
\\
\mathcal{D}P(\boldsymbol{x},t{\mid}\boldsymbol{x}_{0},t_{0}) & = \sum_{i=1}^{N}\sum_{k=1}^{L}\sum_{h=1}^{L}P(\boldsymbol{x}_{1},\ldots,\boldsymbol{x}_{ik}-\boldsymbol{w}_{ikh},\ldots,\boldsymbol{x}_{N},t{\mid}\boldsymbol{x}_{0},t_{0}) \\
& \quad \times b_{ikh}(\boldsymbol{x}-\boldsymbol{w}_{ikh},t)-P(\boldsymbol{x},t{\mid}\boldsymbol{x}_{0},t_{0}) \times b_{ikh}(\boldsymbol{x}_{ik},t)
\end{aligned}
\end{gathered}
\label{eq:rdme}
\end{equation}
Where: $\mathcal{R}P(\boldsymbol{x},t{\mid}\boldsymbol{x}_{0},t_{0})$ and $\mathcal{D}P(\boldsymbol{x},t{\mid}\boldsymbol{x}_{0},t_{0})$ are the reaction and diffusion components of the equation, respectively; $\boldsymbol{x}_{k}$ (or $\boldsymbol{x}_{h}$) is the state vector of the voxel $\Omega_{k}$ (or $\Omega_{h}$); $\boldsymbol{x}_{ik}$ (or $\boldsymbol{x}_{ih}$) is the copy number of $S_{i}$ for $\Omega_{k}$ (or $\Omega_{h}$); there are $L$ voxels ($k,h \in \{1, \ldots, L\}$); $b_{ikh}$ is the nominal diffusion rate of $S_{i}$ from $\Omega_{k}$ to $\Omega_{h}$; $w_{ikh}$ is the state-change vector (set of stoichiometric coefficients) for the diffusion of $S_{i}$ from $\Omega_{k}$ to $\Omega_{h}$.

The RDME framework works at the mesoscopic level, and its simulation schemes are based on custom versions of the SSA tailored to incorporate reaction-diffusion processes. Here, we depart from one of such schemes, the Next Subvolume Method (NSM) \cite{fange_stochastic_2010, erban_stochastic_2020}, which separates events into two distinct kinds: reaction firing inside every cell, and diffusive jumps between cells. For our system model, each cell is treated as a well-mixed voxel/environment, and tissue communication materializes by representing signaling-molecule diffusion as a morphogen-exchange process between neighboring cells; as such, we will commonly refer to this process simply as ``diffusive jump'', ``jump diffuse'', or ``jump diffusion''.

To put it briefly, our event-driven simulator is congruent with the NSM because it involves the SSA, the computational spatial domain is partitioned into artificially well-mixed compartments where only molecules belonging to the same compartment can react, diffusive jumps transport molecules between neighboring voxels, and there are well-defined event queues. Outside these shared features, our simulator allows for complex interactions among voxels or cells, which facilitates the presence of multiple tissue types and the corresponding relabeling of molecules once they undergo jump-diffuse steps. Likewise, the nominal diffusive-jump rates are calculated based on an arbitrary system model geometry and the principle of conservation of (molecular) flow; unlike the NSM, which calculates the nominal jump-diffuse rates based on a regular cubic geometry and the voxel size.

\subsubsection*{Key stages of mouse embryo preimplantation development} \label{subsubsection:preimplantation_stages}

To guide our model construction and in silico analysis, we relied on wet-lab experimental descriptions of core phases in early mouse development. The mouse preimplantation period encompasses a series of morphological and molecular changes which transform the zygote (one totipotent cell) into an approximately 256-cell (7-8 cleavages) embryo at around E4.5; at this point, the embryo comprises three spatially segregated cell types: TE, EPI, and PRE. For a complete recap of the mouse embryo preimplantation development, please see \cite{simon_making_2018, plusa_common_2020, saiz_coordination_2020}.

The first cell-fate decision happens between E2.5 and E3.0 (from 8- to 32-cell stage): cells acquire TE or ICM identities. The second cell-fate decision happens at the ICM between E3.0 and E4.0 (from 32- to 128-cell stage). From E4.0 to 4.5, EPI and PRE populations spatially separate.

\clearpage

While it is customary to define the blastocyst-formation period between E3.0 and E4.5 \cite{nissen_four_2017, allegre_nanog_2022}, these boundaries are ultimately arbitrary as development occurs in a continuum and diverse experimental arrangements/conditions are in use between distinct labs. Moreover, ICM cells adopt their next identities asynchronously as the blastocyst forms \cite{saiz_asynchronous_2016, bessonnard_icm_2017, simon_making_2018}. Together with these aspects, it is also commonly accepted that cells are already coexpressing \textit{Nanog}- and \textit{Gata6}-related factors at around E2.75 \cite{plusa_distinct_2008, plusa_common_2020}, plus \textit{Fgf4} expression is already perceptible at around E3.25 \cite{nissen_four_2017, allegre_nanog_2022}. For these reasons, our standard model simulations target a time-window of 48 hours (E2.75-E4.75); this range allows us to circumvent potential discrepancies among timing annotations and keep a temporally faithful description of the biological system under study.

\subsubsection*{Fundamental interactions among central GRN components}

Many processes coexist during blastocyst development. These processes materialize at multiple temporal/spatial scales and embody the relationships of numerous components operating simultaneously. A vast number of elements conjointly orchestrate developmental progress scaling from cell-level adaptable gene expression mechanisms to tissue-level mechanical/signaling coordination structures. Particularly, the GRN controlling the ICM specification process has a rich collection of components and interactions. Here, we model this GRN by accounting for the key interactions among its main components, as reported by recent experimental studies.

To start, we suppose that our core GRN motif consists of the species and interactions primarily governing the \textit{Nanog}- and \textit{Gata6}-gene expression dynamics. This collection of ingredients only includes \textit{Nanog} mRNA, \textit{Gata6} mRNA, NANOG protein, and GATA6 protein, naturally. As transcription factors (TFs), both NANOG and GATA6 proteins exhibit self-activation and mutual repression \cite{saiz_coordination_2020, thompson_extensive_2022}.

The remainder of the complete GRN encompasses all the species and interactions secondarily governing the \textit{Nanog}- and \textit{Gata6}-gene expression dynamics. This group includes \textit{Fgf4} mRNA, FGF4 protein, and ERK protein. Among these explicit elements, we also implicitly include two FGF receptor (FGFR) complexes, which concertedly facilitate biochemical signal transduction during blastocyst formation \cite{kang_lineage_2017,molotkov_distinct_2018}. Recently, a comprehensive experimental study demonstrated that NANOG and GATA6 proteins are capable of jointly binding to both EPI and PRE \textit{cis}-regulatory modules \cite{thompson_extensive_2022}. This concrete evidence supports the previously proposed direct NANOG activation plus GATA6 repression of the \textit{Fgf4} gene, both in vitro and in vivo \cite{demot_cell_2016, tosenberger_multiscale_2017}. Likewise, ERK has been indicated to play a crucial role for this GRN \cite{simon_live_2020, yeh_capturing_2021}. At transcriptional level, ERK is capable of recruiting diverse repressor TFs to \textit{Nanog}-gene loci \cite{kale_nanog-perk_2022}. For antisymmetry and simplicity, we assumed that ERK is capable of recruiting diverse activator TFs to \textit{Gata6}-gene loci; however, there is indeed some experimental evidence indicating such a motif \cite{meng_gata6_2018}. At post-translational level, NANOG phosphorylation by ERK promotes its instability, which consequently reduces its lifetime \cite{kim_erk1_2014, liu_usp21_2016, kale_nanog-perk_2022}. Contrastively, it has been reported that GATA6 phosphorylation by ERK enhances its stability; nevertheless, the implications of this motif are not completely clear and we exclude it \cite{meng_gata6_2018}.

Finally, the upstream release of FGF4 induces the FGFR-FGF4 monomer complex formation, which successively induces the FGFR-FGF4 dimer complex formation. This FGFR-FGF4 dimer complex ultimately triggers the pathways downstream of ERK \cite{deathridge_live_2019, simon_live_2020, raina_fgf4_2021, yeh_capturing_2021}.

\subsubsection*{Model at cell scale} \label{subsubsection:cell_scale_model}

The core GRN motif is exclusively comprised by \textit{Nanog}- and \textit{Gata6}-related elements. To be more precise, all their directly related species and interactions. The rest of the full GRN is built around the core motif, thus consolidating the remaining elements and their collective effects on the dynamics of the two main players.

\clearpage

Importantly, we have arranged all cell-scale reaction events into several groups as follows: summary of gene expression dynamics; promoter binding and unbinding; mRNA synthesis and degradation; protein synthesis and degradation; FGFR activation and inactivation; ERK activation and inactivation; NANOG phosphorylation and dephosphorylation. In that regard, we report all the particular interactions implemented by our simulator and their respective literature sources.

\underline{Summary of gene expression dynamics.} The only three genes with an explicit mRNA step are \textit{Nanog}, \textit{Gata6}, and \textit{Fgf4}; Table~\ref{table2} summarizes their relationships. The expression of the other two genes (\textit{Fgfr} and \textit{Erk}) is only visible either at an implicit form or at the protein level. FGFR also does not have an explicit protein count as it is available rather uniformly on the cell membrane \cite{ornitz_fibroblast_2015, kang_lineage_2017, molotkov_distinct_2018, ornitz_new_2022, karl_ligand_2023}; instead, FGFR appears as an implicit component of the auxiliary protein variables/species M-FGFR-FGF4 (FGFR-FGF4 monomer complex) and D-FGFR-FGF4 (FGFR-FGF4 dimer complex), which helps reducing the number of reactions as well as alleviating the computational resources. ERK has itself two different protein forms: I-ERK (inactive ERK) is abundant in the cell cytoplasm \cite{lavoie_erk_2020, raina_fgf4_2021}, and it is already present at the start of all the simulations; A-ERK (active ERK) is always inversely proportional to I-ERK, thus it is a product of the action of D-FGFR-FGF4 on I-ERK.

\begin{table}[ht!]
\caption{{\bf Regulation of gene transcription.}}
\centering
\begin{tabular}{ | c | c | c | c | c | }
\hline
\rowcolor{lightgray}
Gene (Promoter) & TF & TFBSs & TF Role & PVSs \\
\hline
\textit{Nanog} & \{NANOG, P-NANOG\} & 4 & Activator & \cite{bessonnard_gata6_2014, tosenberger_multiscale_2017} \\
\textit{Nanog} & GATA6 & 4 & Repressor & \cite{bessonnard_gata6_2014, tosenberger_multiscale_2017} \\
\textit{Gata6} & \{NANOG, P-NANOG\} & 4 & Repressor & \cite{bessonnard_gata6_2014, tosenberger_multiscale_2017} \\
\textit{Gata6} & GATA6 & 4 & Activator & \cite{bessonnard_gata6_2014, tosenberger_multiscale_2017} \\
\hline
\textit{Fgf4} & \{NANOG, P-NANOG\} & 2 & Activator & Self \\
\textit{Fgf4} & GATA6 & 2 & Repressor & Self \\
\hline
\textit{Nanog} & A-ERK & 3 & Repressor & \cite{bessonnard_gata6_2014, tosenberger_multiscale_2017} \\
\textit{Gata6} & A-ERK & 3 & Activator & \cite{bessonnard_gata6_2014, tosenberger_multiscale_2017} \\
\hline
\end{tabular}
\begin{flushleft}
Notation: TF = Transcription Factor; TFBSs = TF Binding Sites; P-NANOG = Phosphorylated NANOG; A-ERK = Active ERK; PVSs = Parameter Value Sources.
\end{flushleft}
\label{table2}
\end{table}

\underline{Promoter binding and unbinding.} Each of the three genes \textit{Nanog}, \textit{Gata6}, and \textit{Fgf4} has a respective promoter with multiple independent binding sites for each of its TFs; check Table~\ref{table3}. Both gene activation and repression are cooperative: exactly \emph{q} TF copies must be simultaneously bound to their particular promoter sites for activation or repression of expression; by default, repression takes precedence over activation. For a given TF \emph{A}, the (time-dependent) effective promoter binding rate is calculated via the formula $\widetilde{k}_{b} = 4{\pi}dDA_{t}/V$ (diffusion-limited regime). Here, $d = 10$ nm is a typical binding-site diameter \cite{sokolowski_mutual_2012}, $D = 10$ \textmu{m}\textsuperscript{2}s\textsuperscript{\textminus{1}} is the cytoplasmic/nuclear TF diffusion coefficient, \emph{A\textsubscript{t}} is the TF copy number at time \emph{t}, and $V = 4200$ \textmu{m}\textsuperscript{3} is a typical mouse blastocyst cell volume \cite{aiken_direct_2004, van_zon_diffusion_2006, vijaykumar_intrinsic_2017}. For our system model, we do not know the diffusion coefficients of all the biochemical species, thereby we simply made an educated guess and assumed the same value for all the TFs based on other representative biological systems \cite{milo_bionumbersdatabase_2010, sokolowski_mutual_2012, coulier_systematic_2022, coulier_multiscale_2021}. Concisely, the nominal promoter binding rate is determined by the equation $k_{b} = \widetilde{k}_{b}/A_{t} = 4{\pi}dD/V$.

We model TF cooperativity by expressly tuning the promoter unbinding rates. This rate tuning influences the promoter regulation model to mimic a Hill-function-like (nonlinear) transcriptional response. The usage of the Hill function is a staple of phenomenological modelling, however it is incompatible with mechanistic modelling; directly using a Hill equation as a reaction propensity function ignores the non-instantaneous (stochastic) nature of delays between biochemical events and introduces several other simulation artifacts \cite{frank_input-output_2013, chen_stochastic_2017, bottani_hill_2017}.

\clearpage

We use the concepts of a half-saturation constant and a cooperativity coefficient to perform promoter unbinding rate tuning; accordingly, both quantities are incorporated into elementary reactions to describe promoter unbinding dynamics. This half-saturation constant (\emph{h\textsubscript{act}} or \emph{h\textsubscript{rep}} depending on TF role) is a free model parameter which dictates the threshold of TF copies needed for reaching 50\% of negative or positive gene transcriptional control; consequently, each gene-TF pair requires its own separate half-saturation threshold. This cooperativity coefficient (\emph{k\textsubscript{coop}}) is an auxiliary variable which adjusts the strength of mutual influence among TF copies; we arbitrarily defined it as $k_{coop} = 5$ to increase TF-cooperativity potency (i.e., $\uparrow{k_{\emph{coop}}} \Rightarrow \downarrow{k_{u}}$). Specifically, we calculate the nominal promoter unbinding rates via the formula $k_{u} = h_{\emph{sat}}k_{b}/k_{\emph{coop}}^{q}$. Where, for a given gene-TF pair: $h_{\emph{sat}} \in \{h_{\emph{rep}}, h_{\emph{act}}\}$; \emph{k\textsubscript{b}} is its nominal promoter binding rate; \emph{q} is its maximal occupancy.

\begin{figure}[hpt!]
\includegraphics[width = 5in, height = 2.5in]{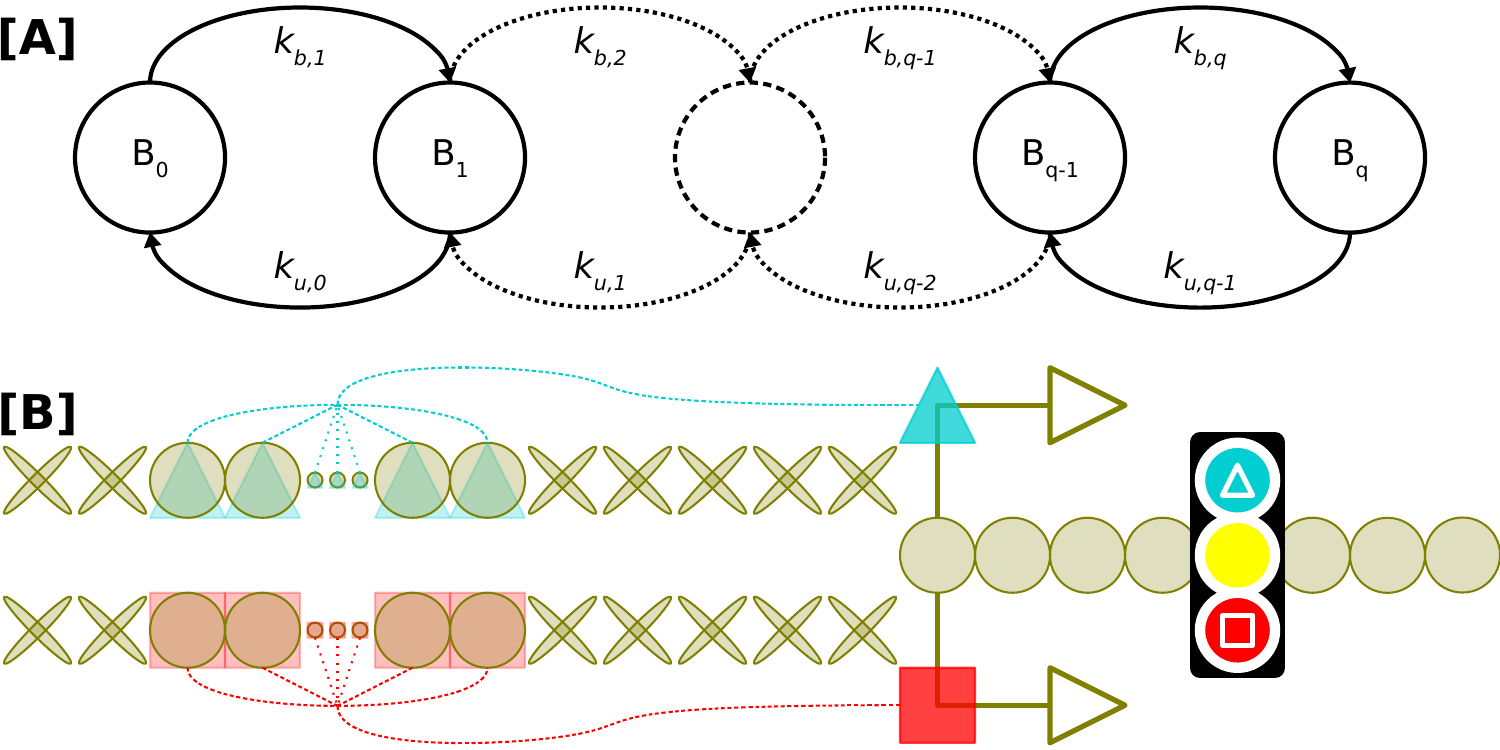} \centering 
\caption{{\bf Architecture of the modeled gene regulatory region.} Transcription factor (TF) \emph{A} binds to regulatory sites \emph{B} at an effective rate proportional to $k_b$ and unbinds from them at an effective rate proportional to $k_u$. Unlike $k_b$, $k_u$ is directly related to \emph{q} (which is the maximal occupancy of \emph{B} by \emph{A}); i.e., $k_{b,1} = \cdots = k_{b,q}$, where \emph{q} is the number of TF binding sites (TFBSs) for \emph{A} at \emph{B}. {\bf [A]} Markov chain transition diagram of TF binding/unbinding dynamics. {\bf [B]} Every gene has a respective regulatory region with multiple independent binding sites for each of its regulating TFs. Gene activation and repression are both cooperative: exactly \emph{q} TF copies must be simultaneously bound to their particular sites for activation (cyan triangles) or repression (red squares) of expression. By default, repression takes precedence over activation.}
\label{fig103}
\end{figure}

To summarize, Eq~(\ref{eq:gtfp_bu}) illustrates the most basic reaction set of the TF promoter binding/unbinding dynamics, plus Fig~\ref{fig103} shows the elementary promoter architecture.

\begin{equation} 
A + B_{0} \xrightleftharpoons[k_{u,0}]{k_{b,1}} B_{1} \quad \cdots \quad A + B_{q-1} \xrightleftharpoons[k_{u,q-1}]{k_{b,q}} B_{q}
\label{eq:gtfp_bu}
\end{equation}
Where: \emph{A} is a given TF; \emph{B\textsubscript{Q}} is the current occupancy of a given gene promoter \emph{B} by \emph{A}; $Q \in \{0, \ldots, q\}$; \emph{q} is the maximum number of binding sites for \emph{A} at \emph{B}; \emph{k\textsubscript{b}} is the nominal promoter binding rate; \emph{k\textsubscript{u}} is the nominal promoter unbinding rate.

\underline{Synthesis and degradation of mRNA.} For the transcription model, we assume that mRNA synthesis occurs as a single-step reaction but it is only possible when the gene promoter is not under control of a repressor TF; recall that we follow the “all-or-nothing” gene activation/repression configuration. We have as well accounted for two concomitant transcription modes: basal and full-induction production. Basal transcription contributes 20\% of the maximal average steady-state mRNA copy number. The remaining 80\% of the maximum mean steady-state mRNA copy number is contributed by full-induction transcription; which is only possible when an activator TF is occupying all of its binding sites at a given promoter. Accordingly, the mRNA synthesis rate is calculated via the formula $k_{m,s} = k_{m,s,\emph{basal}} + k_{m,s,\emph{find}} = \overline{M}/\tau_{m}$. Where: $k_{m,s,\emph{basal}} = c_{\emph{basal}}\overline{M}/\tau_{m}$; $k_{m,s,\emph{find}} = c_{\emph{find}}\overline{M}/\tau_{m}$; \emph{c\textsubscript{basal}} and \emph{c\textsubscript{find}} are the basal and full-induction relative contributions (i.e. $c_{\emph{basal}} + c_{\emph{find}} = 1$), respectively. The symbol $\overline{M} \doteqdot {\langle{M_{t}}\rangle}_{\emph{equilibrium}}$ is a shorthand for the maximum mean steady-state value of mRNA copies for a given gene at full activation (\emph{M\textsubscript{t}} is the number of mRNA molecules at time \emph{t}). The symbol $\tau_{m}$ denotes the lifetime (or half-life $t_{\frac{1}{2},m}$) for a molecule of mRNA.

For \textit{Nanog}, this mean mRNA value has been indicated to reach the order of hundreds of copies; approximately 100-400 molecules \cite{xenopoulos_heterogeneities_2015, feigelman_stochastic_2016, skinner_single-cell_2016, ochiai_genome-wide_2020}. For \textit{Gata6} and \textit{Fgf4}, there are no concrete mean mRNA values reported, but it seems they are similar to the average \textit{Nanog}-expression level \cite{ohnishi_cell--cell_2014}. Analogously, the \textit{Nanog}-mRNA lifetime has been reported to be around 4-5 hours \cite{tan_brf1_2014, abranches_stochastic_2014, ochiai_stochastic_2014, feigelman_stochastic_2016}, the \textit{Gata6}-mRNA lifetime has been reported to be around 3-4 hours \cite{elatmani_rna-binding_2011, chitforoushzadeh_tnf-insulin_2016}, and we have not found concrete reports about the \textit{Fgf4}-mRNA lifetime.

For simplicity, we have considered the mean mRNA values for \textit{Nanog} and \textit{Gata6} to be the same, which classifies them as fixed parameter values (we chose $\overline{M}_{\textit{Nanog}} = \overline{M}_{\textit{Gata6}} = 250$ copies). In the case of \textit{Fgf4}, its mean mRNA value is deemed to be identical to the case of the other two genes and it is also considered a fixed parameter value. However, we additionally impose that any \textit{Fgf4} expression must be entirely regulated by NANOG and GATA6 levels; in other words, \textit{Fgf4} has no basal mRNA production ($\overline{M}_{\textit{Fgf4}} = 200$ copies). Likewise, the mRNA half-lives for \textit{Nanog}, \textit{Gata6}, and \textit{Fgf4} are determined to be the same (we chose $\tau_{m,\textit{Nanog}} = \tau_{m,\textit{Gata6}} = \tau_{m,\textit{Fgf4}} = 4$ hours).

For the mRNA degradation mechanism, we assumed that it is a first-order process: the nominal degradation rate is simply the multiplicative inverse of the lifetime; $k_{m,d} = 1/\tau_{m} = \tau_{m}^{-1}$.

In a nutshell, Eq~(\ref{eq:mrna_sd}) illustrates the reaction set of the mRNA synthesis and degradation dynamics.

\begin{equation}
\begin{gathered}
\emptyset \xrightarrow{k_{\emph{basal}}} M \\
G \xrightarrow{k_{\emph{find}}} G + M \\
M \xrightarrow{k_{m,d}} \emptyset \\
k_{\emph{basal}} = \begin{cases}
    (1-G_{\emph{rep}})k_{m,s,\emph{basal}} \quad & \textrm{if} \quad G \in \{\textit{Nanog}, \textit{Gata6}\}, \\
    0 \quad & \textrm{if} \quad G = \textit{Fgf4}.
\end{cases} \\
k_{\emph{find}} = (1-G_{\emph{rep}})G_{\emph{act}}k_{m,s,\emph{find}}
\end{gathered}
\label{eq:mrna_sd}
\end{equation}
Where: \emph{G} is a particular gene; \emph{M} is the mRNA of \emph{G}; \emph{G\textsubscript{rep}} and \emph{G\textsubscript{act}} are indicator random variables representing the current state of \emph{G}. \emph{G\textsubscript{rep}} indicates a repressed gene (full promoter occupancy by repressor TF) and \emph{G\textsubscript{act}} indicates an activated gene (full promoter occupancy by activator TF), respectively.

\underline{Synthesis and degradation of protein.} The translation model assumes that, once mRNA is available, protein synthesis occurs as a single-step reaction. This assumption holds for NANOG, GATA6, and FGF4 proteins. For ERK, as there is no mRNA step, spontaneous activity produces its inactive protein form (I-ERK), which can undergo phosphorylation and become active (A-ERK). There are several indications for high abundance of ERK during the developing blastocyst, so it is not a limiting factor for the cell signaling process \cite{fujioka_dynamics_2006, tian_mathematical_2012, aoki_quantitative_2013, deathridge_live_2019, simon_live_2020}. It has also been reported that ERK has a long lifetime (48-72 hours) \cite{busca_erk1_2016, saba-el-leil_redundancy_2016, lavoie_erk_2020}. To reflect this strong presence of ERK across the cellular reaction domain, every standard model simulation assigns to each cell an initial high amount of ERK. As well, ERK can be synthesized and degraded at a rate directly proportional to its chosen half-life ($\tau_{p,\emph{ERK}} = 48$ hours) and its chosen maximum mean steady-state protein copy number ($\overline{P}_{\emph{ERK}} = 1000$ copies). Here: $k_{p,s,\emph{ERK}} = \overline{P}_{\emph{ERK}}/\tau_{p,\emph{ERK}}$ is the rate of ERK synthesis; $k_{p,d,\emph{ERK}} = 1/\tau_{p,\emph{ERK}} = \tau_{p,\emph{ERK}}^{-1}$ is the rate of ERK degradation.

For the remaining protein species, NANOG-GATA6-FGF4, an extra assumption was made to incorporate an additional sense of bursty production: every mRNA molecule is capable of synthesizing on average 4 protein molecules before it decays naturally; i.e., the theoretical maximum average steady-state value of protein molecules per cell can be calculated to be $\overline{P}_{\emph{NANOG}} = 1000$, $\overline{P}_{\emph{GATA6}} = 1000$, and $\overline{P}_{\emph{FGF4}} = 800$ copies. Bursty gene expression has been shown to significantly increase cell-cell variability of mRNA and protein levels, which itself has been suggested to enable enhanced adaptation to environmental changes and constraints \cite{zoller_diverse_2018, ochiai_genome-wide_2020}. Thus, the nominal synthesis rates for these proteins can be calculated via the formula $k_{p,s} = \overline{P}/\tau_{p}$. For the cases of NANOG and GATA6, the protein half-lives have been reported to be 2-3 hours \cite{abranches_generation_2013, wu_distinct_2013, bates_auxin-degron_2021} and 1-3 hours \cite{elatmani_rna-binding_2011, chitforoushzadeh_tnf-insulin_2016}, respectively; we chose, for simplicity, $\tau_{p,\emph{NANOG}} = 2$ hours and $\tau_{p,GATA6} = 2$ hours. Just like the previous case, the symbol $\overline{P} \doteqdot {\langle{P_{t}}\rangle}_{\emph{equilibrium}}$ is a shorthand for the maximum mean steady-state value of protein copies for a given gene at full activation (\emph{P\textsubscript{t}} is the number of protein molecules at time \emph{t}).

We assumed that any protein degradation process simply follows first-order dynamics. This condition means that, for a particular protein, its nominal degradation rate is the multiplicative inverse of its half-life: $k_{p,d} = 1/\tau_{p} = \tau_{p}^{-1}$. This assumption not only holds for these primary proteins but also applies to all the derivative molecules: P-NANOG, M-FGFR-FGF4, and D-FGFR-FGF4. For NANOG, phosphorylation by ERK reduces its stability, which in turn accelerates its degradation and essentially halves its lifetime \cite{liu_usp21_2016}. Hence, P-NANOG half-live has been categorized as a fixed parameter value ($\tau_{p,\emph{P-NANOG}} = 1$ hour). For D-FGFR-FGF4, we did not learn any concrete information about its lifetime. However, we determined that any practical decay should occur after monomerization (indirectly via transitions between dimer and monomer configurations); as such the value for D-FGFR-FGF4 half-life is set artificially high and is a fixed parameter value ($\tau_{p,\emph{D-FGFR-FGF4}} = 240$ hour). The lifetime of M-FGFR-FGF4 ($\tau_{p,\emph{M-FGFR-FGF4}}$) is therefore a free model parameter controlling the actual extracellular stability of FGF4-related resources. A similar challenge happens for FGF4 itself, as there are contrasting reports about its lifetime \cite{ding_half-life_2021, daneshpour_macroscopic_2023}, we made its half-life a free model parameter ($\tau_{p,\emph{FGF4}}$).

In summary, maximal average steady-state protein levels for NANOG, GATA6, FGF4, and ERK are fixed model-parameter values. Protein half-lives for NANOG, GATA6, ERK, P-NANOG, and D-FGFR-FGF4 are also fixed model-parameter values. The lifetimes of FGF4 and M-FGFR-FGF4 are free model parameters. Eq~(\ref{eq:protein_sd}) illustrates the reaction set of the protein synthesis and degradation dynamics.

\begin{equation}
\begin{gathered}
M \xrightarrow{k_{p,s}} M + P \quad (P \neq \emph{ERK}) \\
\emptyset \xrightarrow{k_{p,s,\emph{ERK}}} P \quad (P = \emph{ERK}) \\
P \xrightarrow{k_{p,d}} \emptyset
\end{gathered}
\label{eq:protein_sd}
\end{equation}
Where: \emph{M} is a particular mRNA; \emph{P} is the protein of \emph{M}; \emph{k\textsubscript{p,s}} and \emph{k\textsubscript{p,d}} are the protein synthesis and degradation rates, respectively.

\underline{FGFR activation and inactivation.} Multiple experimental studies have demonstrated that FGF signaling is a fundamental coordinator of the ICM specification into EPI and PRE populations \cite{ornitz_fibroblast_2015, kang_lineage_2017, molotkov_distinct_2018, krawczyk_paracrine_2022, ornitz_new_2022}. They have also indicated that FGF4 binds to two receptors: FGFR1 and FGFR2. However, FGFR1 plays the main role in the ICM cell-fate establishment and FGFR2 has a supporting/redundant character in the PRE-lineage regulation \cite{kang_lineage_2017, molotkov_distinct_2018}. Additionally, FGFR1 is expressed abundantly all over the ICM \cite{karl_ligand_2023}, plus FGF4 signaling via FGFR1 is critical for maintaining physiological levels of NANOG in EPI cells to help them reach primed pluripotency \cite{kang_lineage_2017}. Nonetheless, we do not model explicitly any FGF receptor; instead, for our model simulations, once FGF4 is available at the plasma membrane, we simply consider it to be the receptor-ligand complex in its inactive (monomer) form. In other words, whenever FGF4 undergoes a diffusive-jump event we relabel it as the monomer complex M-FGFR-FGF4. Furthermore, FGFR activation requires ligand-receptor dimer assembly, which in turn makes possible biochemical transduction of FGF signaling \cite{sarabipour_mechanism_2016, karl_ligand_2023}. As such, the D-FGFR-FGF4 (dimer) complex represents this active form triggering the FGF/ERK pathway in our system model.

The nominal rates of FGFR dimerization (activation) and monomerization (inactivation) are fixed model-parameter values. For the dimerization case, we follow the theory of diffusion-controlled reactions in a similar manner to the gene promoter scenario \cite{vijaykumar_intrinsic_2017}: we use the same formula as for \emph{k\textsubscript{b}} but we assume a receptor-ligand complex diffusion constant 30 times slower than the typical TF diffusion constant. For the monomerization case, we assume that FGFR1 phosphorylation and dephosphorylation follow similar kinetics, thus we take the average time (approximately 360 seconds) for reaching the half-saturation point based on some experimental receptor-ligand (FGFR-FGF4) response curves \cite{sarabipour_mechanism_2016, karl_ligand_2023}.

In brief, Eq~(\ref{eq:fgfr_ai}) recaps the reaction set of FGFR dimerization and monomerization.

\begin{equation}
\begin{gathered}
\emph{M-FGFR-FGF4} \quad + \quad \emph{M-FGFR-FGF4} \quad \xrightleftharpoons[k_{\emph{mono}}]{k_{\emph{dime}}} \quad \emph{D-FGFR-FGF4}
\end{gathered}
\label{eq:fgfr_ai}
\end{equation}
Where: \emph{k\textsubscript{dime}} and \emph{k\textsubscript{mono}} are both fixed parameter values; $k_{\emph{dime}} = 4{\pi}d(D/30)/V$; $k_{\emph{mono}} = 1/360$ s\textsuperscript{\textminus{1}}.

\underline{ERK activation and inactivation.} Proper stimulation of the FGF/ERK signalling pathway is a requisite for ICM cell differentiation during mouse blastocyst formation \cite{lanner_role_2010, azami_regulation_2019}. This stimulation is also essential for escaping naive pluripotency in mouse embryonic stem cells (ESCs) \cite{lanner_role_2010, deathridge_live_2019}. It has been shown experimentally that activation of ERK occurs via phosphorylation; this mechanism is present at ICM progenitors as well as PRE and EPI tissues \cite{azami_regulation_2019}. After all, ERK is the main FGF-signaling effector, which relays FGF4 fluctuations downstream of FGFR1 and FGFR2. This signaling pathway is a basic component for the regulation of cellular differentiation and homeostasis \cite{saba-el-leil_redundancy_2016}.

Previous reports have suggested that ERK experiences highly heterogeneous dynamics \cite{simon_live_2020, raina_fgf4_2021}. This high variability has been regarded as an additional layer of plasticity, which could enhance the cell-fate decision process by augmenting cellular heterogeneity during ICM/blastocyst development \cite{yeh_capturing_2021}. As such, ERK activity has been extensively examined in recent studies \cite{deathridge_live_2019, simon_live_2020, raina_fgf4_2021, yeh_capturing_2021}. Nevertheless, quantitative information about important reaction rates, relevant molecular concentrations, and many other kinetic parameters remains mostly elusive. Here, we use current in vitro and in silico reports on oscillations of ERK nuclear translocation for related biological systems, as well as experimental descriptions of ERK activity dynamics when exogenous FGF4 is present in mouse ESC systems \cite{fujioka_dynamics_2006, raina_cell-cell_2021}. We gathered all this information in order to establish generous bounds for the free model parameters representing the ERK phosphorylation (activation) and dephosphorylation (inactivation) reaction kinetic rates.

Fortunately, there are several succinct studies indicating that ERK has a long lifetime (48-72 hours) \cite{saba-el-leil_redundancy_2016, busca_erk1_2016}. They also report that there is no concrete evidence for feedback mechanisms (or stimulus-induced changes) regulating its protein expression, plus they indicate that ERK displays high physiological protein levels \cite{aoki_quantitative_2013}. These observations allows us to treat the collection of ERK molecules as a pool of ready-to-use kinases; in other words, the availability of ERK is not a limiting factor for cell-cell communication, rather its influence on the ICM specification process is controlled upstream of the release of FGF4 into the extracellular environment. Therefore, there is a need for tuning the reaction rates of ERK activation and deactivation.

To recap: we chose $\tau_{p,\emph{ERK}} = \tau_{p,\emph{I-ERK}} = \tau_{p,\emph{A-ERK}} = 48$ hours; \emph{k\textsubscript{pho,A-ERK}} (phosphorylation) and \emph{k\textsubscript{doh,I-ERK}} (dephosphorylation) are free parameters. Eq~(\ref{eq:erk_ai}) encapsulates the ERK activation and inactivation reaction dynamics.

\begin{equation}
\begin{gathered}
\emph{D-FGFR-FGF4} \quad + \quad \emph{I-ERK} \quad \xrightarrow{k_{pho,\emph{A-ERK}}} \quad \emph{D-FGFR-FGF4} \quad + \quad \emph{A-ERK} \\
\emph{A-ERK} \quad \xrightarrow{k_{doh,\emph{I-ERK}}} \quad \emph{I-ERK}
\end{gathered}
\label{eq:erk_ai}
\end{equation}

\underline{NANOG phosphorylation and dephosphorylation.} The \textit{Nanog} gene plays a key role in the regulation of pluripotency, self-renewal, and differentiation potential in human/mouse ESCs as well as early-embryo development \cite{kim_erk1_2014, liu_usp21_2016, kale_nanog-perk_2022}. Thus, tight control of \textit{Nanog} expression is of relevance for the correct progression of such developmental systems. This control is partially executed by ERK via two complementary mechanisms working at distinct scales. Firstly, A-ERK can recruit other proteins to the \textit{Nanog} locus and repress its transcriptional activity \cite{kale_nanog-perk_2022}. Along with this inhibition of \textit{Nanog} transcription, A-ERK phosphorylates NANOG, which directly leads to a reduction of its protein stability and an increase of its degradation rate \cite{liu_usp21_2016, kale_nanog-perk_2022}. All together, these observations suggest that this NANOG-regulation motif is indirectly orchestrated by NANOG itself. In this sense, ERK-mediated control of NANOG protein levels can be seen as a negative autocrine feedback loop (indirect autorepression), which might emerge following high FGF signaling. As we mentioned before, NANOG phosphorylation by A-ERK decreases/halves its lifetime, which in consequence implicitly decreases/halves its average steady-state protein copy number ($\overline{P}_{\emph{P-NANOG}} \approx \overline{P}_{\emph{NANOG}}/2$) \cite{kim_erk1_2014}. However, we do not have access to any other concrete data about the kinetic parameters related to this process. For this reason, the transition rates between NANOG and P-NANOG must enter our model as inferable parameters.

In short, $\tau_{p,\emph{P-NANOG}} = 1$ hour is a fixed value, while \emph{k\textsubscript{pho,P-NANOG}} and \emph{k\textsubscript{doh,NANOG}} are both free parameters. Eq~(\ref{eq:nanog_pho_doh}) portrays the NANOG phosphorylation and dephosphorylation reaction set.

\begin{equation}
\begin{gathered}
\emph{A-ERK} \quad + \quad \emph{NANOG} \quad \xrightarrow{k_{pho,\emph{P-NANOG}}} \quad \emph{A-ERK} \quad + \quad \emph{P-NANOG} \\
\emph{P-NANOG} \quad \xrightarrow{k_{doh,\emph{NANOG}}} \quad \emph{NANOG}
\end{gathered}
\label{eq:nanog_pho_doh}
\end{equation}

\subsubsection*{Model at tissue scale} \label{subsubsection:tissue_scale_model}

Technically, all events (reactions or diffusive jumps) share the same essential characteristics under the SSA formulation. However, we make an explicit distinction between reaction events and jump-diffuse events for two reasons: first, it is a convenient abstraction for distinguishing multiple temporal and spatial scales; second, their respective rates are calculated or estimated in contrasting manners because they depend on distinct features of our system model. Consequently, we have separated the focal bulk of the signaling model from the other reaction groups, and we treat it as a cohesive submodel at tissue level.

The most fundamental idea surrounding our signaling-model approach concerns the concept of conservation of molecular flux. In simple words, once the signaling molecule undergoes a jump-diffusion event, the probability of arrival over each of the cell neighbors (which includes the origin cell itself) must be a conserved feature. This conserved feature depends on the neighborhood configuration and has to be calculated for each cell. For our system model, the full ICM neighborhood representation is a rectangular voxel grid with a one-cell thickness; each voxel emulates an embryo cell.

In this sense, two cells (voxels) are categorized as first-degree neighbors only when they share a complete face; they can communicate directly with each other. In other words, if they only share a single edge (2 adjacent vertices), then they are not categorized as first-degree neighbors (any communication occurs indirectly between them); however, they are categorized as second-degree neighbors. This requisite implies that a particular cell must have strictly 2, 3, or 4 members within its first-degree neighborhood (which does not include itself). Thus, by recursion, it is easy to construct high-order neighborhoods; analogously, it is easy to identify low-degree neighborhood decompositions. By definition, each cell by itself forms a zero-degree neighborhood.

We model the release of FGF4 as a jump-diffuse event. When FGF4 molecules experience diffusive jumps, they cooperatively act as a biochemical signal, which is transduced by the respective FGFRs into an intracellular response involving multifold messenger molecules. This signal ultimately stimulates gene expression changes which consequently promote cellular adaptability \cite{nies_fibroblast_2016}. For our system model, FGF4 signaling manifests via autocrine and paracrine feedback loops. We also have included, for completeness, a third FGF4-signaling mode: membrane-level exchange of ligand molecules. Although, to the best of our knowledge, there is no experimental study of this auxiliary signaling mode yet, it is natural to think in terms of our modeling framework that once a molecule of FGF4 is available at the cellular membrane, it can experience ligand internalization together with its receptor, or it can sustain diffusive jumps between neighboring membranes. Accordingly, these two premises enter our final reaction set.

We assume that there is a primary flux transporting/releasing FGF4 molecules into the extracellular domain. For our implementation, this molecular flux is modeled as a first-degree process whose reaction rate depends on factors such as cellular geometry, spatial distribution of molecular escape channels, intracellular diffusion constant of FGF4, and many other features. We decided to treat this transport/release rate as a free model parameter because we only have access to some well-informed estimates for its bounds. These bounds are based on several studies of first-passage time distributions for related theoretical problems \cite{grebenkov_full_2019, grebenkov_distribution_2021}.

Subsequently, the FGF4 stream is split into two secondary fluxes. This split is controlled through the free-parameter relationship $\chi_{\emph{auto}} + \chi_{\emph{para}} = 1$, which dictates the probability ratio of ligand binding to its origin cell ($\chi_{\emph{auto}}$) or one of its neighboring cells ($\chi_{\emph{para}}$). In short, the exit rate of FGF4 molecules ($k_{\emph{escape}}$) eventually gives raise to two complementary signaling channels: autocrine and paracrine loops.

In a similar fashion, the exchange rate of FGF4 between cell membranes (\emph{k\textsubscript{exchange}}) is assumed to be limited by ligand-receptor affinity, just like for the case of FGFR monomerization \cite{sarabipour_mechanism_2016}. As such, generous bounds for this rate were placed and its concrete value is treated as a free parameter.

\underline{Autocrine signaling.} The autocrine feedback loop is thought to play an essential role for the \textit{Nanog} self-regulation. In that regard, the collateral \textit{Nanog} auto-repression is a self-perpetuating process maintaining physiologically-relevant NANOG levels. This process is fundamental because it allows an EPI cell to enter a state of primed pluripotency which supports the correct developmental progression \cite{kale_nanog-perk_2022}. Within our model, the rate of autocrine signaling is denoted by $k_{\emph{escape},\emph{auto}}$, and it is calculated via the formula $k_{\emph{escape},\emph{auto}} = \chi_{\emph{auto}}\rho_{\emph{auto}}k_{\emph{escape}}$. Here, the argument $\rho_{\emph{auto}} = (1-\chi_{\emph{para}}\rho_{\emph{meme}})/\chi_{\emph{auto}}$ involves the variable $\rho_{\emph{meme}}$ which quantifies the fraction of surface area shared between a particular cell and its first-degree neighborhood.

\underline{Paracrine signaling.} The paracrine feedback loop is deemed to have a critical role in inducing the PRE fate. The dose-dependent upregulation of \textit{Fgf4} by NANOG prompts FGF4 paracrine communication, which triggers a reaction cascade concurrently downregulating NANOG and upregulating GATA6 levels \cite{meng_gata6_2018, kale_nanog-perk_2022}. Within our model, the rate of paracrine signaling is denoted by $k_{\emph{escape},\emph{para}}$, and it is calculated via the formula $k_{\emph{escape},\emph{para}} = \chi_{\emph{para}}\rho_{\emph{meme}}k_{\emph{escape}}$.

\underline{Membrane exchange.} We denote the exchange rate of FGF4 molecules between cellular membranes as \emph{k\textsubscript{exchange}}. This FGF4 exchange rate is independent of the primary FGF4 secretion rate (\emph{k\textsubscript{escape}}); however, it must be adjusted relative to the actual number of first-degree neighbors for each particular cell. Thus, it is preferable to think of \emph{k\textsubscript{exchange}} as the maximal membrane-level FG4-exchange rate, which is only possible when a given cell completely shares all of its faces with other neighbors ($\rho_{meme} = 1$). In other words, we assume that the practical FGF4-exchange rate \emph{k\textsubscript{meme}} is simply directly proportional to the maximum FGF4-exchange rate \emph{k\textsubscript{exchange}}, where $0 \leq k_{\emph{meme}} \leq k_{exchange}$. This constraint enters our model via the formula $k_{\emph{meme}} = \rho_{meme}k_{exchange}$.

To finalize, Eq~(\ref{eq:communication}) illustrates the reaction set of FGF4 autocrine, paracrine, and membrane-exchange communication modes.

\begin{equation}
\begin{gathered}
\emph{FGF4} \quad \xrightarrow{k_{\emph{escape},\emph{auto}}} \quad \emph{M-FGFR-FGF4} \quad \textrm{(Self-Cell)} \\
\emph{FGF4} \quad \xrightarrow{k_{\emph{escape},\emph{para}}} \quad \emph{M-FGFR-FGF4} \quad \textrm{(Other-Cell)} \\
\textrm{(Self-Cell)} \quad \emph{M-FGFR-FGF4} \quad \xrightarrow{k_{\emph{meme}}} \quad \textrm{(Other-Cell)} \quad \emph{M-FGFR-FGF4}
\end{gathered}
\label{eq:communication}
\end{equation}

\subsubsection*{Inventory of model parameter values}

We provide here two complementary tables summarizing the most significant model parameters presented so far. Table~\ref{table3} recaps compactly all the fixed values we have either chosen based on well-informed estimates or taken from our literature sources. Table~\ref{table4} presents a quick view of the model parameters we have categorized as free values and their respective ranges. These parameter ranges were derived from data found across all our literature sources, and they generally represent educated guesses made by collecting information about closely related biological systems. Nonetheless, we largely assumed generous bounds for all these ranges, delegating the search for biophysically-relevant values to our parameter inference scheme.

\begin{table}[ht!]
\begin{adjustwidth}{-1.75in}{0in}
\caption{{\bf Summary of fixed model parameters.}}
\centering
\begin{tabular}{ | c | c | c | c | }
\hline
\rowcolor{lightgray}
Name & Alias & Description & Value \\
\hline
\emph{V} &  & Cell volume & 4200 \textmu{m}\textsuperscript{3} \\
\emph{D} &  & Protein diffusion coefficient (cytoplasm or nucleus) & 10 \textmu{m}\textsuperscript{2}s\textsuperscript{-1} \\
\emph{d} &  & Promoter binding-site diameter & 10 nm \\
$k_{b}$ &  & Transcription factor binding rate & $0.3{\cdot}10^{-3}$ s\textsuperscript{-1} \\
$k_{\emph{coop}}$ &  & Cooperativity coefficient & 5 \\
$c_{\emph{basal}}$ &  & Relative contribution of basal mRNA production & 0.2 \\
$c_{\emph{find}}$ &  & Relative contribution of full-induction mRNA production & 0.8 \\
$\overline{M}_{\textit{Nanog}}$ &  & Mean steady-state mRNA copy number at full induction & 250 copies \\
$\overline{M}_{\textit{Gata6}}$ &  & Mean steady-state mRNA copy number at full induction & 250 copies \\
$\overline{M}_{\textit{Fgf4}}$ &  & Mean steady-state mRNA copy number at full induction & 200 copies \\
$\tau_{m,\textit{Nanog}}$ & $t_{\frac{1}{2},m,\textit{Nanog}}$ & Lifetime (or half-life) for a molecule of mRNA & 4 hours \\
$\tau_{m,\textit{Gata6}}$ & $t_{\frac{1}{2},m,\textit{Gata6}}$ & Lifetime (or half-life) for a molecule of mRNA & 4 hours \\
$\tau_{m,\textit{Fgf4}}$ & $t_{\frac{1}{2},m,\textit{Fgf4}}$ & Lifetime (or half-life) for a molecule of mRNA & 4 hours \\
$\overline{P}_{\emph{NANOG}}$ &  & Mean steady-state protein copy number at full induction & 1000 copies \\
$\overline{P}_{\emph{GATA6}}$ &  & Mean steady-state protein copy number at full induction & 1000 copies \\
$\overline{P}_{\emph{FGF4}}$ &  & Mean steady-state protein copy number at full induction & 800 copies \\
$\overline{P}_{\emph{ERK}}$ &  & Mean steady-state protein copy number & 1000 copies \\
$\tau_{p,\emph{NANOG}}$ & $t_{\frac{1}{2},p,\emph{NANOG}}$ & Lifetime (or half-life) for a molecule of protein & 2 hours \\
$\tau_{p,\emph{GATA6}}$ & $t_{\frac{1}{2},p,\emph{GATA6}}$ & Lifetime (or half-life) for a molecule of protein & 2 hours \\
$\tau_{p,\emph{ERK}}$ & $t_{\frac{1}{2},p,\emph{ERK}}$ & Lifetime (or half-life) for a molecule of protein & 48 hours \\
$\tau_{p,\emph{P-NANOG}}$ & $t_{\frac{1}{2},p,\emph{P-NANOG}}$ & Lifetime (or half-life) for a molecule of protein & 1 hour \\
$\tau_{p,\emph{D-FGFR-FGF4}}$ & $t_{\frac{1}{2},p,\emph{D-FGFR-FGF4}}$ & Lifetime (or half-life) for a molecule of protein & 240 hours \\
$k_{\emph{dime}}$ &  & M-FGFR-FGF4 dimerization (activation) & $10{\cdot}10^{-6}$ s\textsuperscript{-1} \\
$k_{\emph{mono}}$ &  & D-FGFR-FGF4 monomerization (inactivation) & $3{\cdot}10^{-3}$ s\textsuperscript{-1} \\
\hline
\end{tabular}
\begin{flushleft}
Note. For complete details about all implemented GRN interactions related to these fixed model parameters, please check \nameref{subsubsection:cell_scale_model}. See also Table~\ref{table2}.
\end{flushleft}
\label{table3}
\end{adjustwidth}
\end{table}

\begin{table}[ht!]
\begin{adjustwidth}{-1.75in}{0in}
\caption{{\bf Summary of inferred (free) model parameters.}}
\centering
\begin{tabular}{ | c | c | c | c | c | c | c | }
\hline
\rowcolor{lightgray}
Name & Alias & Description & \parbox[c][2.5em][c]{3.75em}{\centering Lower \\ Bound} & \parbox[c][2.5em][c]{3.75em}{\centering Inferred \\ Value} & \parbox[c][2.5em][c]{3.75em}{\centering Upper \\ Bound} & Units \\
\hline
$h_{\textrm{act},\textit{Nanog},\textrm{NANOG}}$ & \textit{Nanog}\_NANOG & \parbox[c][2.5em][c]{10em}{\centering Half-saturation level \\ (self-activation)} & 0 & 125 & 1000 & [pc] \\
$h_{\textrm{act},\textit{Gata6},\textrm{GATA6}}$ & \textit{Gata6}\_GATA6 & \parbox[c][2.5em][c]{10em}{\centering Half-saturation level \\ (self-activation)} & 0 & 275 & 1000 & [pc] \\
$h_{\textrm{rep},\textit{Gata6},\textrm{NANOG}}$ & \textit{Gata6}\_NANOG & \parbox[c][2.5em][c]{10em}{\centering Half-saturation level \\ (mutual-repression)} & 0 & 412 & 1000 & [pc] \\
$h_{\textrm{rep},\textit{Nanog},\textrm{GATA6}}$ & \textit{Nanog}\_GATA6 & \parbox[c][2.5em][c]{10em}{\centering Half-saturation level \\ (mutual-repression)} & 0 & 411 & 1000 & [pc] \\
\hline
$h_{\textrm{act},\textit{Fgf4},\textrm{NANOG}}$ & \textit{Fgf4}\_NANOG & \parbox[c][2.5em][c]{10em}{\centering Half-saturation level \\ (activation)} & 0 & 552 & 1000 & [pc] \\
$h_{\textrm{act},\textit{Gata6},\textrm{A-ERK}}$ & \textit{Gata6}\_A-ERK & \parbox[c][2.5em][c]{10em}{\centering Half-saturation level \\ (activation)} & 0 & 615 & 1000 & [pc] \\
$h_{\textrm{rep},\textit{Fgf4},\textrm{GATA6}}$ & \textit{Fgf4}\_GATA6 & \parbox[c][2.5em][c]{10em}{\centering Half-saturation level \\ (repression)} & 0 & 95 & 1000 & [pc] \\
$h_{\textrm{rep},\textit{Nanog},\textrm{A-ERK}}$ & \textit{Nanog}\_A-ERK & \parbox[c][2.5em][c]{10em}{\centering Half-saturation level \\ (repression)} & 0 & 695 & 1000 & [pc] \\
\hline
$\tau_{\textrm{escape}}$ & $k_{\textrm{escape}}^{-1}$ & \parbox[c][2.5em][c]{10em}{\centering Mean escape time \\ (FGF4)} & 300 & 2303 & 4500 & [s] \\
$\tau_{\textrm{exchange}}$ & $k_{\textrm{exchange}}^{-1}$ & \parbox[c][2.5em][c]{10em}{\centering Mean exchange time \\ (FGF4)} & 30 & 1380 & 4200 & [s] \\
$\chi_{\textrm{auto}}$ & $1-\chi_{\textrm{para}}$ & Autocrine signaling fraction & 0 & 0.39 & 1 &  \\
\hline
$\tau_{\textrm{pho},\textrm{ERK}}$ & $k_{\textrm{pho},\textrm{A-ERK}}^{-1}$ & \parbox[c][2.5em][c]{10em}{\centering Half-turnover time \\ (phosphorylation)} & 300 & 31379 & 43200 & [s] \\
$\tau_{\textrm{doh},\textrm{ERK}}$ & $k_{\textrm{doh},\textrm{I-ERK}}^{-1}$ & \parbox[c][2.5em][c]{10em}{\centering Half-turnover time \\ (dephosphorylation)} & 30 & 1025 & 43200 & [s] \\
$\tau_{\textrm{pho},\textrm{NANOG}}$ & $k_{\textrm{pho},\textrm{P-NANOG}}^{-1}$ & \parbox[c][2.5em][c]{10em}{\centering Half-turnover time \\ (phosphorylation)} & 300 & 18944 & 43200 & [s] \\
$\tau_{\textrm{doh},\textrm{NANOG}}$ & $k_{\textrm{doh},\textrm{NANOG}}^{-1}$ & \parbox[c][2.5em][c]{10em}{\centering Half-turnover time \\ (dephosphorylation)} & 30 & 21361 & 43200 & [s] \\
\hline
\parbox[c][2.5em][c]{7.5em}{\centering Mean Initial \\ mRNA Count} & \textit{Nanog}\_\textit{Gata6} & Initial condition & 0 & 117 & 250 & [mc] \\
\parbox[c][2.5em][c]{7.5em}{\centering Mean Initial \\ PROTEIN Count} & NANOG\_GATA6 & Initial condition & 0 & 482 & 1000 & [pc] \\
$\tau_{\textrm{d},\textrm{FGF4}}$ & $k_{\textrm{p},\textrm{d},\textrm{FGF4}}^{-1}$ & Lifetime or half-life & 300 & 13158 & 28800 & [s] \\
$\tau_{\textrm{d},\textrm{M-FGFR-FGF4}}$ & $k_{\textrm{p},\textrm{d},\textrm{M-FGFR-FGF4}}^{-1}$ & Lifetime or half-life & 300 & 3456 & 28800 & [s] \\
\hline
\end{tabular}
\begin{flushleft}
Note. For complete details about all implemented GRN interactions related to these inferred (free) model parameters, please see \nameref{subsubsection:cell_scale_model} and \nameref{subsubsection:tissue_scale_model}. The two parameters ``Mean Initial mRNA Count'' and ``Mean Initial PROTEIN Count'' have not been properly introduced yet, but their usage will be explained at \nameref{subsection:inference_framework}. Notation: act = activation; rep = repression; pho = phosphorylation; doh = dephosphorylation; [pc] = [protein copies]; [mc] = [mRNA copies]; [s] = [seconds]. See also Figs~\ref{fig101} and~\ref{fig102}.
\end{flushleft}
\label{table4}
\end{adjustwidth}
\end{table}

\clearpage 


\subsection*{Model parameter inference framework} \label{subsection:inference_framework} 

The exploration of the immense parameter space of a mechanistic model is a demanding computational task, especially for biophysical problems dealing with spatial-stochastic system representations \cite{wang_massive_2019}. Several classical inference/optimization methods such as heuristic tuning, stochastic gradient descent, simulated annealing, and approximate Bayesian computation (ABC) have been traditionally used for finding reasonable parameter sets of biophysical mechanistic models \cite{schnoerr_approximation_2017, cranmer_frontier_2020, franzin_landscape-based_2023, stillman_generative_2023, prescott_efficient_2024}. Nonetheless, all of these aforementioned methods are often computationally inefficient because they require many expensive simulations for properly scanning the model parameter space, and they commonly lack powerful parameter-space interpolation strategies. As a consequence, these inference methods become prohibitively ineffective when they are applied to high-dimensional spatial-stochastic models, which are nowadays usually employed to represent complex biological systems.

Here, by leveraging our access to high-performance computing (HPC) resources, we take advantage of the sequential neural posterior estimation (SNPE) algorithm and combine it with classical ML concepts, in order to implement a comprehensive model parameter inference framework. The SNPE algorithm is part of a ground-breaking family of likelihood-free inference techniques \cite{greenberg_automatic_2019, cranmer_frontier_2020, deistler_truncated_2022}. These state-of-art simulation-based inference (SBI) algorithms fundamentally rely on artificial neural networks (ANNs) to approximate model parameter probability distributions; the respective ANNs are trained with simulation/synthetic data, but are conditioned on target/experimental observations \cite{cranmer_frontier_2020, stillman_generative_2023}.

In our case, this original framework was applied to infer two individual parameter sets for two separate models, which are capable of recapitulating the most fundamental characteristics of the fully-formed mouse blastocyst. The first model represents the wild-type variant of the underlying biophysical system, and it is our principal system model; in other relevant contexts, we also refer to it as the ``inferred-theoretical wild-type'' or ITWT. This system has a functional cell-cell communication. The second model is an auxiliary system; all the parameter values of the ITWT are reused for this supplementary system, except for the core GRN components which are reinferred in a loss-of-function mutant setting: this system model has a nonfunctional cell-cell communication. We consequently refer to it as the ``reinferred-theoretical mutant'' or RTM.

To illustrate the key stages of our exploratory Bayesian inferential framework, we present a concise list of the most important ideas of our workflow. For additional information concerning some general considerations of our model parameter inference framework, please see Fig~\ref{fig1} and \nameref{section:supporting_information}.

\subsubsection*{Constructing the prior distribution}

A key stage of our workflow is the construction of the model parameter prior distribution. Once every suitable model parameter has an appropriate fixed value (see Table~\ref{table2} and Table~\ref{table3}), it is necessary to condense every belief/intuition about the remaining free model parameters into a reasonable prior distribution. In this work, this prior is a multivariate uniform distribution with 19 dimensions (vector components). Every such vector component has a predefined range; those ranges are derived from educated guesses informed by data on closely related biological systems or found across all our literature sources. We largely assume rich but realistic bounds for all of those ranges (see Table~\ref{table4}), as ultimately the aim is to delegate the search for biophysically-relevant values to our parameter inference scheme. But while the general idea applies to both system models (ITWT and RTM), the RTM prior distribution has only 4 dimensions corresponding to the core GRN motif.

Two of the freely-varying model parameters fulfill a particular role, namely ``Mean Initial mRNA Count'' and ``Mean Initial PROTEIN Count''. These parameters are used to define meaningful initial condition distributions (ICDs), as there is no detailed experimental information currently available about them. In the typical setting of our simulation procedure, they operate together as the mean-value vector of an ICD which is itself a composition of Poissonian and binomial distributions (see \nameref{subsection:computational_experiments}); as such, they also dictate the variance matrix of this ICD. From that ICD, we sample the starting \textit{Nanog}-\textit{Gata6} mRNA and NANOG-GATA6 (protein) copy numbers; accordingly, this sampling is performed per simulated cell. In this regard, the only stipulated hard molecular constraint follows from imposing the maximal mean mRNA and protein copy numbers reached at full induction: 250 and 1000 molecules, respectively.

\subsubsection*{Simulation data generation}

At this stage, we explicitly incorporate two of the basic empirical observations we are aiming at recapitulating into the simulation procedure itself. Highly reproducible ratio of $2:3$ for EPI-PRE lineages \cite{bessonnard_gata6_2014, saiz_asynchronous_2016}; absence of FGF4-mediated signaling (no spatial coupling) forces the ICM to almost exclusively adopt the EPI fate \cite{bessonnard_icm_2017, thompson_extensive_2022}. See also \nameref{section:supporting_information}. Via parallel computing, our simulator generates after each run a composite trajectory for two complementary configurations of our system model: functioning cell signaling, and nonfunctioning cell signaling. When cell signaling is functioning, the targeted proportions are 40\% for the EPI population and 60\% for the PRE population. Whereas, when cell signaling is nonfunctioning, the targeted proportions are 100\% for the EPI population and 0\% for the PRE population. While these two configurations are necessary for correctly deriving the ITWT system, the valid derivation of the RTM system requires only one configuration: as cell-cell communication is always turned off for the RTM, here the targeted proportions are 40\% for the EPI lineage and 60\% for the PRE lineage.

Furthermore, each run should produce at least 48 hours of simulated time to sufficiently capture the pertinent model dynamics; for additional information, see \nameref{subsubsection:preimplantation_stages}. By design, the simulation data generation can be executed independently of any other stage, which allows the production of a sufficiently large trajectory batch per run. This simulation batch size is arbitrary and totally influenced by the available computational resources; in our case, the simulation batch size is generally 100 thousand (composite) trajectories per each ANN training round.

\subsubsection*{Training data generation}

The raw simulation data is ineffective for training the ANN, because of the high dimensionality and the multiscale character of the generated trajectories. Via sequential data transformations, we create features that can be used for successfully training the ANN. In practical terms, these features are low-dimensional projections of high-dimensional temporal-spatial stochastic dynamics data.

Our simulations are genuinely event-driven and therefore produce trajectories irregularly spaced in time. As a preparative step, we accordingly first resample the simulated data onto a regular time grid. For simplicity, we always generate a resampled time series with a sampling period equal to 0.25 hours (or 15 minutes); we deemed that sampling period to be sufficient for fully characterizing the underlying temporal dynamics on the tissue scale.

All subsequent steps are performed on that regularized time series. Per simulated cell, the resampled time series represents the dynamics of all the involved biochemical species. In order to avoid tracking all of the many biochemical species in our model, we focus on three system observables that characterize the specification process at cell scale: total NANOG, which is the aggregate sum of the available unphosphorylated NANOG and phosphorylated NANOG protein molecular counts; total GATA6, which is simply the GATA6 protein molecular count; total FGF4, which is the aggregate sum of all the available FGF4 proteins at both cytoplasm and membrane levels (including receptor-bound FGF4 molecules). While total FGF4 is only useful for analyzing our system under perturbed conditions, total NANOG and total GATA6 are the guiding drivers of our training data generation: these are the main markers determining the lineage for every cell at each simulated time point; for additional details, check \nameref{subsection:computational_experiments}.

We next further group the cell-level observables into a tissue-level one. The key tissue-scale observable variable is the total count for each of the three possible cell fates at each simulated time point. As such, this tissue-level observable is used to define a ``pattern score'' that meaningfully discriminates the targeted/idealized patterning behavior from undesired patterning behaviors of our system. This constitutes the most important part of our data-transformation pipeline, and it is fully described in the next part.

\subsubsection*{Constructing the pattern score (objective) function} \label{subsubsection:score}

To generate the actual ANN training dataset, two closely connected steps are required for bringing it to fruition. First, at every time point for each possible fate, the total cell count is compared against the target cell count. Moreover, there is a specific target cell count per each prescribed configuration of the respective system: for the ITWT, the EPI-PRE-UND lineage target proportions are 40-60-0 percent with functioning spatial coupling, and 100-0-0 percent when spatial coupling is inactivated; the RTM (in which spatial coupling is always inactivated) has the EPI-PRE-UND lineage target proportions of 40-60-0 percent. The resultant is a set of marginal scores, one per system configuration; i.e., two marginal scores for the ITWT, and one for the RTM. Second, all these resulting (marginal) configuration scores must be combined into a discriminatory pattern score, which measures the distance between a particular simulation score and the idealized behavior score.

The (joint) pattern score time series performs as the outcome of a vector-valued objective function whose main inputs are the total cell count and the target cell count per system configuration, for each possible fate at every time point. Therefore, our essential goal is to intelligently optimize (maximize) this objective function. As such, the joint pattern score is the foundational element of the ANN training dataset.

For the first step, in formal terms, let $\boldsymbol{Z}_{t} = (Z_{t,0}, Z_{t,1}, Z_{t,2})$ be a discrete random vector taking values in $\mathbb{N}^{3}$ (naturals or nonnegative integers), which represents the total cell count for each lineage at a given discrete time point $t \in \mathbb{N}$. For simplicity, the three possible cell lineages are arbitrarily indexed as 0 (EPI), 1 (PRE), and 2 (UND). Also, let $\boldsymbol{w}_{m} = (w_{m,0}, w_{m,1}, w_{m,2}) \in \mathbb{N}^3$ be a discrete vector representing the target cell count for each lineage, and for a given system configuration which is arbitrarily indexed by $m \in \{0,1\}$. The configuration score is thus the continuous random variable $S_{t,m}$ taking values in $\mathbb{R}$ (reals), which is a nonlinear transformation mapping from an absolute-difference vector $|\boldsymbol{z}_{t}-\boldsymbol{w}_{m}| = (|z_{t,0}-w_{m,0}|, |z_{t,1}-w_{m,1}|, |z_{t,2}-w_{m,2}|)$ to a point $s_{t,m}$ in the closed interval $[0, 1]$; i.e.:

\begin{equation} 
S_{t,m} = \frac{\exp{\left(-{\|\boldsymbol{Z}_{t}-\boldsymbol{w}_{m}\|}_{1}/{\|\boldsymbol{w}_{m}\|}_{1}\right)}-\overset{\star}{w}}{1-\overset{\star}{w}}
\label{eq:marginal_score}
\end{equation}
Here $\overset{\star}{w} = \exp{\left(-2\max{\left(\boldsymbol{w}_{m}\right)}/{\|\boldsymbol{w}_{m}\|}_{1}\right)}$, ${\|\boldsymbol{w}_{m}\|}_{1}/{\|\boldsymbol{Z}_{t}\|}_{1} = 1$, and the expression ${\|\cdot\|}_{1}$ indicates the $\ell^{1}$ vector norm.

For the second step, in formal terms, let $S_{t,0}$ and $S_{t,1}$ be the marginal scores for the prescribed configurations of the ITWT; clearly, this step is not applicable to the RTM, as it has only one configuration. While this step is far from being a trivial process, to combine these two marginal scores into one pattern score, we employed an $\ell^{1}$-inspired penalty method. This penalty directly affects the sum of the configuration scores, and it is intended to increase the discriminatory power of the joint score by favouring similarly high marginal scores, as described by the following formula:

\begin{equation} 
S_{t} = \frac{\left(S_{t,0}+S_{t,1}\right)-\overset{\star}{S_{t}}}{2}
\label{eq:joint_score}
\end{equation}
Here, $\overset{\star}{S_{t}} = |S_{t,0}-S_{t,1}|$ is the penalty term.

By applying $S_{t}$, we construct a time series spanning the last 12-hour window of the predetermined simulation period (from 0 hours to 48 hours), and by producing many such time series, we generate the effective ANN training dataset.

\subsubsection*{Training the ANN}

The key idea of training an ANN is to use a relatively small number of simulations. The exact number of trajectories generated by the simulator should be determined by aiming at meaningfully balancing both the algorithm's computational feasibility and the ANN training reliability. This balance enables an effective exploration of the massive parameter space for the proposed system model via the trained ANN \cite{wang_massive_2019}.

To this end, we exploited a state-of-art SBI toolbox which allowed us to easily integrate these novel AI-powered algorithms into our workflow \cite{tejero-cantero_sbi_2020}. Specifically, we employed the SNPE procedure to train a deep neural density estimator which directly estimates the model parameter posterior distribution conditional on a goal observation. As our training dataset is already a convenient latent representation of the simulation data, this stage was relatively straightforward. Furthermore, there was no need to tune the algorithm hyperparameters; the applied SBI toolbox has reasonable preset values for them.

Altogether, we trained ANNs to successfully predict model parameter sets capable of recapitulating the targeted patterning behaviors of the systems under study.

\subsubsection*{Constructing a synthetic target/goal observation}

At heart, our data-transformation pipeline generates a suitable latent (feature) space representation of the simulated dataset for training the ANN, which strongly facilitates predicting inferred parameter distributions that agree with a prescribed effective behavior in that latent space.

Ultimately, our (synthetic) goal observation is simply a score time series whose support spans the last 12-hour window of the prescribed simulation period; in agreement with one of the basic empirical observations: EPI and PRE populations should reach their expected fate proportions roughly 8 or 12 hours before the end of the preimplantation period \cite{plusa_common_2020, saiz_coordination_2020, yanagida_cell_2022, allegre_nanog_2022}. See also \nameref{section:supporting_information}. Moreover, each entry/value of this time series is a ``1'': the ideal score of the target patterning behavior.

\subsubsection*{Selecting a posterior distribution}

To account for the inherent stochasticity of the ANN training algorithm (mini-batch stochastic gradient descent) and the implicit randomness of the pattern score trajectories, we train multiple ANNs with the same dataset. Each trained ANN produces a posterior distribution, and from it we compute the maximum-a-posteriori (MAP) estimate of all the model parameters. Using each distinct MAP estimate, we generate an extra simulation batch with many fresh pattern score time series. Moreover, we construct a pattern score distribution for each of these extra batches. The idea is not only to maximize the accuracy of the underlying system behavior, but also to increase its precision, optimizing its robustness. As such, we measure the quality of each separate MAP estimate and summarize it into a single number, in order to select the best learned posterior distribution of the model parameters conditional on the target observation.

This quality measure of the MAP estimate is described by Eq~(\ref{eq:meta_score}), and it is accordingly referred to as the ``meta score'' $\overline{S}$ (see also Fig~\ref{fig1}).

\begin{equation}
\overline{S} = \operatorname{mean}\left(\max{\left(\boldsymbol{\alpha}_{50}-\boldsymbol{\beta},\vec{0}\right)}\right)
\label{eq:meta_score}
\end{equation}
\clearpage

Where $\overline{s} \in [0,1] \subset \mathbb{R}$ ($\overline{s}$ is the realization of the random variable $\overline{S}$), $\alpha_{50}$ denotes the 50-percentile (or median) vector of a given pattern score distribution at each simulated time point, $\boldsymbol{\beta} = ((\boldsymbol{\alpha}_{95}-\boldsymbol{\alpha}_{5})/2)^{2}$ is an element-wise penalty vector favouring high accuracy together with high precision, $\vec{0}$ denotes the corresponding zero vector, and the maximum operator is applied to each component of the input vector pair. Thus, the selected model parameter set has the highest associated (MAP estimate) meta score among the available ones.

\subsubsection*{Performing multiple rounds of inference}

In general terms, single-round inference is not enough for learning truly-applicable model parameter values, even with a significantly high simulation budget. The problem of amortized inference is the wasteful use of the simulated trajectories: the ANN effectively estimates the posterior distribution for all the possible goal observations across the entire prior space of the model parameters.

If there is only one prominent target observation (just like our case), then it is advantageous to perform multiple sequential rounds of (non-amortized) inference. The model parameter search will be focused on this single target observation. This process can be performed arbitrarily many times until the current-round score time series matches as close as possible the target score time series. In this study, we performed 8 consecutive rounds of inference producing 800 thousand (composite) simulations in total for the ITWT system, and we performed 4 consecutive rounds of inference producing 400 thousand simulations in total for the RTM system.

In this context, it is also worth mentioning that the next-round prior does not necessarily need to be the current-round posterior. The proposal distribution can be iteratively adapted to fully leverage the power of multi-round inference, which is all based on the earliest prior and the latest posterior; as such, it is possible to obtain a well-informed next-round mixture distribution \cite{deistler_truncated_2022}. While we do not exploit this technique, it can be easily integrated into our workflow.

\subsection*{Computational experiments} \label{subsection:computational_experiments}

The initial gene expression profile per cell is commonly sampled from a composition of two distributions: the (first) Poissonian part whose mean-value vector construction is dictated by the two parameters ``Mean Initial mRNA Count'' and ``Mean Initial PROTEIN Count''; the (second) binomial part which fairly splits the respective cellular resources. While it is easy to change the goal biochemical species, we generally target only the ICDs of the two main players: \textit{Nanog}-\textit{Gata6} mRNA and NANOG-GATA6 (protein) copy numbers. As such, every typical simulation starts with reasonably balanced cell resources, which guarantees the UND state across the tissue despite the inherent randomness of the initial conditions.

For the starting test of robustness to initial condition perturbations (ICPs), one extra layer of variability is necessary. Instead of using only a two-element composition, the original ICD now integrates an additional layer which performs uniform sampling of mRNA and protein copy numbers per cell. This uniform perturbation of cellular resources is thus the ICD root component for this testing case, whereas the original layers are the other two remaining elements. Simply put, firstly a random number is uniformly sampled from a particular discrete range ($[0, 250]$ for mRNA and $[0, 1000]$ for protein species), secondly this preceding number is employed as the mean value of a Poisson distribution which is sampled accordingly, and thirdly the eventual molecular count for a particular biochemical species is sampled from a fair binomial distribution whose (independent) trial-number parameter is dictated by the precursory Poissonian layer. It is hence easy to see that this ICP greatly increases the early variance of the goal cell resources.

The cell-fate classification thresholds were purposefully chosen to ease the ICM cellular categorization based on a proper evaluation of the most important GRN relationships. This cell-lineage categorization is performed by comparing NANOG and GATA6 (protein) levels against constant threshold levels. Assuming Poissonian noise, a low NANOG/GATA6 cell classification occurs when the respective protein level is below a threshold of approximately 329 copies (mean basal protein level plus five times its standard deviation), a high NANOG cell classification occurs when the respective protein level is above a threshold of approximately 388 copies (mean full-induction-phosphorylation protein level minus five times its standard deviation), and a high GATA6 cell classification occurs when the respective protein level is above a threshold of approximately 842 copies (mean full-induction protein level minus five times its standard deviation). Thus, the EPI categorization occurs when a cell simultaneously displays the low-GATA6 alongside high-NANOG states, the PRE classification occurs when a cell simultaneously displays the low-NANOG alongside high-GATA6 states, and the UND categorization occurs when a cell displays any other combination of states.

\begin{figure}[hpt!]
\includegraphics[width = 5.5in, height = 2.75in]{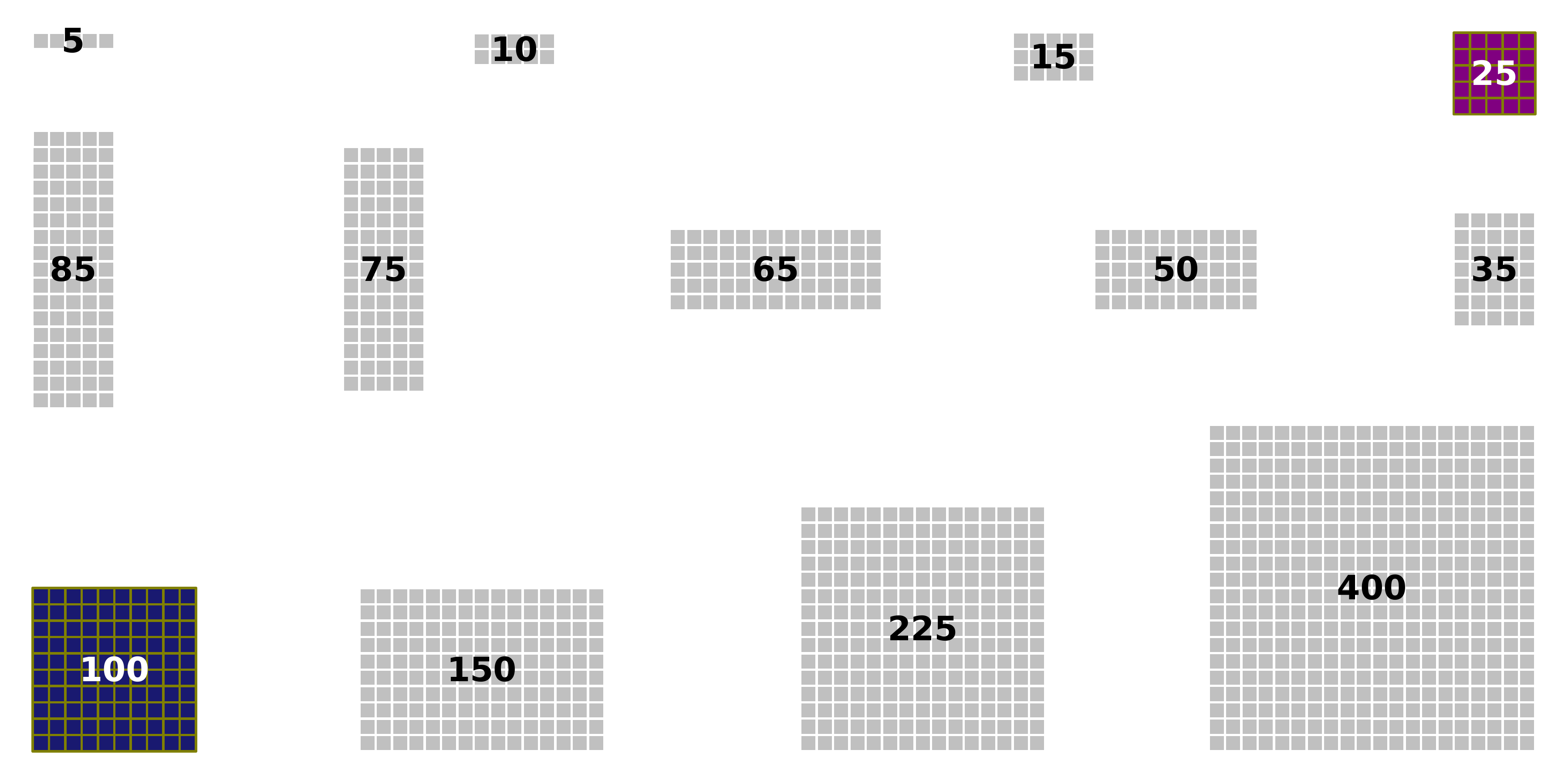} \centering 
\caption{{\bf Overview of cell-grid sizes employed for inference and simulations.} The 25-cell grid size (purple highlighting) was used invariably for the actual model parameter inference scheme. All test simulations (assessment of system properties such as robustness in wild-type-like and mutated conditions) employed the 100-cell tissue size (dark-blue highlighting). The remaining cell-grid sizes (gray) were solely used to study tissue-level noise (see also Fig~\ref{fig5}).}
\label{fig104}
\end{figure}

For the actual model parameter inference procedure, a 25-cell grid size was employed consistently. The 100-cell tissue size was used for all the test simulations. Any other cell-grid size was exclusively employed to study noise at tissue level (Fig~\ref{fig5}). To see a graphical comparison of every tissue size, please check Fig~\ref{fig104}.

\subsubsection*{Computational simulation details}

The simulator was fully implemented in the \href{https://www.python.org/about/}{Python} programming language. There are multiple such software packages supporting the simulator design: \href{https://numpy.org/about/}{NumPy}, \href{https://numba.pydata.org/}{Numba}, \href{https://scipy.org/about/}{SciPy}, \href{https://pytorch.org/}{PyTorch}, \href{https://sbi-dev.github.io/sbi/}{SBI}, \href{https://matplotlib.org/}{Matplotlib}, \href{https://seaborn.pydata.org/}{Seaborn}, among many others. All the computational simulations were performed in the general-purpose HPC cluster \href{https://csc.uni-frankfurt.de/wiki/doku.php?id=public:service:goethe-hlr}{Goethe-HLR} belonging to the Goethe-Universität Frankfurt am Main.

\newpage 




\section*{Supporting information} \label{section:supporting_information}

\subsection*{Summary of inferred model parameter interactions (posterior distribution)}

We summarize the model parameter values estimated using our SNPE-powered workflow. We present the full inferred model parameter distribution, as well as the first estimate of the model parameter sensitivity. In Fig~\ref{fig101} and Fig~\ref{fig102}, the unconditional (odd rows) and the conditional (even rows) posterior parameter distributions are split into multiple distinct blocks grouping parameters according to their particular regulatory roles in the ICM development process. This split is helpful for showing the critical model components and the correlations among the most concomitant inferred parameters. For each block, the diagonal entries are the marginal (one-dimensional) distributions, and the upper-diagonal entries are the joint (two-dimensional) distributions between successive parameters. For obvious reasons, other higher-dimensional projections are not displayed. See Table~\ref{table4} for a recap of the inferred model parameter values and their corresponding notation.

Concentrating on the original inferred system (ITWT), we now examine the posterior distribution of its model parameters. The first pair of parameter blocks (Fig~\ref{fig101}[A, C]) displays two closely associated distributions for the core GRN motif components: the unconditional posterior (top row) shows the comparatively short span over the parameter space which enables the system to display the postulated target behavior; the conditional posterior (bottom row) shows at marginal and joint levels the sensitivity to variations of these parameter values. To correctly analyze this last posterior, we must initially recognize that this distribution is indeed the raw inferred posterior, but it is now conditioned on the maximum-a-posteriori (MAP) estimate of all the model parameters. In that sense, every diagonal entry or marginal distribution reflects the tolerance to fluctuations of a single value, given that all the other parameter values are fixed at its corresponding MAP estimate. In other words, we automatically get an approximation of the sensitivity for every inferred model parameter conditional on the MAP estimate of the remaining values; this feature is inherent to the SBI technique used here, without the need for additional expensive simulations. Likewise, we can perform an analogous assessment for every upper-diagonal entry or joint distribution based on the same criteria. In this case, we see that the core GRN motif values are constrained to a small parameter space region, which emphasizes their sensitive balancing act and their importance for attaining the ideal score.

Similarly, the second pair of parameter blocks (Fig~\ref{fig102}[A, E]) shows the distributions linked to the interactions among NANOG, GATA6, FGF4, and ERK. Especially for the \textit{Ffg4} gene, the span of possible parameter values pertaining to its regulation is relatively broad. Nevertheless, when the other parameters are fixed appropriately, the sensitivities of \textit{Fgf4}\_NANOG (half-saturation threshold for transcriptional activation of \textit{Fgf4} promoter by NANOG) and \textit{Fgf4}\_GATA6 (half-saturation threshold for transcriptional repression of \textit{Fgf4} promoter by GATA6) increase considerably, stressing their significance for the target system-model behavior. In contrast, for the case of \textit{Gata6}\_A-ERK (half-saturation threshold for transcriptional activation of \textit{Gata6} promoter by A-ERK) and \textit{Nanog}\_A-ERK (half-saturation threshold for transcriptional repression of \textit{Nanog} promoter by A-ERK), we see relatively low sensitivities; hence, these parameters could potentially be adjusted easily without affecting the ideal score and the final cell-fate ratio.

In the same manner, the third pair of parameter blocks (Fig~\ref{fig102}[B, F]) shows that the values for signaling components can be drawn from a rich and large parameter space region without affecting the target behavior of the underlying system. This property clearly only holds provided that cell-cell communication is fully operational, because its absence will cause the system model to materialize almost exclusively high-NANOG- and low-GATA6-expressing cells.

\begin{adjustwidth}{-1.75in}{0in} 
\includegraphics[width = 7in, height = 7in]{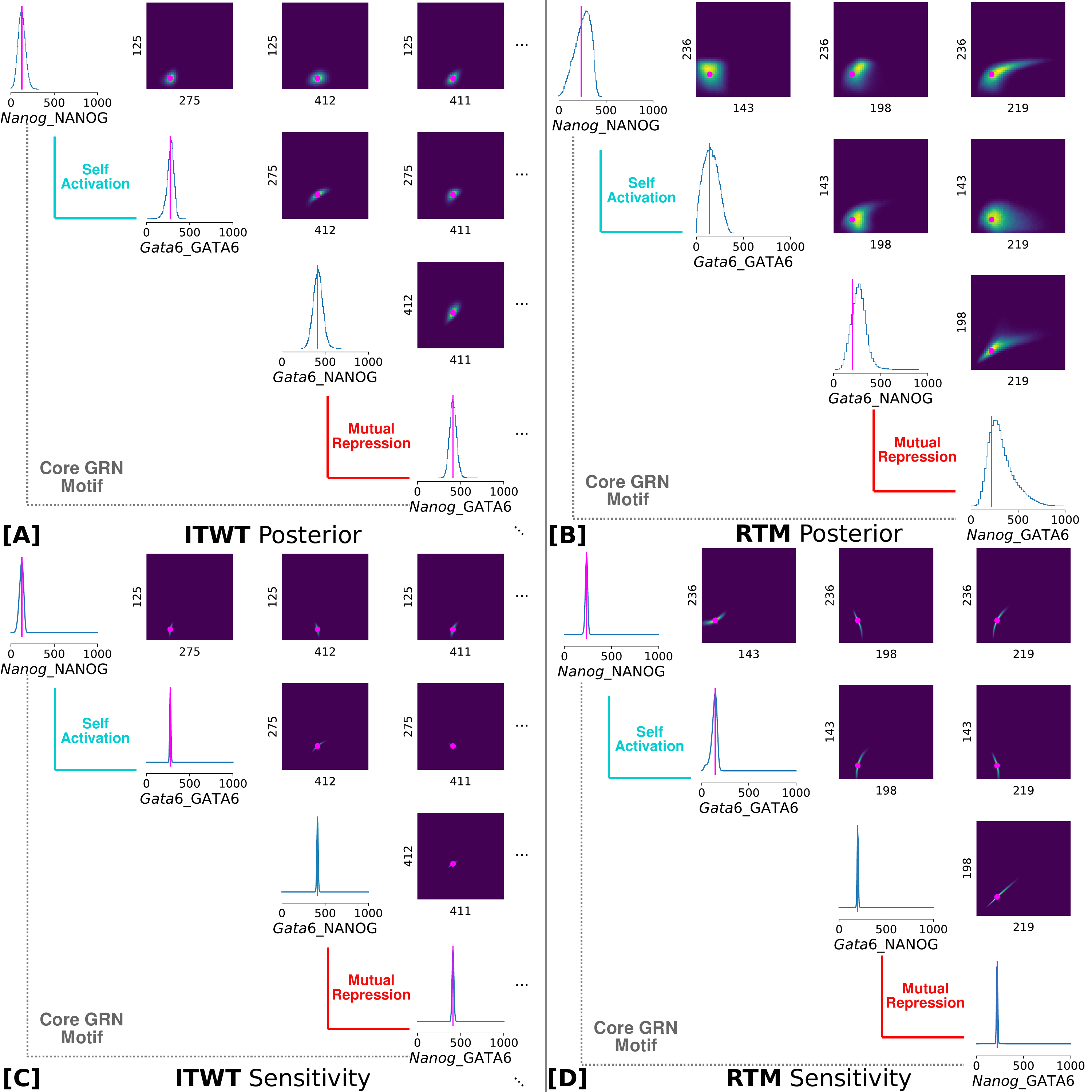} \centering 
\end{adjustwidth} 
\begin{figure}[hpt!]
\begin{adjustwidth}{-1.75in}{0in} 
\caption{{\bf Summary of inferred core GRN motif interactions (ITWT versus RTM).} The central component of our inference scheme is the sequential neural posterior estimation (SNPE) algorithm. Both unconditional (top row [A, B]) and conditional (bottom row [C, D]) posterior parameter distributions were obtained following 8 consecutive rounds of inference. 800 thousand composite simulations were performed for the ITWT system (left column [A, C]). For the RTM system (right column [B, D]), 4 consecutive rounds of inference were performed producing 400 thousand simulations. For complete details of the model parameter inference procedure, see \nameref{subsection:inference_framework}. {\bf [A, B]} Model parameter posterior distribution. For ease of visualization, we only show the one-dimensional projection of all posterior components representing the core GRN motif interactions. {\bf [C, D]} First assessment of model parameter sensitivity. These panels show the same components as in [A, B] but the posterior is now conditioned on the MAP estimate of the model parameters.}
\label{fig101}
\end{adjustwidth} 
\end{figure}
\clearpage

Comparably, the fourth pair of parameter blocks (Fig~\ref{fig102}[C, G]) displays a set of rather broad spans for the phosphorylation and dephosphorylation parameters relevant to the FGF-ERK pathway that propagates the external FGF4 signal back into the cell, except for $\tau_{\textrm{doh},\textrm{ERK}}$, the half-turnover time for the inactivation of ERK. The fast inactivation rate that we find here might indicate that the system relies on quick adaptations to the amount of extracellular signaling.

Lastly, the fifth pair of parameter blocks (Fig~\ref{fig102}[D, H]) shows several interesting relationships. The two parameters ``Mean Initial mRNA Count'' and ``Mean Initial PROTEIN Count'' are noticeably anti-correlated in a linear fashion. This fact plausibly indicates that the neural network learnt the complementary interaction between these two parameters, which have a conjoint effect on the initial expression dynamics of the three main genes (\textit{Nanog}, \textit{Gata6}, and \textit{Fgf4}). This complementary interaction itself reflects the need for constraining these two values to a region where they will not exceed the theoretical maximum mean copy numbers per cell of mRNA and protein molecules for the key players at the start of the simulations, in our case 250 mRNA (\textit{Nanog}-\textit{Gata6}) and 1000 protein (NANOG-GATA6) copies (see subsection \nameref{subsubsection:cell_scale_model} of ``Materials and Methods'') . For information about the definite usage of these two parameters to create the initial condition distributions for model simulations, see \nameref{subsection:computational_experiments}. In the same way, we see that the restricted value range of the parameter $\tau_{\textrm{d},\textrm{M-FGFR-FGF4}}$ might indicate that, once a substantial amount of FGF4 escapes the cytoplasm and binds to the available FGFR population on the cellular membrane, the FGFR-FGF4 complex-monomer lifetime must be tightly controlled as to not over-amplify the signal or to compromise the buffering of FGF4 fluctuations.

\subsection*{General considerations for model parameter inference}

Biophysically-realistic mechanistic modeling is integral to achieving a quantitative understanding of the behavior of complex biological systems. These biophysical representations are fundamentally generative models: they aim at providing a mechanistic description of the underlying biological phenomenon, enabling the generation of temporally-faithful synthetic trajectories of the given modeled dynamics, which potentially resemble empirical findings or target observations. As such, these models require a rich collection of experimental data detailing the principal mechanisms and physical processes triggering the phenomenon under study \cite{torregrosa_mechanistic_2021}.

For early developmental biology systems, in particular, experimental studies typically can not simultaneously measure all pivotal biophysical variables, and they generally can not capture enough granular features to facilitate comprehensive mechanistic modeling; especially, considering the complex GRNs and signaling pathways coordinating the proper progression of these systems, as well as their inherent nonlinear-multiscale dynamics. While these models are challenging to construct, we here demonstrate that despite the lack of detailed quantitative measurements, empirical qualitative observations can provide suitable bases for biophysical mechanistic modeling.

In that respect, formulating such a nonlinear-multiscale representation of the underlying biological system is only the preparatory step for model building. Arguably, the greatest challenge of biophysical mechanistic modeling is the estimation of suitable parameter values which allow the inferred model to recapitulate the most fundamental characteristics of the studied system behavior.

A popular method known as approximate Bayesian computation (ABC) has been widely used in computational biology for inferring sensible model parameter sets within the context of biophysical mechanistic modeling \cite{cranmer_frontier_2020}. ABC is itself a collection of modern algorithms which essentially compare experimental (target) and synthetic (simulation) data based on a given metric or distance function, measuring discrepancies between these two data vectors via the involvement of predetermined summary statistics. What is more, the family of ABC algorithms empowers the applicability of statistical inference to situations where the likelihood function of a stated model is intractable, enabling an approximation of the true posterior distribution of the respective parameter values.

\clearpage
\begin{adjustwidth}{-1.75in}{0in} 
\includegraphics[width = 4.375in, height = 8.75in]{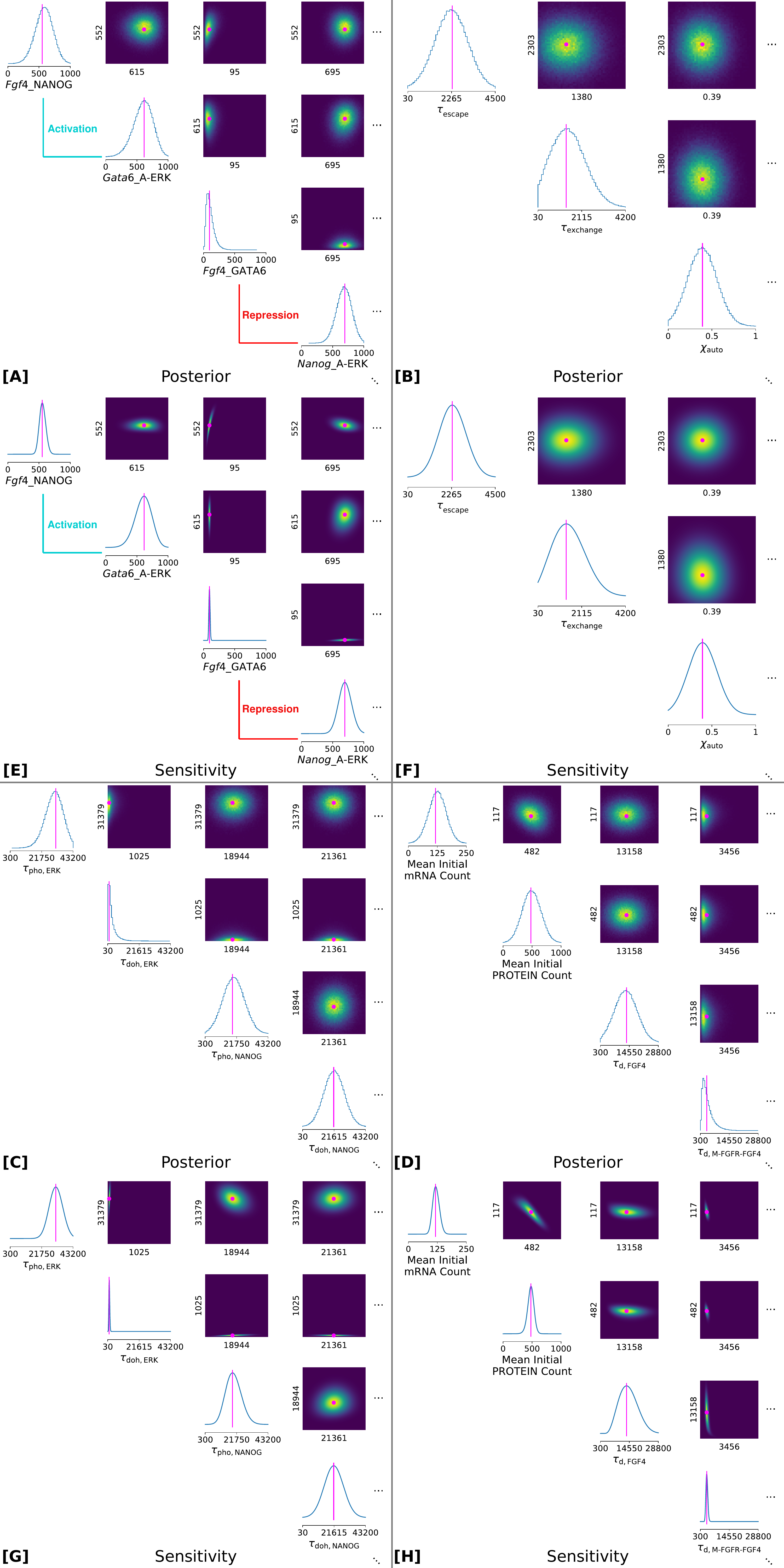} \centering 
\end{adjustwidth} 
\begin{figure}[hpt!]
\begin{adjustwidth}{-1.75in}{0in} 
\caption{{\bf Summary of inferred signaling and other model parameter interactions (ITWT only).} The central component of our inference scheme is the sequential neural posterior estimation (SNPE) algorithm. Both unconditional (top rows [A-D]) and conditional (bottom rows [E-H]) posterior parameter distributions were obtained following 8 consecutive rounds of inference. 800 thousand composite simulations were performed for the ITWT system. For complete details of the model parameter inference procedure, see \nameref{subsection:inference_framework}. {\bf [A-D]} Model parameter posterior distribution. For ease of visualization, the posterior was arbitrarily partitioned into four distinctive groups. We emphasize the top-right group [B, F], which displays the most important signaling model parameter interactions. {\bf [E-H]} First assessment of model parameter sensitivity. These panels show the same components as in [A-D] but the posterior is now conditioned on the maximum-a-posteriori probability (MAP) estimate of the model parameters.}
\label{fig102}
\end{adjustwidth} 
\end{figure}

However, an AI-powered SBI framework has recently emerged as an influential technique to tackle any likelihood-free inferential challenge \cite{cranmer_frontier_2020}. In general, SBI does not attempt to retrieve the single best parameter set. Instead, SBI helps to discover high-probability parameter space regions capable of explaining the target observations, while potentially facilitating the quantification of parameter uncertainty \cite{tejero-cantero_sbi_2020}.

Even though these AI-centered techniques are similar to the ABC methods, they provide significant benefits over their prevalent counterparts: they are innately flexible, which allow them to tackle a broad range of problems; they are structurally designed to learn low-dimensional representations of high-dimensional datasets, but trading off interpretability against predictive power of the stated model; they can in principle dramatically improve inference efficiency because they demand relatively small training sets, generating additional synthetic data from the learned latent space of the ANN; they automatically offer an estimate of the model parameter uncertainty, which does not need supplemental simulations. Particularly, the sequential neural posterior estimation (SNPE) procedure completely skips the creation of a virtual likelihood function, and it explicitly approximates the model parameter posterior distribution \cite{greenberg_automatic_2019, deistler_truncated_2022}.

In this work, to overcome our challenging inferential task and inspired by the simulation-based inference (SBI) framework, we exploit this novel AI-powered parameter estimation approach and combine it with various classical ML techniques. At its core, our inference approach directly exploits simulation data for exploring parameter space, yet it indirectly relies on primarily qualitative data to inform/constrain the parameter values; the simulation data is accordingly generated via the proposed spatial-stochastic model of mouse ICM lineage differentiation. Our model parameter estimation scheme is thus guided by the SBI paradigm \cite{cranmer_frontier_2020}, and its centerpiece is the sequential neural posterior estimation (SNPE) algorithm \cite{greenberg_automatic_2019, deistler_truncated_2022}.

While this novel family of Bayesian inference algorithms has a huge application potential, these schemes are typically difficult to implement, requiring access to big computational resources. These contemporary methods have been mostly adopted in neuroscience research \cite{cranmer_frontier_2020, deistler_energy-efficient_2022, kaiser_simulation-based_2023, tolley_methods_2023}, so far having a limited reach within the broad field of computational biology (and developmental biophysics in particular).

\subsubsection*{Preliminary model design}

The preparatory stage is establishing the elementary structure of the model. Our primary goal is to integrate two basic building blocks: a cell-scale submodel which consists of the GRN coordinating the mouse ICM lineage specification process; a tissue-scale submodel which describes the cell-cell signaling interactions. Another pivotal goal is to create a minimal spatial-stochastic description of the developing mouse ICM. This minimal model should enable exact simulation of the proposed system dynamics. Therefore, only the most relevant biochemical species prevailing in the underlying biological system should be part of this minimal model (see Table~\ref{table1}). On the flip side, other purely-computational species might be necessary for correctly analyzing and tracing the intricate simulated dynamics.

\clearpage

An additional basic assumption concerns the ICM spatial configuration. In our case, the full neighborhood representation is a rectangular voxel grid with a one-cell thickness, where each voxel emulates an embryo cell. This static cell arrangement isolates our analysis from other dimensions of variability, and it focuses the study to the question of how a cell-cell signaling mechanism might create the so-called ``salt-and-pepper'' pattern \cite{fischer_transition_2020, liebisch_cell_2020, forsyth_iven_2021, fischer_salt-and-pepper_2023}.

Altogether, biophysical mechanistic modeling is expected to give us a panoramic insight into the central mechanisms causing the phenomena under study. Ideally, this enriched understanding will generate testable predictions, guiding innovative experiments which pursue related processes. Although, systematically developing these models is a complex and nontrivial problem in computational biology \cite{massonis_distilling_2023}.

\subsubsection*{Basic theoretical assumptions}

To quantitatively model processes such as gene expression dynamics and cell signaling pathways, the Chemical Master Equation (CME) is a popular approach. Furthermore, to incorporate reaction-diffusion processes, these biochemical networks are preferably abstracted using the Reaction-Diffusion Master Equation (RDME) formalism. In that regard, these two frameworks represent biochemical phenomena as a network system involving multiple reactions (edges) and species (vertices or nodes), where each edge has an associated rate related to the propensity function of the given reaction \cite{pessoa_accelerating_2023}. Along with it, while balancing the relationship between computational efficiency and biophysical realism, the RDME fuses the notion of spatial heterogeneity or partitioning with the ingrained stochasticity of the CME.

Such stochasticity or noise plays a focal role for multifold cellular processes \cite{eldar_functional_2010, munsky_using_2012}. One particularly important example is the cell-fate decision-making process: even when genetically identical cells are subject to homogeneous (equivalent) experimental conditions, they manifest significant gene-expression variability \cite{schnoerr_approximation_2017, pang_probability_2023}. In fact, this cell-differentiation process is deemed to be a truthful reflection of the underlying gene regulatory network (GRN) dynamics. In a general sense, a central GRN not only must be capable of perpetuating a stable/attractor functional state, but it also must be sensitive to external stimuli in order to adapt the genetic program of a cell according to its environment, allowing it to reach other stable/attractor functional states \cite{bonnaffoux_wasabi_2019}. These factors ultimately permit a cellular population to coordinately achieve some globally-conserved pattern of lineages.

Within this context, two contrasting perspectives exist to interpret cellular patterning heterogeneity. The classical perspective views gene expression and cell-fate decision making as fundamentally deterministic mechanisms where noise has a mainly passive role; in this case, the phenomenon of random molecular/cellular count fluctuations is assumed to be an additive (or sometimes multiplicative) noise process. Instead, we adopt in this work the other ``contemporary'' perspective: cell-cell variability encapsulates highly-valuable information about the genetic programs orchestrating cellular-function specification, which itself manifests the inherently noisy nature of the underlying gene expression regulation \cite{bonnaffoux_wasabi_2019}. This stochasticity-controlled perspective (where noise has a structurally active role) is thus attained via the RDME formalism, which in turn facilitates the inclusion of spatial/compartmental modeling elements.

\subsubsection*{Basic empirical observations}

A central idea of biochemical network model building is to reproduce some particularly interesting behaviors, phenotypes, or empirical observations \cite{stillman_generative_2023}. Nevertheless, because of highly-nonlinear interactions among genetic regulatory components, a single set of experimental data by itself can not uncover the full functional dynamics of such biological systems \cite{schnoerr_approximation_2017}. It is hence normally necessary to consolidate several datasets coming from multiple related systems under diverse experimental conditions; for this reason, coherent and viable data analysis becomes a considerably difficult endeavor on its own.

What is more, these quantitative observations often do not provide enough details for spatial-stochastic modeling, and they are typically incomplete: from the practical point of view, it is extremely challenging to provide simultaneous measurements for all the key elements dictating the underlying multiscale dynamics.

However, we can exploit any (high-level) mostly-qualitative experimental observation for successfully constraining/informing our model parameter inference approach. For the developing mouse blastocyst and its related experimental systems, there are three determining characteristics which describe the ICM fate-specification process: (1) the two cell lineages emerging from the ICM exhibit highly-reproducible proportions (ratio of $2:3$ for EPI and PRE fates) \cite{bessonnard_gata6_2014, saiz_asynchronous_2016}; (2) the formation of the blastocyst takes approximately 1.5-2 days of embryonic development \cite{plusa_common_2020, saiz_coordination_2020, allegre_nanog_2022}, and (more crucially) EPI-PRE populations should reach their expected fate proportions by the 40-hour time mark (roughly 8 or 12 hours before the end of the preimplantation period) \cite{yanagida_cell_2022, allegre_nanog_2022}; (3) the absence of FGF4 signaling forces the ICM to almost exclusively adopt an EPI fate (the ICM population is naturally biased towards a naive cellular pluripotency) \cite{bessonnard_icm_2017, thompson_extensive_2022}.

The last characteristic is notably important because it underlines the key role that cell-cell signaling plays for correct ICM patterning formation. This feature also highlights a critical difference between the two inferred system models: for replicating the target behavior of the underlying biological system (final EPI-PRE proportions), the ITWT model should involve a cell-nonautonomous mechanism and the RTM model should depend on a cell-autonomous process. In other words, within the whole ICM population, the probability of a given cell adopting a specific lineage should be conditional on the fates taken by other cells for the ITWT, and it should be independent of the fates taken by other cells for the RTM.




\section*{Acknowledgments}

The successful completion of this research project owes much to the collaboration and support of esteemed colleagues and collaborators. We extend our deepest gratitude to Sabine Fischer, Tim Liebisch, Franziska Matthäus, and Simon Schardt for their pivotal roles in fostering insightful discussions and providing constructive feedback.

We also express our appreciation to Roberto Covino and his lab for their invaluable guidance and expertise throughout the project, especially for pointing us in the direction of the powerful SBI framework. Their thoughtful insights and constructive critiques have enriched the quality of our work.

Furthermore, we would like to acknowledge the CMMS project and FIAS for their financial support, which has been instrumental in facilitating the progress of this research. We also thank the Center for Scientific Computing (CSC) at Goethe University Frankfurt for granting us access to the Goethe-HLR cluster.



\bibliography{Manuscript_Catalog.bib}

\begin{thebibliography}{100}

\bibitem{cang_multiscale_2021}
Cang Z, Wang Y, Wang Q, Cho KWY, Holmes W, Nie Q.
\newblock A multiscale model via single-cell transcriptomics reveals robust patterning mechanisms during early mammalian embryo development.
\newblock PLOS Computational Biology. 2021;17(3):e1008571.
\newblock doi:{10.1371/journal.pcbi.1008571}.

\bibitem{krupinski_simulating_2011}
Krupinski P, Chickarmane V, Peterson C.
\newblock Simulating the {Mammalian} {Blastocyst} - {Molecular} and {Mechanical} {Interactions} {Pattern} the {Embryo}.
\newblock PLOS Computational Biology. 2011;7(5):e1001128.
\newblock doi:{10.1371/journal.pcbi.1001128}.

\bibitem{rossant_blastocyst_2009}
Rossant J, Tam PPL.
\newblock Blastocyst lineage formation, early embryonic asymmetries and axis patterning in the mouse.
\newblock Development. 2009;136(5):701--713.
\newblock doi:{10.1242/dev.017178}.

\bibitem{li_maternal_2010}
Li L, Zheng P, Dean J.
\newblock Maternal control of early mouse development.
\newblock Development (Cambridge, England). 2010;137(6):859--870.
\newblock doi:{10.1242/dev.039487}.

\bibitem{tosenberger_multiscale_2017}
Tosenberger A, Gonze D, Bessonnard S, Cohen-Tannoudji M, Chazaud C, Dupont G.
\newblock A multiscale model of early cell lineage specification including cell division.
\newblock npj Systems Biology and Applications. 2017;3(1):1--11.
\newblock doi:{10.1038/s41540-017-0017-0}.

\bibitem{tosenberger_computational_2019}
Tosenberger A, Gonze D, Chazaud C, Dupont G.
\newblock Computational models for the dynamics of early mouse embryogenesis.
\newblock International Journal of Developmental Biology. 2019;63(3-4-5):131--142.
\newblock doi:{10.1387/ijdb.180418gd}.

\bibitem{habibi_transcriptional_2017}
Habibi E, Stunnenberg HG.
\newblock Transcriptional and epigenetic control in mouse pluripotency: lessons from in vivo and in vitro studies.
\newblock Current Opinion in Genetics \& Development. 2017;46:114--122.
\newblock doi:{10.1016/j.gde.2017.07.005}.

\bibitem{arias_molecular_2013}
Arias AM, Nichols J, Schröter C.
\newblock A molecular basis for developmental plasticity in early mammalian embryos.
\newblock Development. 2013;140(17):3499--3510.
\newblock doi:{10.1242/dev.091959}.

\bibitem{herberg_computational_2015}
Herberg M, Roeder I.
\newblock Computational modelling of embryonic stem-cell fate control.
\newblock Development. 2015;142(13):2250--2260.
\newblock doi:{10.1242/dev.116343}.

\bibitem{miyamoto_pluripotency_2015}
Miyamoto T, Furusawa C, Kaneko K.
\newblock Pluripotency, {Differentiation}, and {Reprogramming}: {A} {Gene} {Expression} {Dynamics} {Model} with {Epigenetic} {Feedback} {Regulation}.
\newblock PLOS Computational Biology. 2015;11(8):e1004476.
\newblock doi:{10.1371/journal.pcbi.1004476}.

\bibitem{nissen_four_2017}
Nissen SB, Perera M, Gonzalez JM, Morgani SM, Jensen MH, Sneppen K, et~al.
\newblock Four simple rules that are sufficient to generate the mammalian blastocyst.
\newblock PLOS Biology. 2017;15(7):e2000737.
\newblock doi:{10.1371/journal.pbio.2000737}.

\bibitem{bessonnard_icm_2017}
Bessonnard S, Coqueran S, Vandormael-Pournin S, Dufour A, Artus J, Cohen-Tannoudji M.
\newblock {ICM} conversion to epiblast by {FGF}/{ERK} inhibition is limited in time and requires transcription and protein degradation.
\newblock Scientific Reports. 2017;7(1):12285.
\newblock doi:{10.1038/s41598-017-12120-0}.

\bibitem{stanoev_robustness_2021}
Stanoev A, Schröter C, Koseska A.
\newblock Robustness and timing of cellular differentiation through population-based symmetry breaking.
\newblock Development. 2021;148(3):dev197608.
\newblock doi:{10.1242/dev.197608}.

\bibitem{robert_initial_2022}
Robert C, Prista~von Bonhorst F, De~Decker Y, Dupont G, Gonze D.
\newblock Initial source of heterogeneity in a model for cell fate decision in the early mammalian embryo.
\newblock Interface Focus. 2022;12(4):20220010.
\newblock doi:{10.1098/rsfs.2022.0010}.

\bibitem{mathew_mouse_2019}
Mathew B, Muñoz-Descalzo S, Corujo-Simon E, Schröter C, Stelzer EHK, Fischer SC.
\newblock Mouse {ICM} {Organoids} {Reveal} {Three}-{Dimensional} {Cell} {Fate} {Clustering}.
\newblock Biophysical Journal. 2019;116(1):127--141.
\newblock doi:{10.1016/j.bpj.2018.11.011}.

\bibitem{liebisch_cell_2020}
Liebisch T, Drusko A, Mathew B, Stelzer EHK, Fischer SC, Matthäus F.
\newblock Cell fate clusters in {ICM} organoids arise from cell fate heredity and division: a modelling approach.
\newblock Scientific Reports. 2020;10(1):22405.
\newblock doi:{10.1038/s41598-020-80141-3}.

\bibitem{zhu_synthetic_2022}
Zhu R, del Rio-Salgado JM, Garcia-Ojalvo J, Elowitz MB.
\newblock Synthetic multistability in mammalian cells.
\newblock Science. 2022;375(6578):eabg9765.
\newblock doi:{10.1126/science.abg9765}.

\bibitem{zhu_principles_2020}
Zhu M, Zernicka-Goetz M.
\newblock Principles of {Self}-{Organization} of the {Mammalian} {Embryo}.
\newblock Cell. 2020;183(6):1467--1478.
\newblock doi:{10.1016/j.cell.2020.11.003}.

\bibitem{plusa_common_2020}
Płusa B, Piliszek A.
\newblock Common principles of early mammalian embryo self-organisation.
\newblock Development. 2020;147(dev183079).
\newblock doi:{10.1242/dev.183079}.

\bibitem{zernicka-goetz_cleavage_2005}
Zernicka-Goetz M.
\newblock Cleavage pattern and emerging asymmetry of the mouse embryo.
\newblock Nature Reviews Molecular Cell Biology. 2005;6(12):919--928.
\newblock doi:{10.1038/nrm1782}.

\bibitem{saiz_coordination_2020}
Saiz N, Hadjantonakis AK.
\newblock Coordination between patterning and morphogenesis ensures robustness during mouse development.
\newblock Philosophical Transactions of the Royal Society B. 2020;doi:{10.1098/rstb.2019.0562}.

\bibitem{ryan_lumen_2019}
Ryan AQ, Chan CJ, Graner F, Hiiragi T.
\newblock Lumen {Expansion} {Facilitates} {Epiblast}-{Primitive} {Endoderm} {Fate} {Specification} during {Mouse} {Blastocyst} {Formation}.
\newblock Developmental Cell. 2019;51(6):684--697.e4.
\newblock doi:{10.1016/j.devcel.2019.10.011}.

\bibitem{saiz_growth-factor-mediated_2020}
Saiz N, Mora-Bitria L, Rahman S, George H, Herder JP, Garcia-Ojalvo J, et~al.
\newblock Growth-factor-mediated coupling between lineage size and cell fate choice underlies robustness of mammalian development.
\newblock eLife. 2020;9.
\newblock doi:{10.7554/eLife.56079}.

\bibitem{simon_making_2018}
Simon CS, Hadjantonakis AK, Schröter C.
\newblock Making lineage decisions with biological noise: {Lessons} from the early mouse embryo.
\newblock WIREs Developmental Biology. 2018;7(4):e319.
\newblock doi:{https://doi.org/10.1002/wdev.319}.

\bibitem{allegre_nanog_2022}
Allègre N, Chauveau S, Dennis C, Renaud Y, Meistermann D, Estrella LV, et~al.
\newblock {NANOG} initiates epiblast fate through the coordination of pluripotency genes expression.
\newblock Nature Communications. 2022;13(1):3550.
\newblock doi:{10.1038/s41467-022-30858-8}.

\bibitem{meilhac_active_2009}
Meilhac SM, Adams RJ, Morris SA, Danckaert A, Le~Garrec JF, Zernicka-Goetz M.
\newblock Active cell movements coupled to positional induction are involved in lineage segregation in the mouse blastocyst.
\newblock Developmental Biology. 2009;331(2):210--221.
\newblock doi:{10.1016/j.ydbio.2009.04.036}.

\bibitem{chen_tracing_2018}
Chen Q, Shi J, Tao Y, Zernicka-Goetz M.
\newblock Tracing the origin of heterogeneity and symmetry breaking in the early mammalian embryo.
\newblock Nature Communications. 2018;9(1):1819.
\newblock doi:{10.1038/s41467-018-04155-2}.

\bibitem{schultz_oocyte--embryo_2018}
Schultz RM, Stein P, Svoboda P.
\newblock The oocyte-to-embryo transition in mouse: past, present, and future.
\newblock Biology of Reproduction. 2018;99(1):160--174.
\newblock doi:{10.1093/biolre/ioy013}.

\bibitem{niwayama_tug--war_2019}
Niwayama R, Moghe P, Liu YJ, Fabrèges D, Buchholz F, Piel M, et~al.
\newblock A {Tug}-of-{War} between {Cell} {Shape} and {Polarity} {Controls} {Division} {Orientation} to {Ensure} {Robust} {Patterning} in the {Mouse} {Blastocyst}.
\newblock Developmental Cell. 2019;51(5):564--574.e6.
\newblock doi:{10.1016/j.devcel.2019.10.012}.

\bibitem{fischer_transition_2020}
Fischer SC, Corujo-Simon E, Lilao-Garzon J, Stelzer EHK, Muñoz-Descalzo S.
\newblock The transition from local to global patterns governs the differentiation of mouse blastocysts.
\newblock PLoS ONE. 2020;15(5).
\newblock doi:{10.1371/journal.pone.0233030}.

\bibitem{chan_integration_2020}
Chan CJ, Hiiragi T.
\newblock Integration of luminal pressure and signalling in tissue self-organization.
\newblock Development. 2020;147(5):dev181297.
\newblock doi:{10.1242/dev.181297}.

\bibitem{yanagida_cell_2022}
Yanagida A, Corujo-Simon E, Revell CK, Sahu P, Stirparo GG, Aspalter IM, et~al.
\newblock Cell surface fluctuations regulate early embryonic lineage sorting.
\newblock Cell. 2022;185(5):777--793.e20.
\newblock doi:{10.1016/j.cell.2022.01.022}.

\bibitem{bessonnard_gata6_2014}
Bessonnard S, De~Mot L, Gonze D, Barriol M, Dennis C, Goldbeter A, et~al.
\newblock Gata6, {Nanog} and {Erk} signaling control cell fate in the inner cell mass through a tristable regulatory network.
\newblock Development. 2014;141(19):3637--3648.
\newblock doi:{10.1242/dev.109678}.

\bibitem{schrode_regulation_2015}
Schrode N.
\newblock Regulation of cell fate choice in the mouse blastocyst stage embryo [{PhD} {Thesis}].
\newblock Ludwig-Maximilians-Universität München; 2015.
\newblock Available from: \url{https://edoc.ub.uni-muenchen.de/18938/}.

\bibitem{schroter_fgfmapk_2015}
Schröter C, Rué P, Mackenzie JP, Martinez~Arias A.
\newblock {FGF}/{MAPK} signaling sets the switching threshold of a bistable circuit controlling cell fate decisions in embryonic stem cells.
\newblock Development. 2015;142(24):4205--4216.
\newblock doi:{10.1242/dev.127530}.

\bibitem{raina_cell-cell_2021}
Raina D, Bahadori A, Stanoev A, Protzek M, Koseska A, Schröter C.
\newblock Cell-cell communication through {FGF4} generates and maintains robust proportions of differentiated cell types in embryonic stem cells.
\newblock Development. 2021;148(21):dev199926.
\newblock doi:{10.1242/dev.199926}.

\bibitem{raina_intermittent_2022}
Raina D, Fabris F, Morelli LG, Schröter C.
\newblock Intermittent {ERK} oscillations downstream of {FGF} in mouse embryonic stem cells.
\newblock Development. 2022;149(4):dev199710.
\newblock doi:{10.1242/dev.199710}.

\bibitem{plachta_oct4_2011}
Plachta N, Bollenbach T, Pease S, Fraser SE, Pantazis P.
\newblock Oct4 kinetics predict cell lineage patterning in the early mammalian embryo.
\newblock Nature Cell Biology. 2011;13(2):117--123.
\newblock doi:{10.1038/ncb2154}.

\bibitem{krawczyk_paracrine_2022}
Krawczyk K, Wilczak K, Szczepańska K, Maleszewski M, Suwińska A.
\newblock Paracrine interactions through {FGFR1} and {FGFR2} receptors regulate the development of preimplantation mouse chimaeric embryo.
\newblock Open Biology. 2022;12(11):220193.
\newblock doi:{10.1098/rsob.220193}.

\bibitem{burda_motifs_2011}
Burda Z, Krzywicki A, Martin OC, Zagorski M.
\newblock Motifs emerge from function in model gene regulatory networks.
\newblock Proceedings of the National Academy of Sciences. 2011;108(42):17263--17268.
\newblock doi:{10.1073/pnas.1109435108}.

\bibitem{sokolowski_mutual_2012}
Sokolowski TR, Erdmann T, Wolde PRt.
\newblock Mutual {Repression} {Enhances} the {Steepness} and {Precision} of {Gene} {Expression} {Boundaries}.
\newblock PLOS Computational Biology. 2012;8(8):e1002654.
\newblock doi:{10.1371/journal.pcbi.1002654}.

\bibitem{sokolowski_deriving_2023}
Sokolowski TR, Gregor T, Bialek W, Tkačik G. Deriving a genetic regulatory network from an optimization principle; 2023.
\newblock Available from: \url{http://arxiv.org/abs/2302.05680}.

\bibitem{majka_stability_2023}
Majka M, Ho RDJG, Zagorski M.
\newblock Stability of pattern formation in systems with dynamic source regions.
\newblock Physical Review Letters. 2023;130(9):098402.
\newblock doi:{10.1103/PhysRevLett.130.098402}.

\bibitem{majka_stable_2023}
Majka M, Becker NB, Wolde PRt, Zagorski M, Sokolowski TR. Stable developmental patterns of gene expression without morphogen gradients; 2023.
\newblock Available from: \url{http://arxiv.org/abs/2306.00537}.

\bibitem{gregor_stability_2007}
Gregor T, Wieschaus EF, McGregor AP, Bialek W, Tank DW.
\newblock Stability and {Nuclear} {Dynamics} of the {Bicoid} {Morphogen} {Gradient}.
\newblock Cell. 2007;130(1):141--152.
\newblock doi:{10.1016/j.cell.2007.05.026}.

\bibitem{bollenbach_precision_2008}
Bollenbach T, Pantazis P, Kicheva A, Bökel C, González-Gaitán M, Jülicher F.
\newblock Precision of the {Dpp} gradient.
\newblock Development. 2008;135(6):1137--1146.
\newblock doi:{10.1242/dev.012062}.

\bibitem{little_formation_2011}
Little SC, Tkačik G, Kneeland TB, Wieschaus EF, Gregor T.
\newblock The {Formation} of the {Bicoid} {Morphogen} {Gradient} {Requires} {Protein} {Movement} from {Anteriorly} {Localized} {mRNA}.
\newblock PLOS Biology. 2011;9(3):e1000596.
\newblock doi:{10.1371/journal.pbio.1000596}.

\bibitem{richards_spatiotemporal_2015}
Richards DM, Saunders TE.
\newblock Spatiotemporal {Analysis} of {Different} {Mechanisms} for {Interpreting} {Morphogen} {Gradients}.
\newblock Biophysical Journal. 2015;108(8):2061--2073.
\newblock doi:{10.1016/j.bpj.2015.03.015}.

\bibitem{smith_role_2016}
Smith T, Fancher S, Levchenko A, Nemenman I, Mugler A.
\newblock Role of spatial averaging in multicellular gradient sensing.
\newblock Physical Biology. 2016;13(3):035004.
\newblock doi:{10.1088/1478-3975/13/3/035004}.

\bibitem{ellison_cellcell_2016}
Ellison D, Mugler A, Brennan MD, Lee SH, Huebner RJ, Shamir ER, et~al.
\newblock Cell–cell communication enhances the capacity of cell ensembles to sense shallow gradients during morphogenesis.
\newblock Proceedings of the National Academy of Sciences. 2016;113(6):E679--E688.
\newblock doi:{10.1073/pnas.1516503113}.

\bibitem{zagorski_decoding_2017}
Zagorski M, Tabata Y, Brandenberg N, Lutolf MP, Tkačik G, Bollenbach T, et~al.
\newblock Decoding of position in the developing neural tube from antiparallel morphogen gradients.
\newblock Science. 2017;356(6345):1379--1383.
\newblock doi:{10.1126/science.aam5887}.

\bibitem{verd_dynamic_2017}
Verd B, Crombach A, Jaeger J.
\newblock Dynamic {Maternal} {Gradients} {Control} {Timing} and {Shift}-{Rates} for {Drosophila} {Gap} {Gene} {Expression}.
\newblock PLOS Computational Biology. 2017;13(2):e1005285.
\newblock doi:{10.1371/journal.pcbi.1005285}.

\bibitem{vakulenko_size_2009}
Vakulenko S, {Manu}, Reinitz J, Radulescu O.
\newblock Size {Regulation} in the {Segmentation} of {Drosophila}: {Interacting} {Interfaces} between {Localized} {Domains} of {Gene} {Expression} {Ensure} {Robust} {Spatial} {Patterning}.
\newblock Physical Review Letters. 2009;103(16):168102.
\newblock doi:{10.1103/PhysRevLett.103.168102}.

\bibitem{kicheva_coordination_2014}
Kicheva A, Bollenbach T, Ribeiro A, Valle HP, Lovell-Badge R, Episkopou V, et~al.
\newblock Coordination of progenitor specification and growth in mouse and chick spinal cord.
\newblock Science. 2014;345(6204):1254927.
\newblock doi:{10.1126/science.1254927}.

\bibitem{raspopovic_digit_2014}
Raspopovic J, Marcon L, Russo L, Sharpe J.
\newblock Digit patterning is controlled by a {Bmp}-{Sox9}-{Wnt} {Turing} network modulated by morphogen gradients.
\newblock Science. 2014;345(6196):566--570.
\newblock doi:{10.1126/science.1252960}.

\bibitem{almuedo-castillo_scale-invariant_2018}
Almuedo-Castillo M, Bläßle A, Mörsdorf D, Marcon L, Soh GH, Rogers KW, et~al.
\newblock Scale-invariant patterning by size-dependent inhibition of {Nodal} signalling.
\newblock Nature Cell Biology. 2018;20(9):1032--1042.
\newblock doi:{10.1038/s41556-018-0155-7}.

\bibitem{verd_damped_2018}
Verd B, Clark E, Wotton KR, Janssens H, Jiménez-Guri E, Crombach A, et~al.
\newblock A damped oscillator imposes temporal order on posterior gap gene expression in {Drosophila}.
\newblock PLOS Biology. 2018;16(2):e2003174.
\newblock doi:{10.1371/journal.pbio.2003174}.

\bibitem{morales_embryos_2021}
Morales JS, Raspopovic J, Marcon L.
\newblock From embryos to embryoids: {How} external signals and self-organization drive embryonic development.
\newblock Stem Cell Reports. 2021;16(5):1039--1050.
\newblock doi:{10.1016/j.stemcr.2021.03.026}.

\bibitem{nikolic_scale_2023}
Nikolić M, Antonetti V, Liu F, Muhaxheri G, Petkova MD, Scheeler M, et~al.. Scale invariance in early embryonic development; 2023.
\newblock Available from: \url{http://arxiv.org/abs/2312.17684}.

\bibitem{ochiai_stochastic_2014}
Ochiai H, Sugawara T, Sakuma T, Yamamoto T.
\newblock Stochastic promoter activation affects {Nanog} expression variability in mouse embryonic stem cells.
\newblock Scientific Reports. 2014;4(1):7125.
\newblock doi:{10.1038/srep07125}.

\bibitem{ochiai_genome-wide_2020}
Ochiai H, Hayashi T, Umeda M, Yoshimura M, Harada A, Shimizu Y, et~al.
\newblock Genome-wide kinetic properties of transcriptional bursting in mouse embryonic stem cells.
\newblock Science Advances. 2020;6(25):eaaz6699.
\newblock doi:{10.1126/sciadv.aaz6699}.

\bibitem{thompson_extensive_2022}
Thompson JJ, Lee DJ, Mitra A, Frail S, Dale RK, Rocha PP.
\newblock Extensive co-binding and rapid redistribution of {NANOG} and {GATA6} during emergence of divergent lineages.
\newblock Nature Communications. 2022;13(1):4257.
\newblock doi:{10.1038/s41467-022-31938-5}.

\bibitem{gonze_modeling-based_2018}
Gonze D, Gérard C, Wacquier B, Woller A, Tosenberger A, Goldbeter A, et~al.
\newblock Modeling-{Based} {Investigation} of the {Effect} of {Noise} in {Cellular} {Systems}.
\newblock Frontiers in Molecular Biosciences. 2018;5.
\newblock doi:{10.3389/fmolb.2018.00034}.

\bibitem{tkacik_many_2021}
Tkačik G, Gregor T.
\newblock The many bits of positional information.
\newblock Development. 2021;148(dev176065).
\newblock doi:{10.1242/dev.176065}.

\bibitem{lin_stochastic_2018}
Lin YT, Hufton PG, Lee EJ, Potoyan DA.
\newblock A stochastic and dynamical view of pluripotency in mouse embryonic stem cells.
\newblock PLoS Computational Biology. 2018;14(2).
\newblock doi:{10.1371/journal.pcbi.1006000}.

\bibitem{lin_central_2016}
Lin Y, Elowitz M.
\newblock Central {Dogma} {Goes} {Digital}.
\newblock Molecular Cell. 2016;61(6):791--792.
\newblock doi:{10.1016/j.molcel.2016.03.005}.

\bibitem{vandevenne_rna_2019}
Vandevenne M, Delmarcelle M, Galleni M.
\newblock {RNA} {Regulatory} {Networks} as a {Control} of {Stochasticity} in {Biological} {Systems}.
\newblock Frontiers in Genetics. 2019;10:403.
\newblock doi:{10.3389/fgene.2019.00403}.

\bibitem{pantazis_transcription_2012}
Pantazis P, Bollenbach T.
\newblock Transcription factor kinetics and the emerging asymmetry in the early mammalian embryo.
\newblock Cell Cycle. 2012;11(11):2055--2058.
\newblock doi:{10.4161/cc.20118}.

\bibitem{bezeljak_stochastic_2020}
Bezeljak U, Loya H, Kaczmarek B, Saunders TE, Loose M.
\newblock Stochastic activation and bistability in a {Rab} {GTPase} regulatory network.
\newblock Proceedings of the National Academy of Sciences. 2020;117(12):6540--6549.
\newblock doi:{10.1073/pnas.1921027117}.

\bibitem{dirk_recognition_2023}
Dirk R, Fischer JL, Schardt S, Ankenbrand MJ, Fischer SC.
\newblock Recognition and reconstruction of cell differentiation patterns with deep learning.
\newblock PLOS Computational Biology. 2023;19(10):e1011582.
\newblock doi:{10.1371/journal.pcbi.1011582}.

\bibitem{greenberg_automatic_2019}
Greenberg DS, Nonnenmacher M, Macke JH. Automatic {Posterior} {Transformation} for {Likelihood}-{Free} {Inference}; 2019.
\newblock Available from: \url{http://arxiv.org/abs/1905.07488}.

\bibitem{deistler_truncated_2022}
Deistler M, Goncalves PJ, Macke JH. Truncated proposals for scalable and hassle-free simulation-based inference; 2022.
\newblock Available from: \url{http://arxiv.org/abs/2210.04815}.

\bibitem{baker_mechanistic_2018}
Baker RE, Peña JM, Jayamohan J, Jérusalem A.
\newblock Mechanistic models versus machine learning, a fight worth fighting for the biological community?
\newblock Biology Letters. 2018;14(5):20170660.
\newblock doi:{10.1098/rsbl.2017.0660}.

\bibitem{cranmer_frontier_2020}
Cranmer K, Brehmer J, Louppe G.
\newblock The frontier of simulation-based inference.
\newblock Proceedings of the National Academy of Sciences. 2020;117(48):30055--30062.
\newblock doi:{10.1073/pnas.1912789117}.

\bibitem{lagergren_biologically-informed_2020}
Lagergren JH, Nardini JT, Baker RE, Simpson MJ, Flores KB.
\newblock Biologically-informed neural networks guide mechanistic modeling from sparse experimental data.
\newblock PLOS Computational Biology. 2020;16(12):e1008462.
\newblock doi:{10.1371/journal.pcbi.1008462}.

\bibitem{seyboldt_latent_2022}
Seyboldt R, Lavoie J, Henry A, Vanaret J, Petkova MD, Gregor T, et~al.
\newblock Latent space of a small genetic network: {Geometry} of dynamics and information.
\newblock Proceedings of the National Academy of Sciences. 2022;119(26):e2113651119.
\newblock doi:{10.1073/pnas.2113651119}.

\bibitem{perez_efficient_2022}
Perez SM, Sailem H, Baker RE.
\newblock Efficient {Bayesian} inference for mechanistic modelling with high-throughput data.
\newblock PLOS Computational Biology. 2022;18(6):e1010191.
\newblock doi:{10.1371/journal.pcbi.1010191}.

\bibitem{tolley_methods_2023}
Tolley N, Rodrigues PLC, Gramfort A, Jones S. Methods and considerations for estimating parameters in biophysically detailed neural models with simulation based inference; 2023.
\newblock Available from: \url{https://www.biorxiv.org/content/10.1101/2023.04.17.537118v1}.

\bibitem{hashemi_amortized_2023}
Hashemi M, Vattikonda AN, Jha J, Sip V, Woodman MM, Bartolomei F, et~al.
\newblock Amortized {Bayesian} inference on generative dynamical network models of epilepsy using deep neural density estimators.
\newblock Neural Networks. 2023;doi:{10.1016/j.neunet.2023.03.040}.

\bibitem{stillman_generative_2023}
Stillman NR, Mayor R.
\newblock Generative models of morphogenesis in developmental biology.
\newblock Seminars in Cell \& Developmental Biology. 2023;147:83--90.
\newblock doi:{10.1016/j.semcdb.2023.02.001}.

\bibitem{schnoerr_approximation_2017}
Schnoerr D, Sanguinetti G, Grima R.
\newblock Approximation and inference methods for stochastic biochemical kinetics—a tutorial review.
\newblock Journal of Physics A: Mathematical and Theoretical. 2017;50(9):093001.
\newblock doi:{10.1088/1751-8121/aa54d9}.

\bibitem{franzin_landscape-based_2023}
Franzin A, Stützle T.
\newblock A landscape-based analysis of fixed temperature and simulated annealing.
\newblock European Journal of Operational Research. 2023;304(2):395--410.
\newblock doi:{10.1016/j.ejor.2022.04.014}.

\bibitem{ramirez-sierra_comparing_2024}
Ramirez-Sierra MA, Sokolowski TR.
\newblock Comparing {AI} and {Optimization} {Workflows} for {Simulation}-{Based} {Inference} of {Spatial}-{Stochastic} {Systems}.
\newblock In Preparation. 2024;doi:{Forthcoming}.

\bibitem{saiz_asynchronous_2016}
Saiz N, Williams KM, Seshan VE, Hadjantonakis AK.
\newblock Asynchronous fate decisions by single cells collectively ensure consistent lineage composition in the mouse blastocyst.
\newblock Nature Communications. 2016;7.
\newblock doi:{10.1038/ncomms13463}.

\bibitem{abranches_stochastic_2014}
Abranches E, Guedes AMV, Moravec M, Maamar H, Svoboda P, Raj A, et~al.
\newblock Stochastic {NANOG} fluctuations allow mouse embryonic stem cells to explore pluripotency.
\newblock Development. 2014;141(14):2770--2779.
\newblock doi:{10.1242/dev.108910}.

\bibitem{xenopoulos_heterogeneities_2015}
Xenopoulos P, Kang M, Puliafito A, Di Talia S, Hadjantonakis AK.
\newblock Heterogeneities in {Nanog} {Expression} {Drive} {Stable} {Commitment} to {Pluripotency} in the {Mouse} {Blastocyst}.
\newblock Cell Reports. 2015;10(9):1508--1520.
\newblock doi:{10.1016/j.celrep.2015.02.010}.

\bibitem{kale_nanog-perk_2022}
Kale HT, Rajpurohit RS, Jana D, Vishnu VV, Srivastava M, Mourya PR, et~al.
\newblock A {NANOG}-{pERK} reciprocal regulatory circuit regulates {Nanog} autoregulation and {ERK} signaling dynamics.
\newblock EMBO reports. 2022;23(11):e54421.
\newblock doi:{10.15252/embr.202154421}.

\bibitem{demot_cell_2016}
De Mot L, Gonze D, Bessonnard S, Chazaud C, Goldbeter A, Dupont G.
\newblock Cell {Fate} {Specification} {Based} on {Tristability} in the {Inner} {Cell} {Mass} of {Mouse} {Blastocysts}.
\newblock Biophysical Journal. 2016;110(3):710--722.
\newblock doi:{10.1016/j.bpj.2015.12.020}.

\bibitem{erdmann_role_2009}
Erdmann T, Howard M, ten Wolde PR.
\newblock Role of {Spatial} {Averaging} in the {Precision} of {Gene} {Expression} {Patterns}.
\newblock Physical Review Letters. 2009;103(25):258101.
\newblock doi:{10.1103/PhysRevLett.103.258101}.

\bibitem{sokolowski_optimizing_2015}
Sokolowski TR, Tkačik G.
\newblock Optimizing information flow in small genetic networks. {IV}. {Spatial} coupling.
\newblock Physical Review E. 2015;91(6):062710.
\newblock doi:{10.1103/PhysRevE.91.062710}.

\bibitem{fancher_fundamental_2017}
Fancher S, Mugler A.
\newblock Fundamental {Limits} to {Collective} {Concentration} {Sensing} in {Cell} {Populations}.
\newblock Physical Review Letters. 2017;118(7):078101.
\newblock doi:{10.1103/PhysRevLett.118.078101}.

\bibitem{yeh_capturing_2021}
Yeh CY, Huang WH, Chen HC, Meir YJJ.
\newblock Capturing {Pluripotency} and {Beyond}.
\newblock Cells. 2021;10(12):3558.
\newblock doi:{10.3390/cells10123558}.

\bibitem{stanoev_robust_2022}
Stanoev A, Koseska A.
\newblock Robust cell identity specifications through transitions in the collective state of growing developmental systems.
\newblock Current Opinion in Systems Biology. 2022;31:100437.
\newblock doi:{10.1016/j.coisb.2022.100437}.

\bibitem{kim_erk1_2014}
Kim SH, Kim MO, Cho YY, Yao K, Kim DJ, Jeong CH, et~al.
\newblock {ERK1} phosphorylates {Nanog} to regulate protein stability and stem cell self-renewal.
\newblock Stem Cell Research. 2014;13(1):1--11.
\newblock doi:{10.1016/j.scr.2014.04.001}.

\bibitem{deathridge_live_2019}
Deathridge J, Antolović V, Parsons M, Chubb JR.
\newblock Live imaging of {ERK} signalling dynamics in differentiating mouse embryonic stem cells.
\newblock Development. 2019;146(12):dev172940.
\newblock doi:{10.1242/dev.172940}.

\bibitem{schardt_adjusting_2023}
Schardt S, Fischer SC.
\newblock Adjusting the range of cell–cell communication enables fine-tuning of cell fate patterns from checkerboard to engulfing.
\newblock Journal of Mathematical Biology. 2023;87(4):54.
\newblock doi:{10.1007/s00285-023-01959-9}.

\bibitem{forsyth_iven_2021}
Forsyth JE, Al-Anbaki AH, Fuente Rdl, Modare N, Perez-Cortes D, Rivera I, et~al.
\newblock {IVEN}: {A} quantitative tool to describe {3D} cell position and neighbourhood reveals architectural changes in {FGF4}-treated preimplantation embryos.
\newblock PLOS Biology. 2021;19(7):e3001345.
\newblock doi:{10.1371/journal.pbio.3001345}.

\bibitem{fischer_salt-and-pepper_2023}
Fischer SC, Schardt S, Lilao-Garzón J, Muñoz-Descalzo S.
\newblock The salt-and-pepper pattern in mouse blastocysts is compatible with signaling beyond the nearest neighbors.
\newblock iScience. 2023;26(11).
\newblock doi:{10.1016/j.isci.2023.108106}.

\bibitem{paulsson_stochastic_2000}
Paulsson J, Berg OG, Ehrenberg M.
\newblock Stochastic focusing: {Fluctuation}-enhanced sensitivity of intracellular regulation.
\newblock Proceedings of the National Academy of Sciences. 2000;97(13):7148--7153.
\newblock doi:{10.1073/pnas.110057697}.

\bibitem{swain_intrinsic_2002}
Swain PS, Elowitz MB, Siggia ED.
\newblock Intrinsic and extrinsic contributions to stochasticity in gene expression.
\newblock Proceedings of the National Academy of Sciences. 2002;99(20):12795--12800.
\newblock doi:{10.1073/pnas.162041399}.

\bibitem{eldar_functional_2010}
Eldar A, Elowitz MB.
\newblock Functional roles for noise in genetic circuits.
\newblock Nature. 2010;467(7312):167--173.
\newblock doi:{10.1038/nature09326}.

\bibitem{tsimring_noise_2014}
Tsimring LS.
\newblock Noise in {Biology}.
\newblock Reports on progress in physics Physical Society (Great Britain). 2014;77(2):026601.
\newblock doi:{10.1088/0034-4885/77/2/026601}.

\bibitem{cuesta_bernoulli_2022}
Cuesta FA, Guerberoff G, Rojo AL.
\newblock Bernoulli and binomial proliferation on evolutionary graphs.
\newblock Journal of Theoretical Biology. 2022;534:110942.
\newblock doi:{10.1016/j.jtbi.2021.110942}.

\bibitem{benzinger_synthetic_2022}
Benzinger D, Ovinnikov S, Khammash M.
\newblock Synthetic gene networks recapitulate dynamic signal decoding and differential gene expression.
\newblock Cell Systems. 2022;13(5):353--364.e6.
\newblock doi:{10.1016/j.cels.2022.02.004}.

\bibitem{frei_genetic_2022}
Frei T, Chang CH, Filo M, Arampatzis A, Khammash M.
\newblock A genetic mammalian proportional–integral feedback control circuit for robust and precise gene regulation.
\newblock Proceedings of the National Academy of Sciences. 2022;119(24):e2122132119.
\newblock doi:{10.1073/pnas.2122132119}.

\bibitem{briat_noise_2023}
Briat C, Khammash M.
\newblock Noise in {Biomolecular} {Systems}: {Modeling}, {Analysis}, and {Control} {Implications}.
\newblock Annual Review of Control, Robotics, and Autonomous Systems. 2023;6(1):283--311.
\newblock doi:{10.1146/annurev-control-042920-101825}.

\bibitem{raina_fgf4_2021}
Raina D.
\newblock {FGF4} drives intermittent oscillations of {ERK} activity in mouse embryonic stem cells [{PhD} {Thesis}].
\newblock Technische Universität Dortmund; 2021.
\newblock Available from: \url{https://eldorado.tu-dortmund.de/handle/2003/40484}.

\bibitem{wang_massive_2019}
Wang S, Fan K, Luo N, Cao Y, Wu F, Zhang C, et~al.
\newblock Massive computational acceleration by using neural networks to emulate mechanism-based biological models.
\newblock Nature Communications. 2019;10(1):4354.
\newblock doi:{10.1038/s41467-019-12342-y}.

\bibitem{tejero-cantero_sbi_2020}
Tejero-Cantero A, Boelts J, Deistler M, Lueckmann JM, Durkan C, Gonçalves PJ, et~al.
\newblock sbi: {A} toolkit for simulation-based inference.
\newblock Journal of Open Source Software. 2020;5(52):2505.
\newblock doi:{10.21105/joss.02505}.

\bibitem{torregrosa_mechanistic_2021}
Torregrosa G, Garcia-Ojalvo J.
\newblock Mechanistic models of cell-fate transitions from single-cell data.
\newblock Current Opinion in Systems Biology. 2021;26:79--86.
\newblock doi:{10.1016/j.coisb.2021.04.004}.

\bibitem{shahbazi_mechanisms_2020}
Shahbazi MN.
\newblock Mechanisms of human embryo development: from cell fate to tissue shape and back.
\newblock Development. 2020;147(14):dev190629.
\newblock doi:{10.1242/dev.190629}.

\bibitem{kar_exploring_2009}
Kar S, Baumann WT, Paul MR, Tyson JJ.
\newblock Exploring the roles of noise in the eukaryotic cell cycle.
\newblock Proceedings of the National Academy of Sciences of the United States of America. 2009;106(16):6471--6476.
\newblock doi:{10.1073/pnas.0810034106}.

\bibitem{ng_stochastic_2018}
Ng KK, Yui MA, Mehta A, Siu S, Irwin B, Pease S, et~al.
\newblock A stochastic epigenetic switch controls the dynamics of {T}-cell lineage commitment.
\newblock eLife. 2018;7:e37851.
\newblock doi:{10.7554/eLife.37851}.

\bibitem{sherman_cell--cell_2015}
Sherman MS, Lorenz K, Lanier MH, Cohen BA.
\newblock Cell-to-{Cell} {Variability} in the {Propensity} to {Transcribe} {Explains} {Correlated} {Fluctuations} in {Gene} {Expression}.
\newblock Cell Systems. 2015;1(5):315--325.
\newblock doi:{10.1016/j.cels.2015.10.011}.

\bibitem{justman_explicit_2015}
Justman QA.
\newblock An {Explicit} {Source} for {Extrinsic} {Noise}.
\newblock Cell Systems. 2015;1(5):308--309.
\newblock doi:{10.1016/j.cels.2015.11.003}.

\bibitem{de_jong_gene_2019}
de~Jong TV, Moshkin YM, Guryev V.
\newblock Gene expression variability: the other dimension in transcriptome analysis.
\newblock Physiological Genomics. 2019;51(5):145--158.
\newblock doi:{10.1152/physiolgenomics.00128.2018}.

\bibitem{bartz_progress_2023}
Bartz J, Jung H, Wasiluk K, Zhang L, Dong X.
\newblock Progress in {Discovering} {Transcriptional} {Noise} in {Aging}.
\newblock International Journal of Molecular Sciences. 2023;24(4):3701.
\newblock doi:{10.3390/ijms24043701}.

\bibitem{gillespie_exact_1977}
Gillespie DT.
\newblock Exact stochastic simulation of coupled chemical reactions.
\newblock The Journal of Physical Chemistry. 1977;81(25):2340--2361.
\newblock doi:{10.1021/j100540a008}.

\bibitem{elf_fast_2003}
Elf J, Ehrenberg M.
\newblock Fast {Evaluation} of {Fluctuations} in {Biochemical} {Networks} {With} the {Linear} {Noise} {Approximation}.
\newblock Genome Research. 2003;13(11):2475--2484.
\newblock doi:{10.1101/gr.1196503}.

\bibitem{munsky_multiple_2007}
Munsky B, Khammash M.
\newblock A multiple time interval finite state projection algorithm for the solution to the chemical master equation.
\newblock Journal of Computational Physics. 2007;226(1):818--835.
\newblock doi:{10.1016/j.jcp.2007.05.016}.

\bibitem{anderson_modified_2007}
Anderson DF.
\newblock A modified next reaction method for simulating chemical systems with time dependent propensities and delays.
\newblock The Journal of Chemical Physics. 2007;127(21):214107.
\newblock doi:{10.1063/1.2799998}.

\bibitem{gillespie_perspective_2013}
Gillespie DT, Hellander A, Petzold LR.
\newblock Perspective: {Stochastic} algorithms for chemical kinetics.
\newblock The Journal of Chemical Physics. 2013;138(17):170901.
\newblock doi:{10.1063/1.4801941}.

\bibitem{simoni_stochastic_2019}
Simoni G, Reali F, Priami C, Marchetti L.
\newblock Stochastic simulation algorithms for computational systems biology: {Exact}, approximate, and hybrid methods.
\newblock Wiley Interdisciplinary Reviews Systems Biology and Medicine. 2019;11(6):e1459.
\newblock doi:{10.1002/wsbm.1459}.

\bibitem{erban_stochastic_2020}
Erban R, Chapman SJ.
\newblock Stochastic {Modelling} of {Reaction}–{Diffusion} {Processes}.
\newblock Cambridge {Texts} in {Applied} {Mathematics}. Cambridge: Cambridge University Press; 2020.
\newblock Available from: \url{https://www.cambridge.org/core/books/stochastic-modelling-of-reactiondiffusion-processes/9BB8B46DE0B898FC019AFBEA95608FAE}.

\bibitem{gupta_deepcme_2021}
Gupta A, Schwab C, Khammash M.
\newblock {DeepCME}: {A} deep learning framework for computing solution statistics of the chemical master equation.
\newblock PLOS Computational Biology. 2021;17(12):e1009623.
\newblock doi:{10.1371/journal.pcbi.1009623}.

\bibitem{coulier_multiscale_2021}
Coulier A, Hellander S, Hellander A.
\newblock A multiscale compartment-based model of stochastic gene regulatory networks using hitting-time analysis.
\newblock The Journal of Chemical Physics. 2021;154(18):184105.
\newblock doi:{10.1063/5.0010764}.

\bibitem{gillespie_rigorous_1992}
Gillespie DT.
\newblock A rigorous derivation of the chemical master equation.
\newblock Physica A: Statistical Mechanics and its Applications. 1992;188(1):404--425.
\newblock doi:{10.1016/0378-4371(92)90283-V}.

\bibitem{andrews_detailed_2010}
Andrews SS, Addy NJ, Brent R, Arkin AP.
\newblock Detailed {Simulations} of {Cell} {Biology} with {Smoldyn} 2.1.
\newblock PLOS Computational Biology. 2010;6(3):e1000705.
\newblock doi:{10.1371/journal.pcbi.1000705}.

\bibitem{gupta_spatial_2018}
Gupta S, Czech J, Kuczewski R, Bartol TM, Sejnowski TJ, Lee REC, et~al.. Spatial {Stochastic} {Modeling} with {MCell} and {CellBlender}; 2018.
\newblock Available from: \url{http://arxiv.org/abs/1810.00499}.

\bibitem{engblom_stochastic_2019}
Engblom S.
\newblock Stochastic {Simulation} of {Pattern} {Formation} in {Growing} {Tissue}: {A} {Multilevel} {Approach}.
\newblock Bulletin of Mathematical Biology. 2019;81(8):3010--3023.
\newblock doi:{10.1007/s11538-018-0454-y}.

\bibitem{sokolowski_egfrd_2019}
Sokolowski TR, Paijmans J, Bossen L, Miedema T, Wehrens M, Becker NB, et~al.
\newblock {eGFRD} in all dimensions.
\newblock The Journal of Chemical Physics. 2019;150(5):054108.
\newblock doi:{10.1063/1.5064867}.

\bibitem{hellander_reaction-diffusion_2012}
Hellander S, Hellander A, Petzold L.
\newblock Reaction-diffusion master equation in the microscopic limit.
\newblock Physical Review E. 2012;85(4):042901.
\newblock doi:{10.1103/PhysRevE.85.042901}.

\bibitem{barrows_parameter_2023}
Barrows D, Ilie S.
\newblock Parameter estimation for the reaction–diffusion master equation.
\newblock AIP Advances. 2023;13(6):065318.
\newblock doi:{10.1063/5.0150292}.

\bibitem{fange_stochastic_2010}
Fange D, Berg OG, Sjöberg P, Elf J.
\newblock Stochastic reaction-diffusion kinetics in the microscopic limit.
\newblock Proceedings of the National Academy of Sciences. 2010;107(46):19820--19825.
\newblock doi:{10.1073/pnas.1006565107}.

\bibitem{plusa_distinct_2008}
Plusa B, Piliszek A, Frankenberg S, Artus J, Hadjantonakis AK.
\newblock Distinct sequential cell behaviours direct primitive endoderm formation in the mouse blastocyst.
\newblock Development. 2008;135(18):3081--3091.
\newblock doi:{10.1242/dev.021519}.

\bibitem{kang_lineage_2017}
Kang M, Garg V, Hadjantonakis AK.
\newblock Lineage {Establishment} and {Progression} within the {Inner} {Cell} {Mass} of the {Mouse} {Blastocyst} {Requires} {FGFR1} and {FGFR2}.
\newblock Developmental Cell. 2017;41(5):496--510.e5.
\newblock doi:{10.1016/j.devcel.2017.05.003}.

\bibitem{molotkov_distinct_2018}
Molotkov A, Soriano P.
\newblock Distinct mechanisms for {PDGF} and {FGF} signaling in primitive endoderm development.
\newblock Developmental Biology. 2018;442(1):155--161.
\newblock doi:{10.1016/j.ydbio.2018.07.010}.

\bibitem{simon_live_2020}
Simon CS, Rahman S, Raina D, Schröter C, Hadjantonakis AK.
\newblock Live {Visualization} of {ERK} {Activity} in the {Mouse} {Blastocyst} {Reveals} {Lineage}-{Specific} {Signaling} {Dynamics}.
\newblock Developmental Cell. 2020;55(3):341--353.e5.
\newblock doi:{10.1016/j.devcel.2020.09.030}.

\bibitem{meng_gata6_2018}
Meng Y, Moore R, Tao W, Smith ER, Tse JD, Caslini C, et~al.
\newblock {GATA6} phosphorylation by {Erk1}/2 propels exit from pluripotency and commitment to primitive endoderm.
\newblock Developmental Biology. 2018;436(1):55--65.
\newblock doi:{10.1016/j.ydbio.2018.02.007}.

\bibitem{liu_usp21_2016}
Liu X, Yao Y, Ding H, Han C, Chen Y, Zhang Y, et~al.
\newblock {USP21} deubiquitylates {Nanog} to regulate protein stability and stem cell pluripotency.
\newblock Signal Transduction and Targeted Therapy. 2016;1(1):1--10.
\newblock doi:{10.1038/sigtrans.2016.24}.

\bibitem{ornitz_fibroblast_2015}
Ornitz DM, Itoh N.
\newblock The {Fibroblast} {Growth} {Factor} signaling pathway.
\newblock WIREs Developmental Biology. 2015;4(3):215--266.
\newblock doi:{10.1002/wdev.176}.

\bibitem{ornitz_new_2022}
Ornitz DM, Itoh N.
\newblock New developments in the biology of fibroblast growth factors.
\newblock WIREs Mechanisms of Disease. 2022;14(4):e1549.
\newblock doi:{10.1002/wsbm.1549}.

\bibitem{karl_ligand_2023}
Karl K, Piccolo ND, Light T, Roy T, Dudeja P, Ursachi VC, et~al.. Ligand bias underlies differential signaling of multiple {FGFs} via {FGFR1}; 2023.
\newblock Available from: \url{https://www.biorxiv.org/content/10.1101/2022.01.06.475273v4}.

\bibitem{lavoie_erk_2020}
Lavoie H, Gagnon J, Therrien M.
\newblock {ERK} signalling: a master regulator of cell behaviour, life and fate.
\newblock Nature Reviews Molecular Cell Biology. 2020;21(10):607--632.
\newblock doi:{10.1038/s41580-020-0255-7}.

\bibitem{aiken_direct_2004}
Aiken CEM, Swoboda PPL, Skepper JN, Johnson MH.
\newblock The direct measurement of embryogenic volume and nucleo-cytoplasmic ratio during mouse pre-implantation development.
\newblock Reproduction. 2004;128(5):527--535.
\newblock doi:{10.1530/rep.1.00281}.

\bibitem{van_zon_diffusion_2006}
van Zon JS, Morelli MJ, Tănase-Nicola S, ten Wolde PR.
\newblock Diffusion of {Transcription} {Factors} {Can} {Drastically} {Enhance} the {Noise} in {Gene} {Expression}.
\newblock Biophysical Journal. 2006;91(12):4350--4367.
\newblock doi:{10.1529/biophysj.106.086157}.

\bibitem{vijaykumar_intrinsic_2017}
Vijaykumar A, Bolhuis PG, Wolde PRt.
\newblock The intrinsic rate constants in diffusion-influenced reactions.
\newblock Faraday Discussions. 2017;195(0):421--441.
\newblock doi:{10.1039/C6FD00104A}.

\bibitem{milo_bionumbersdatabase_2010}
Milo R, Jorgensen P, Moran U, Weber G, Springer M.
\newblock {BioNumbers}—the database of key numbers in molecular and cell biology.
\newblock Nucleic Acids Research. 2010;38(suppl\_1):D750--D753.
\newblock doi:{10.1093/nar/gkp889}.

\bibitem{coulier_systematic_2022}
Coulier A, Singh P, Sturrock M, Hellander A.
\newblock Systematic comparison of modeling fidelity levels and parameter inference settings applied to negative feedback gene regulation.
\newblock PLOS Computational Biology. 2022;18(12):e1010683.
\newblock doi:{10.1371/journal.pcbi.1010683}.

\bibitem{frank_input-output_2013}
Frank SA.
\newblock Input-output relations in biological systems: measurement, information and the {Hill} equation.
\newblock Biology Direct. 2013;8(1):31.
\newblock doi:{10.1186/1745-6150-8-31}.

\bibitem{chen_stochastic_2017}
Chen M, Li F, Wang S, Cao Y.
\newblock Stochastic modeling and simulation of reaction-diffusion system with {Hill} function dynamics.
\newblock BMC Systems Biology. 2017;11(3):21.
\newblock doi:{10.1186/s12918-017-0401-9}.

\bibitem{bottani_hill_2017}
Bottani S, Veitia RA.
\newblock Hill function-based models of transcriptional switches: impact of specific, nonspecific, functional and nonfunctional binding.
\newblock Biological Reviews. 2017;92(2):953--963.
\newblock doi:{10.1111/brv.12262}.

\bibitem{feigelman_stochastic_2016}
Feigelman J.
\newblock Stochastic and deterministic methods for the analysis of {Nanog} dynamics in mouse embryonic stem cells [{PhD} {Thesis}].
\newblock Technische Universität München; 2016.
\newblock Available from: \url{https://mediatum.ub.tum.de/1279519}.

\bibitem{skinner_single-cell_2016}
Skinner SO, Xu H, Nagarkar-Jaiswal S, Freire PR, Zwaka TP, Golding I.
\newblock Single-cell analysis of transcription kinetics across the cell cycle.
\newblock eLife. 2016;5:e12175.
\newblock doi:{10.7554/eLife.12175}.

\bibitem{ohnishi_cell--cell_2014}
Ohnishi Y, Huber W, Tsumura A, Kang M, Xenopoulos P, Kurimoto K, et~al.
\newblock Cell-to-cell expression variability followed by signal reinforcement progressively segregates early mouse lineages.
\newblock Nature cell biology. 2014;16(1):27--37.
\newblock doi:{10.1038/ncb2881}.

\bibitem{tan_brf1_2014}
Tan FE, Elowitz MB.
\newblock Brf1 posttranscriptionally regulates pluripotency and differentiation responses downstream of {Erk} {MAP} kinase.
\newblock Proceedings of the National Academy of Sciences. 2014;111(17):E1740--E1748.
\newblock doi:{10.1073/pnas.1320873111}.

\bibitem{elatmani_rna-binding_2011}
Elatmani H, Dormoy-Raclet V, Dubus P, Dautry F, Chazaud C, Jacquemin-Sablon H.
\newblock The {RNA}-{Binding} {Protein} {Unr} {Prevents} {Mouse} {Embryonic} {Stem} {Cells} {Differentiation} {Toward} the {Primitive} {Endoderm} {Lineage}.
\newblock Stem Cells. 2011;29(10):1504--1516.
\newblock doi:{10.1002/stem.712}.

\bibitem{chitforoushzadeh_tnf-insulin_2016}
Chitforoushzadeh Z, Ye Z, Sheng Z, LaRue S, Fry RC, Lauffenburger DA, et~al.
\newblock {TNF}-insulin crosstalk at the transcription factor {GATA6} is revealed by a model that links signaling and transcriptomic data tensors.
\newblock Science signaling. 2016;9(431):ra59.
\newblock doi:{10.1126/scisignal.aad3373}.

\bibitem{fujioka_dynamics_2006}
Fujioka A, Terai K, Itoh RE, Aoki K, Nakamura T, Kuroda S, et~al.
\newblock Dynamics of the {Ras}/{ERK} {MAPK} {Cascade} as {Monitored} by {Fluorescent} {Probes}*.
\newblock Journal of Biological Chemistry. 2006;281(13):8917--8926.
\newblock doi:{10.1074/jbc.M509344200}.

\bibitem{tian_mathematical_2012}
Tian T, Song J.
\newblock Mathematical {Modelling} of the {MAP} {Kinase} {Pathway} {Using} {Proteomic} {Datasets}.
\newblock PLOS ONE. 2012;7(8):e42230.
\newblock doi:{10.1371/journal.pone.0042230}.

\bibitem{aoki_quantitative_2013}
Aoki K, Takahashi K, Kaizu K, Matsuda M.
\newblock A {Quantitative} {Model} of {ERK} {MAP} {Kinase} {Phosphorylation} in {Crowded} {Media}.
\newblock Scientific Reports. 2013;3(1):1541.
\newblock doi:{10.1038/srep01541}.

\bibitem{busca_erk1_2016}
Buscà R, Pouysségur J, Lenormand P.
\newblock {ERK1} and {ERK2} {Map} {Kinases}: {Specific} {Roles} or {Functional} {Redundancy}?
\newblock Frontiers in Cell and Developmental Biology. 2016;4.

\bibitem{saba-el-leil_redundancy_2016}
Saba-El-Leil MK, Frémin C, Meloche S.
\newblock Redundancy in the {World} of {MAP} {Kinases}: {All} for {One}.
\newblock Frontiers in Cell and Developmental Biology. 2016;4.

\bibitem{zoller_diverse_2018}
Zoller B, Little SC, Gregor T.
\newblock Diverse {Spatial} {Expression} {Patterns} {Emerge} from {Unified} {Kinetics} of {Transcriptional} {Bursting}.
\newblock Cell. 2018;175(3):835--847.e25.
\newblock doi:{10.1016/j.cell.2018.09.056}.

\bibitem{abranches_generation_2013}
Abranches E, Bekman E, Henrique D.
\newblock Generation and {Characterization} of a {Novel} {Mouse} {Embryonic} {Stem} {Cell} {Line} with a {Dynamic} {Reporter} of {Nanog} {Expression}.
\newblock PLOS ONE. 2013;8(3):e59928.
\newblock doi:{10.1371/journal.pone.0059928}.

\bibitem{wu_distinct_2013}
Wu J, Tzanakakis ES.
\newblock Distinct {Allelic} {Patterns} of {Nanog} {Expression} {Impart} {Embryonic} {Stem} {Cell} {Population} {Heterogeneity}.
\newblock PLOS Computational Biology. 2013;9(7):e1003140.
\newblock doi:{10.1371/journal.pcbi.1003140}.

\bibitem{bates_auxin-degron_2021}
Bates LE, Alves MRP, Silva JCR.
\newblock Auxin-degron system identifies immediate mechanisms of {OCT4}.
\newblock Stem Cell Reports. 2021;16(7):1818--1831.
\newblock doi:{10.1016/j.stemcr.2021.05.016}.

\bibitem{ding_half-life_2021}
Ding I, Peterson AM.
\newblock Half-life modeling of basic fibroblast growth factor released from growth factor-eluting polyelectrolyte multilayers.
\newblock Scientific Reports. 2021;11(1):9808.
\newblock doi:{10.1038/s41598-021-89229-w}.

\bibitem{daneshpour_macroscopic_2023}
Daneshpour H, van~den Bersselaar P, Chao CH, Fazzio TG, Youk H.
\newblock Macroscopic quorum sensing sustains differentiating embryonic stem cells.
\newblock Nature Chemical Biology. 2023;19(5):596--606.
\newblock doi:{10.1038/s41589-022-01225-x}.

\bibitem{sarabipour_mechanism_2016}
Sarabipour S, Hristova K.
\newblock Mechanism of {FGF} receptor dimerization and activation.
\newblock Nature Communications. 2016;7(1):10262.
\newblock doi:{10.1038/ncomms10262}.

\bibitem{lanner_role_2010}
Lanner F, Rossant J.
\newblock The role of {FGF}/{Erk} signaling in pluripotent cells.
\newblock Development. 2010;137(20):3351--3360.
\newblock doi:{10.1242/dev.050146}.

\bibitem{azami_regulation_2019}
Azami T, Bassalert C, Allègre N, Valverde~Estrella L, Pouchin P, Ema M, et~al.
\newblock Regulation of the {ERK} signalling pathway in the developing mouse blastocyst.
\newblock Development. 2019;146(14):dev177139.
\newblock doi:{10.1242/dev.177139}.

\bibitem{nies_fibroblast_2016}
Nies VJM, Sancar G, Liu W, van Zutphen T, Struik D, Yu RT, et~al.
\newblock Fibroblast {Growth} {Factor} {Signaling} in {Metabolic} {Regulation}.
\newblock Frontiers in Endocrinology. 2016;6.

\bibitem{grebenkov_full_2019}
Grebenkov DS, Metzler R, Oshanin G.
\newblock Full distribution of first exit times in the narrow escape problem.
\newblock New Journal of Physics. 2019;21(12):122001.
\newblock doi:{10.1088/1367-2630/ab5de4}.

\bibitem{grebenkov_distribution_2021}
Grebenkov DS, Metzler R, Oshanin G.
\newblock Distribution of first-reaction times with target regions on boundaries of shell-like domains.
\newblock New Journal of Physics. 2021;23(12):123049.
\newblock doi:{10.1088/1367-2630/ac4282}.

\bibitem{prescott_efficient_2024}
Prescott TP, Warne DJ, Baker RE.
\newblock Efficient multifidelity likelihood-free {Bayesian} inference with adaptive computational resource allocation.
\newblock Journal of Computational Physics. 2024;496:112577.
\newblock doi:{10.1016/j.jcp.2023.112577}.

\bibitem{deistler_energy-efficient_2022}
Deistler M, Macke JH, Gonçalves PJ.
\newblock Energy-efficient network activity from disparate circuit parameters.
\newblock Proceedings of the National Academy of Sciences. 2022;119(44):e2207632119.
\newblock doi:{10.1073/pnas.2207632119}.

\bibitem{kaiser_simulation-based_2023}
Kaiser J, Stock R, Müller E, Schemmel J, Schmitt S. Simulation-based {Inference} for {Model} {Parameterization} on {Analog} {Neuromorphic} {Hardware}; 2023.
\newblock Available from: \url{http://arxiv.org/abs/2303.16056}.

\bibitem{massonis_distilling_2023}
Massonis G, Villaverde AF, Banga JR. Distilling identifiable and interpretable dynamic models from biological data; 2023.
\newblock Available from: \url{https://www.biorxiv.org/content/10.1101/2023.03.13.532340v2}.

\bibitem{pessoa_accelerating_2023}
Pessoa P, Schweiger M, Sgouralis I, Pressé S.
\newblock Accelerating likelihood calculations for biochemical network discovery.
\newblock Biophysical Journal. 2023;122(3):539a.
\newblock doi:{10.1016/j.bpj.2022.11.2856}.

\bibitem{munsky_using_2012}
Munsky B, Neuert G, van Oudenaarden A.
\newblock Using {Gene} {Expression} {Noise} to {Understand} {Gene} {Regulation}.
\newblock Science. 2012;336(6078):183--187.
\newblock doi:{10.1126/science.1216379}.

\bibitem{pang_probability_2023}
Pang Y, Liang J.
\newblock Probability landscape of a stochastic model of gene expression in single cells through exact solution of chemical master equation.
\newblock Biophysical Journal. 2023;122(3):539a.
\newblock doi:{10.1016/j.bpj.2022.11.2857}.

\bibitem{bonnaffoux_wasabi_2019}
Bonnaffoux A, Herbach U, Richard A, Guillemin A, Gonin-Giraud S, Gros PA, et~al.
\newblock {WASABI}: a dynamic iterative framework for gene regulatory network inference.
\newblock BMC Bioinformatics. 2019;20(1):220.
\newblock doi:{10.1186/s12859-019-2798-1}.

\end{thebibliography}




\end{document}